\documentclass[twocolumn,twocolappendix,tighten]{aastex7}

\usepackage[normalem]{ulem} 
\usepackage{physics,bm}	
\usepackage[mathscr]{euscript}
\usepackage{xcolor}
\usepackage{chemformula} 
\usepackage{multirow}
\shorttitle{Interstellar Polarization Survey V}
\shortauthors{Angarita et al.}
\received{November 11, 2024}
\revised{December 9, 2024}
\accepted{May 28, 2025}

\begin{document}


\title{Interstellar Polarization Survey. V. Galactic magnetic field tomography in the spiral arms
using optical and near-infrared starlight polarization}

\author[orcid=0000-0001-5016-5645,gname=Yenifer,sname=Angarita]{Y.Angarita}
\altaffiliation{Corresponding author}
\affiliation{Department of Astrophysics/IMAPP, Radboud University, PO Box 9010, 6500 GL Nijmegen, The Netherlands}
\affiliation{Department of Space, Earth \& Environment, Chalmers University of Technology, 412 93 Gothenburg, Sweden}
\email[show]{yenifer.angarita@chalmers.se, y.angarita@astro.ru.nl}  

\author[orcid=0000-0003-0400-8846,gname=Jos\'e,sname=Versteeg]{M.J.F. Versteeg} 
\affiliation{Department of Astrophysics/IMAPP, Radboud University, PO Box 9010, 6500 GL Nijmegen, The Netherlands}
\email{J.Versteeg-Veltkamp@astro.ru.nl}

\author[orcid=0000-0002-5288-312X,gname=Marijke,sname=Haverkorn]{M. Haverkorn}
\affiliation{Department of Astrophysics/IMAPP, Radboud University, PO Box 9010, 6500 GL Nijmegen, The Netherlands}
\email{m.haverkorn@astro.ru.nl}

\author[orcid=0000-0002-5053-3847,gname=Vincent,sname=Pelgrims]{V. Pelgrims}
\affiliation{Universit\'e Libre de Bruxelles, Science Faculty CP230, B-1050 Brussels, Belgium}
\email{vincent.pelgrims@ulb.be}

\author[orcid=0000-0002-9459-043X,gname=Claudia,sname=Rodrigues]{C.V. Rodrigues}
\affiliation{Divis\~ao de Astrof\'isica, Instituto Nacional de Pesquisas Espaciais (INPE/MCTI), Av. dos Astronautas, 1758, S\~ao Jos\'e dos Campos, SP, Brazil}
\email{claudia.rodrigues@inpe.br}

\author[orcid=0000-0002-1580-0583,gname=Antonio,sname=Magalh\~aes]{A.M. Magalh\~aes}
\affiliation{IAG, Universidade de S\~ao Paulo, Brazil}
\email{antonio.mario@iag.usp.br}

\author[orcid=0000-0001-6880-4468,gname=Reinaldo,sname=Santos-Lima]{R. Santos-Lima}
\affiliation{IAG, Universidade de S\~ao Paulo, Brazil}
\email{reinaldo.lima@iag.usp.br}

\author[orcid=0000-0001-6099-9539,gname=Koji,sname=Kawabata]{Koji S. Kawabata}
\affiliation{Hiroshima Astrophysical Science Center, Hiroshima University, Kagamiyama, Higashi-Hiroshima, Hiroshima, 739-8526, Japan}
\email{kawabtkj@hiroshima-u.ac.jp}

\begin{abstract}

    Interstellar linear polarization occurs when starlight passes through elongated dust grains aligned by interstellar magnetic fields. The observed polarization can come from different dust structures along the line of sight (LOS). By combining polarization measurements with stellar distances, we can study the plane-of-sky Galactic magnetic field (GMF) between the observer and the star and separate the contributions of clouds with different GMF properties.    
    We used optical and near-infrared (NIR) polarization data from three regions in the Galactic plane ($|b|<1^{\circ}$ and \mbox{$19.\!\!^{\circ}8<l<25.\!\!^{\circ}5$}) to perform a polarization decomposition across the Galactic arms. 
    A comparison between optical and NIR data showed an optical-to-NIR polarization ratio of 2 to 3 along the LOS and a consistent polarization angle across both wavelengths in all studied regions, within measurement uncertainties.
    We applied the Bayesian Inference of Starlight Polarization in one dimension and the Gaussian Mixture Model methods to decompose the polarization in the three regions.
    Optical and NIR observations complemented each other, consistently identifying nearby ($d\lesssim143$~pc), intermediate \mbox{($0.47$~kpc $< d < 1.2$~kpc)}, and distant \mbox{($1.5$~kpc $< d < 2.5$~kpc)} polarizing clouds, in agreement with previous findings in the Local Bubble wall, the Local arm, and the Sagittarius arm dust structures. The results from both polarization decomposition methods agree and complement each other. Polarization tomography revealed significant LOS variations in the plane-of-sky magnetic field orientation in two of the three regions. 
    The relative alignment between the magnetic fields traced by starlight polarization and Planck's polarized thermal dust emission at 353 GHz reaffirmed these variations.

\end{abstract}

\keywords{\uat{Starlight polarization}{1571} --- \uat{Interstellar medium}{847} --- \uat{Galaxy: structure}{622} --- \uat{Interstellar magnetic fields}{845} --- \uat{Milky Way magnetic fields}{1057}}


\section{Introduction} 
\label{sec:intro}

    Linear starlight polarization is produced in the interstellar medium (ISM) when unpolarized light from stars is scattered by elongated dust grains aligned with the Galactic magnetic field (GMF) \citep{Hiltner_1949, Hall_1949}. The direction of the linear polarization indicates the magnetic field orientation on the plane of the sky, while the intensity is associated with the amount of polarizing dust and the magnetic field's inclination relative to the line of sight (LOS) \citep{Serkowski_1962}. Therefore, starlight polarization provides insights into the plane-of-sky magnetic field component and dust properties along the LOS up to the position of the stars.

    The magnetized ISM is a dynamic and heterogeneous environment where the properties of dust and magnetic field vary spatially \citep[e.g.,~see][and references therein]{Ferriere_2015, Haverkorn_2015}. Consequently, the observed starlight polarization may contain contributions of multiple structures characterized by distinct magnetic field and dust properties \citep[see, e.g.,][]{Panopoulou_tomography_2019}. When combined with accurate stellar distance measurements, starlight polarization becomes a powerful tool for localizing and characterizing the different magnetized dust structures found between the observer and the star. 
    
    Starlight polarization has been used alongside stellar distance measurements to study structures in the ISM and large-scale GMF \citep[e.g.,][]{Lloyd_Harwit_1973, Fowler_1974, Ellis_Axon_1978, Heiles_1996, Fosalba_2002, Berdyugin_2014, Versteeg_2023}. For example, \cite{Leroy_1999} combined interstellar dust polarization and Hipparcos parallaxes to estimate the distance to the Local Bubble wall, revealing its irregular shape, with the nearest boundary around 70~pc and the farthest extending beyond 150~pc. More recently, the development of accurate stellar distance catalogs \citep[e.g.,][]{Bailer_Jones_2021, Anders_2022} and three-dimensional (3D) dust extinction maps \citep[e.g.,][]{Lenz_2017, Green_2019, Leike_2020, Lallement_2022, Vergely_2022, Edenhofer_dustmap_2023}, based on precise parallax and photometry observations from the Gaia satellite \citep{Gaia_Collaboration_2018, Gaia_Collaboration_2021b}, has led to advancements in investigating the 3D magnetic field structure in the Galaxy. 
    For instance, \cite{Panopoulou_tomography_2019} demonstrated the feasibility of tomographic reconstruction of the interstellar magnetic field using stellar distance, \ion{H}{1} emission spectra, and $R$-band starlight polarization. 
    
    Investigating the 3D structure of the GMF at both small and large scales enables us to understand the dynamics of important processes in the ISM, such as the formation and evolution of structures \citep[e.g.,~see][]{Pattle_2023}, the propagation of charged particles \citep[e.g.,][]{Beck_review_2003}, and the magnetic field's contributions to turbulence and gas dynamics \citep[e.g.,~see][]{Beresnyak_2019}. Models and observations suggest that the large-scale GMF follows the shape of the spiral arms, thereby aligning nearly parallel to the Galactic thin disk \citep{Beck_review_2003, Ferriere_2015, Haverkorn_2015}, while our understanding of the GMF in interarm regions remains limited. Starlight polarization data generally confirms this alignment \citep{Fosalba_2002, Nishiyama_2009, Clemens_GPIPS-DR4_2020} and can test large-scale GMF models \citep{Pavel_2012, Pavel_2014}. Moreover, starlight polarization orientations probe the local plane-of-sky direction of the GMF along the Galactic arms \citep{Berdyugin_1995, Heiles_1996}. On smaller scales, the GMF properties vary significantly, particularly in regions of high column density \citep{Soler_2016, Pattle_2023}. However, in the nearby diffuse ISM, the small-scale magnetic field appears to align with the structure of the large-scale GMF \citep{Angarita_2024}. Furthermore, polarization efficiency analyses further reveal that the turbulent, or random, component of the magnetic field may dominate over the regular, ordered component \citep{Hatano_2013}.
    
    Many polarization tomography techniques have been developed, providing insights into the 3D structure of the GMF at different scales \citep[e.g.,][]{Pavel_2014, Medan_Andersson_2019, Panopoulou_tomography_2019, Pelgrims_2023, Pelgrims_2024, Doi_2024, Versteeg_2024}. For example, \cite{Doi_2024} probed multiple polarizing clouds in the Sagittarius arm between $1.23$ and $2.23$~kpc, using $R$-band polarization. The study revealed different magnetic field orientations deviating notably from the direction parallel to the Galactic thin disk (i.e.,~$90\degr$). Furthermore, \cite{Andersson_Potter_CoalSack_2005} and \cite{Versteeg_2024} examined the magnetic field properties of the Coalsack region, located about $200$~pc away, using optical polarization. They found that, in addition to the polarization caused by the Coalsack cloud, the farthest stars revealed a second polarizing cloud roughly between $1.3$ and $1.5$~kpc, likely in the Carina-Sagittarius spiral arm. Finally, \cite{Uppal_2024} found signatures of the Local arm, the Perseus arm, and the Outer arm using optical linear polarization data of open clusters toward the Galactic anticenter.
    
    In this work, we present a tomographic analysis of the GMF across the spiral arms, from the Local Bubble wall to the Sagittarius arm. To this end, we use optical and near-infrared (NIR) starlight polarization observations from the Interstellar Polarization Survey (IPS, \citealt{Magalhaes_2005}) and the Galactic Plane Infrared Polarization Survey (GPIPS, \citealt{Clemens_GPIPS_2012}), respectively. In Section~\ref{sec:observ_data}, we present the polarization catalogs and ancillary data, such as distance and three-dimensional dust maps, used for the analysis. In Section~\ref{sec:pol_tomography}, we introduce the polarization decomposition methods used in the tomography analysis. We describe our results in Section~\ref{sec:results}, starting with a comparison between optical and NIR polarization observations, followed by the results of the magnetic field tomography. We discuss these results in Section~\ref{sec:discu}, comparing them with previous findings in similar regions, and present our conclusions and summary in Section~\ref{sec:summary}.

\section{Polarization and ancillary data} 
\label{sec:observ_data}

\begin{figure*}[ht!]
        \epsscale{1.18}
        \plotone{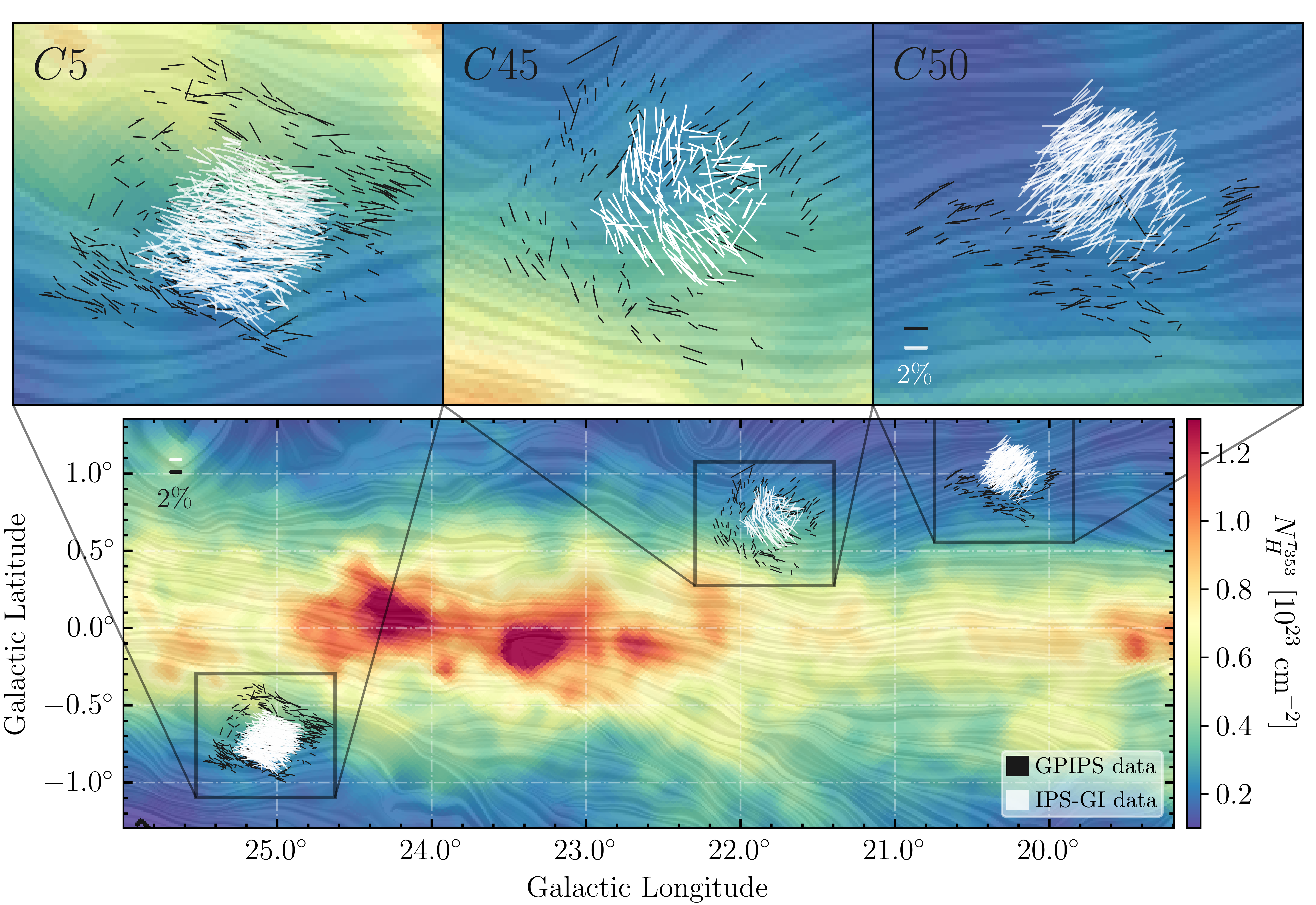}
        \caption{
        Starlight polarization observations from IPS-GI (in white) and GPIPS (in black) catalogs, after filtering, in the regions \textit{C5}, \textit{C45}, and \textit{C50}. The colored background is the column density obtained using the thermal dust optical depth maps at 353~GHz from \cite{Planck-Collaboration_2016}. The white texture in the background represents the plane-of-sky GMF orientation as observed by Planck \citep{Planck-Collaboration_2018_20}, and it is generated using line integral convolution (LIC, \citealt{Cabral_LIC_1993}).
        \label{fig:FoV_data}}
    \end{figure*}

    We use three regions, identified as \textit{C5}, \textit{C45}, and \textit{C50}, from the IPS observations (\citealt{Magalhaes_2005} and A.~M.~Magalh\~aes et al. 2025, in preparation), located within the Galactic thin disk at $-1\fdg0 < b < 1\fdg3$ and $20\fdg0 < l < 25\fdg5$ (Figure~\ref{fig:FoV_data}). These regions, with an area of $0\fdg3\times0\fdg3$ each, were chosen for their low Galactic latitude, their proximity in the sky, and the large number of objects with optical and NIR polarization observations, all of which facilitate the multiwavelength polarization study of the Galactic arms. The following sections describe the starlight polarization observations in the $V$-band and $H$-band used in our study, along with ancillary data. Additionally, we outline the quality filters and pre-processing steps employed to prepare our datasets for analysis. 
    %

    \subsection{Optical polarization data}  
    \label{subsec:data_IPS-GI}

        The $V$-band polarization data were extracted from the IPS-GI catalog \citep{Versteeg_2023}. The $V$-band photometric and polarimetric parameters were processed using the SOLVEPOL pipeline \citep{Ramirez_2017}. The catalog includes relevant parameters such as the 84th, 50th, and 16th percentiles of the optical extinction posterior distributions \citep{Anders_2022}, and the unique Gaia-EDR3 source identifier \citep{Gaia_Collaboration_2021b}.

        We kept optical polarization measurements with a signal-to-noise ratio (SNR) above three, as we aim to retain nearby low-polarization values often discarded by higher SNR cuts \citep{Versteeg_2023, Angarita_2023}. At SNR $> 3$, the degree of polarization measured may be affected by Ricean bias \citep[e.g.,][]{Ramirez_2017}. Therefore, we debiased the polarization degree with the generalized modified asymptotic estimator of \cite{Plaszczynski_2014},
        \begin{equation}
        \label{eq:PV_debias}
            p_{\mathrm{v},d} = p_\mathrm{v} - \sigma_{p_\mathrm{v}}^2 \frac{1 - e^{-p_\mathrm{v}^2/\sigma_{p_\mathrm{v}}^2}}{2p_\mathrm{v}}~,
        \end{equation}
        where $p_{\mathrm{v},d}$ is the debiased polarization, $p_\mathrm{v}$ is the observed polarization degree, and $\sigma_{p_\mathrm{v}}$ is its uncertainty.
        
        We do not correct for the estimated instrumental polarization of $0.07\%$ since it is below the median fractional polarization error measured \citep[see][for more details]{Versteeg_2023}. Furthermore, following the quality filters suggested by \cite{Versteeg_2023}, we excluded stars with $\sigma_{p_\mathrm{v}}>0.8\%$ to avoid untrustworthy polarization values. Additionally, we discarded the duplicates generated during the cross-match with the Gaia-EDR3 catalog \citep{Gaia_Collaboration_2021b}; see details in \cite{Versteeg_2023}. The optical polarization maps of each region are presented in Figure~\ref{fig:FoV_data} (the white pseudo-vectors). 

    \subsection{NIR polarization data}  
    \label{subsec:data_GPIPS}
    
        We use NIR polarization observations in the $H$-band ($1.6~\mu$m) from the GPIPS data release 4 (DR4) presented by \cite{Clemens_GPIPS-DR4_2020}. We queried the GPIPS catalog through the TAP VizieR service\footnote{\url{https://tapvizier.u-strasbg.fr/adql/?J/ApJS/249/23}}.~We selected rectangular regions centered in the IPS-GI fields, each with twice the size of the IPS-GI field of view, i.e.,~\mbox{$0\fdg6\times0\fdg6$}. GPIPS observations surveyed Galactic latitudes between $-1^\circ$ and $+1^\circ$. As a result, the IPS-GI region \textit{C50} is covered only partially by the NIR polarization catalog. 

        The $H$-band polarization degree is already debiased in \cite{Clemens_GPIPS-DR4_2020} following 
        \begin{equation}
        \label{eq:PH_debias}
            p_{\mathrm{h},d} = \sqrt{p_{\mathrm{h}}^2 - \sigma_{p_{\mathrm{h}}}^2}~, 
        \end{equation}
        where $\sigma_{p_{\mathrm{h}}}$ is the polarization degree uncertainty. We kept the NIR polarization SNR above three. Additionally, we filtered the GPIPS dataset by the magnitude and polarization uncertainty, retaining bright stars with $m_{\mathrm{h}} < 12.5$ mag and $\sigma_{p_{\mathrm{h}}} < 2\%$, as suggested by \cite{Clemens_GPIPS-DR4_2020}. 
        
        We converted the normalized Stokes parameters $q_{\mathrm{h}}^{eq} = (Q_{\mathrm{h}}/I_{\mathrm{h}})_{eq}$ and $u_{\mathrm{h}}^{eq} = (U_{\mathrm{h}}/I_{\mathrm{h}})_{eq}$ in the Equatorial reference frame to the International Astronomical Union (IAU) Galactic reference frame, i.e.,~$q_{\mathrm{h}}^{gal}$ and $u_{\mathrm{h}}^{gal}$, to facilitate the analysis and comparison with optical polarization. From here on, we will refer to the fractional Stokes parameters in Galactic coordinates as $q_{\mathrm{h}}$ and $u_{\mathrm{h}}$ for simplicity. The NIR polarization maps of each region are presented in Figure~\ref{fig:FoV_data} (the black pseudo-vectors).

    \subsection{Parallax and photo-geometric distance}  
    \label{subsec:data_distance}

        We cross-matched the IPS-GI and GPIPS datasets separately with the Gaia-EDR3 catalog \citep{Gaia_Collaboration_2021b} to obtain the renormalized unit weight error (RUWE, which indicates the goodness of the astrometric solution), the parallax ($\varpi$), and photometry parameters, along with their uncertainties. We employ Gaia's parameters to define quality filters when using parallax measurements. We also cross-matched the individual polarization datasets with the \cite{Bailer_Jones_2021} distance catalog to obtain the median stellar photo-geometric distance, including the 84th (high) and 16th (low) posterior percentiles. 
        
        The catalogs were cross-matched using the unique Gaia-EDR3 source identifier. In the case of the GPIPS dataset, we also used the precomputed cross-matched catalog available in Gaia's archive to relate the source identifier of Gaia-DR2 to Gaia-EDR3. The reliability of Gaia parameters, including the photo-geometric distance, is assessed with the parallax SNR, $\varpi/\sigma_{\varpi}$, and the RUWE value associated with each Gaia source.  We adopted $\varpi/\sigma_{\varpi} > 2$ and RUWE $< 1.4$, as suggested in \cite{Gaia_Collaboration_2021b}, \cite{GaiaDR3_Collaboration_2023} and \cite{Bailer_Jones_2021}.

    \subsection{Cross-matching polarization datasets}  
    \label{subsec:data_cross-match}

        Next, we cross-matched the IPS-GI datasets with the GPIPS datasets using the unique Gaia-EDR3 source identifier. We relaxed the NIR polarization SNR filter to two in \textit{C45} and \textit{C50} due to the limited number of common stars available for analysis. The cross-matched dataset ended up with 47, 26, and 12 measurements in the regions of \textit{C5}, \textit{C45}, and \textit{C50}, respectively, after filtering. The small number of common stars in \textit{C50} is due to GPIPS only partially covering the IPS-GI region. Therefore, the statistical results are less robust in this area than in the other two regions, which have more measurements.   

    \subsection{Non-ISM polarization}  
    \label{subsec:data_nonISM_pol}

        Despite the quality filters applied (Sections~\ref{subsec:data_IPS-GI} and~\ref{subsec:data_GPIPS}), atypical polarization measurements may remain. These observations may be caused by intrinsically polarized sources. Additionally, in the case of IPS-GI data, odd polarization measurements might be due to mismatches or superposition of the ordinary and extraordinary images created by the Savart prism in a crowded field of view \citep{Magalhaes_2005, Versteeg_2023}.
        
        %
        \begin{figure*}[htb]
            \epsscale{1.18}
            \plotone{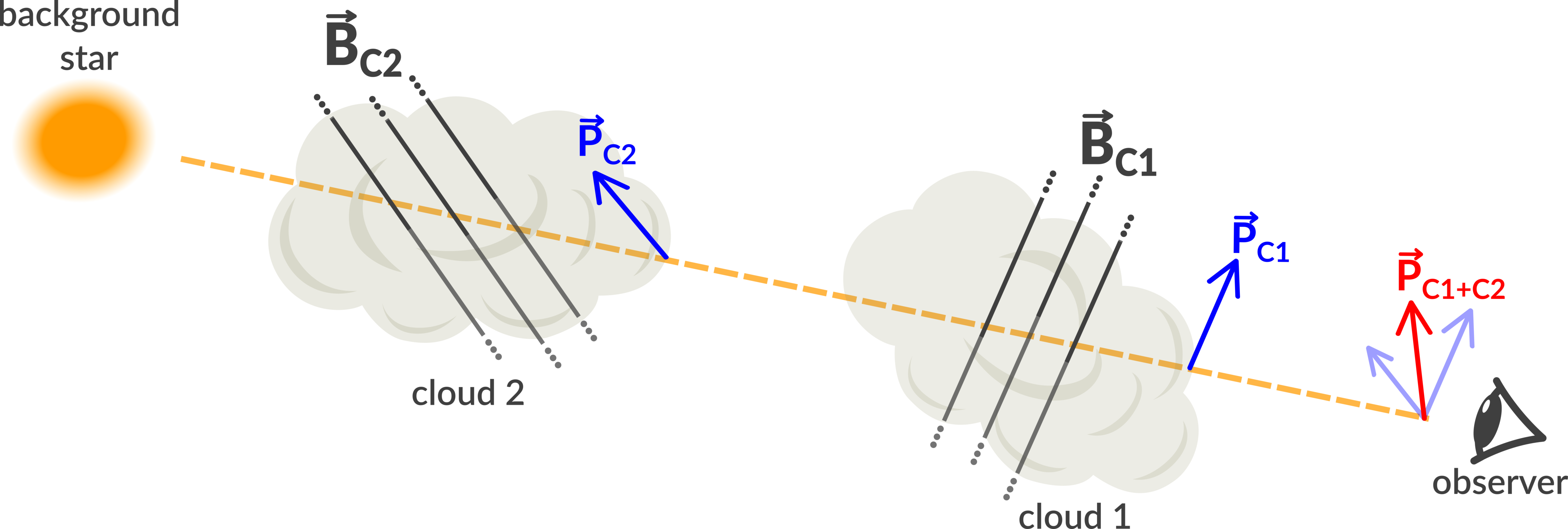}
            \caption{Illustration of the observed starlight polarization produced by two clouds with different magnetic field properties.
            \label{fig:pol_tomo_cartoon}}
        \end{figure*}

        To remove potential non-ISM polarization, we analyzed the correlation of the distributions of the normalized Stokes parameters $q$ and $u$ with distance and performed a sigma clipping on atypical values. We adopted the method from \citealt[][]{Pelgrims_2024} as follows. First, we defined a set of nearby neighbors for each data point in a 3D spatial reference frame, typically between 10 and 15, depending on the number of measurements available in the region. Then, we computed the mean of the $q$ and $u$ measurements and the covariance matrix from their uncertainties, respectively, for each group of neighbors. We did not consider distance uncertainties here. Subsequently, we evaluated the deviations of each data point from the neighborhood mean value, weighting by the uncertainties (Equation~(8) from \citealt{Pelgrims_2024}). Finally, we removed values with deviations above a $3\sigma$ threshold.   
        
        In the data processing for the decomposition with BISP-1 (Section~\ref{subsec:tomo_BISP1}), we utilized the inverse parallax in the sigma clipping, while for the clustering analysis (Section~\ref{subsec:tomo_GMM}), we employed photo-geometric distances. Furthermore, we did not apply sigma clipping to the cross-matched polarization datasets (Section~\ref{subsec:data_cross-match}) due to the limited number of common stars in each region.

    \subsection{3D dust maps of the local Galaxy}  
    \label{subsec:data_dust_maps}

        We use three-dimensional (3D) dust maps from \cite{Vergely_2022} to track and compare the possible origin of the starlight polarization within large-scale dust structures along the sightlines. We retrieved the 3D maps through the G-Tomo platform\footnote{\url{https://explore-platform.eu/sdas}} using the central coordinates of the IPS-GI fields. Since GPIPS data span a wide range of distances, we use 3D dust maps with resolutions of $10$~pc, $25$~pc, and $50$~pc, extending to radii of ${\sim}1.5$~kpc, ${\sim}3$~kpc, and ${\sim}5$~kpc, respectively. The low-resolution maps often contain significant uncertainties, especially at large distances. Therefore, the dust structures interpretation must be done with caution.

\section{Starlight polarization tomography} 
\label{sec:pol_tomography}

    The starlight polarization carries information on the plane-of-sky magnetic field component and dust properties of the ISM between the observer and the star \citep{Hiltner_1949, Hall_1949, Serkowski_1962}. This signal can include contributions from distinct dust structures with varying magnetic field properties, as illustrated in Figure~\ref{fig:pol_tomo_cartoon}. The variations in magnetic field and dust properties along the LOS are reflected in changes in the normalized Stokes parameters $q=Q/I$ and $u=U/I$, as a function of distance \citep[e.g., see][]{Panopoulou_tomography_2019, Pelgrims_2023, Doi_2024}. 
    In the following section, we present two methods that use stellar polarization and distance information to localize and characterize the main polarizing screens along the LOS.

    \subsection{Bayesian approach to find polarizing screens}  
    \label{subsec:tomo_BISP1}

        The Bayesian Inference of Starlight Polarization in one dimension (BISP-1,\footnote{\url{https://github.com/vpelgrims/BISP_1/}} \citealt{Pelgrims_2023}) performs a tomographic decomposition of the plane-of-sky magnetic field properties in the ISM along the LOS. BISP-1 models dust clouds as thin polarizing layers that independently contribute to the total observed polarization of background stars. Thus, it looks for changes in the Stokes parameters $q$ and $u$ as a function of distance obtained from the parallax. 
        
        This Bayesian method considers all sources of uncertainty in the data and modeling. It accounts for uncertainties in stellar parallax, assessing the probability of stars being in front or behind a polarizing cloud, thereby determining its polarization properties \citep{Pelgrims_2023}. To this end, it uses the Gaussian nature of the parallax error. Furthermore, the approach considers the polarization covariance matrix $\Sigma_i$ of the $i$-th measurement as the combination of the intrinsic scatter, $C_{\mathrm{int}}$, primarily attributed to ISM turbulence, and the observational errors, $C_{\mathrm{obs}}$. If the star is in the foreground of all the clouds, then the polarization is assumed to be zero, and the uncertainty is only due to the observational error. 
    
        We tested models with one to five polarizing layers for each dataset considered. We used the Akaike Information Criterion (AIC, \citealt{Akaike_1998}) to identify the model that minimizes the loss of information. The AIC uses the maximized value of the likelihood function and the number of parameters to balance the goodness of the model fit with the complexity of the model, penalizing complex models to avoid overfitting. 
        
        The solution provided posterior distributions of the distance, Stokes parameters, and elements of the covariance ($C_{\mathrm{int}}$) for each cloud, obtained after $40{,}000$ nested sampling iterations \citep[see][for more details]{Pelgrims_2023}.
        We calculated the polarization degree and orientation posteriors from the cumulative posterior of $q_{(\mathrm{v,h})}$ and $u_{(\mathrm{v,h})}$.  
        From these posteriors and their spread, we derive the mean and median values, along with uncertainties, for the distance, polarization degree, and orientation.

        BISP-1 is a tool developed primarily for polarization tomography analysis at high Galactic latitudes, where the ISM is usually diffuse \citep[e.g.,][]{Lenz_2017}, and the majority of the polarization is created in the nearby ISM \citep[e.g.,~see][]{Berdyugin_2014, Alves_2018}. Therefore, the use of the inverse parallax as a proxy for the stellar distance is justified if the low-extinction stars have high parallax SNR and their measured Gaia parallaxes are positive \citep{Bailer_Jones_2021, GaiaDR3_Fouesneau_StellarParams_2023}. However, as we move closer to the Galactic thin disk, the inverse parallax becomes a poor distance estimate, especially for distant stars with high column density, whose parallax SNR is often low (i.e.,~$d\gtrsim 2$~kpc, see Section~\ref{subsec:Disc_dist_Model}). Consequently, the modeled distances of the clouds, derived using stellar parallax and polarization observations in the Galactic thin disk, are often overestimated and should be interpreted carefully. In Section~\ref{subsec:Disc_dist_Model}, we propose a method to correct \mbox{BISP-1's} distance estimates in our samples. The correction is based on models of the inverse parallax-distance relation generated using our observations and photo-geometric distance measurements from \cite{Bailer_Jones_2021}. 
        
        Furthermore, we show in Appendix~\ref{sec:Append_A} that faint red stars (those with $G>18$~mag) have spurious parallax values. We give some possible explanations for this issue that affects only NIR observations. To avoid these unreliable measurements, we discard faint red stars with $G>18$~mag for the polarization tomography analysis with BISP-1 using GPIPS data.        
        
    \subsection{Gaussian Mixture Models}  
    \label{subsec:tomo_GMM}

        The Gaussian Mixture Model (GMM, \citealt{McLachlan_1988}) is a versatile probabilistic method used to represent a dataset as a combination of a finite number of Gaussian distributions. The method estimates the mean, covariance, and mixing coefficients that best fit the data using the Expectation-Maximization (EM) algorithm\footnote{EM is an iterative method to find (local) maximum likelihood estimates of parameters in statistical models.}. GMMs can be particularly useful when the underlying data distribution is not well-defined or contains overlapping clusters.  

        We used the Gaussian mixture method, implemented through the scikit-learn Python library \citep{Pedregosa_Scikitlearn_2011}, with optical and NIR starlight polarization, separately.
        We use the photo-geometric distance and the normalized Stokes parameters $q_{(\mathrm{v,h})}$ and $u_{(\mathrm{v,h})}$ as the principal clustering features to find groups of sightlines sharing magnetic field properties. The three variables are normalized with the MinMax scaler tool\footnote{\url{https://scikit-learn.org/stable/modules/generated/sklearn.preprocessing.MinMaxScaler.html}}.        
        GMM relies solely on these observables to identify groups, ignoring the uncertainties. Then, it assigns a label and membership probability to each element in the dataset. The labels identify group members, allowing us to estimate the group's mean polarization properties and roughly determine the location of the polarizing clouds. This approach has been successfully applied by \cite{Versteeg_2024} toward the Coalsack region, demonstrating multiple polarizing components along the LOS.  
        
        We tested four types of covariance (i.e.,~four mixture models): full, tied, diagonal, and spherical. These options offer flexibility in modeling the data. Their differences are described in Appendix~\ref{sec:Append_C}. The selection of the model and the number of Gaussian components is eased by the Bayesian Information Criterion (BIC, \citealt{Schwarz_1978}). The BIC is a statistical metric used for model selection among a group of potential models. It balances model fit and complexity by penalizing models with a higher component number. The lower BIC score represents the model that best captures the data patterns while avoiding overfitting. The number of clusters and the covariance model that minimizes the BIC in each case are presented in Table~\ref{tab:GMM_params} (Appendix~\ref{sec:Append_C}). We ran the scikit-learn GMM algorithm\footnote{\url{https://scikit-learn.org/stable/modules/mixture.html}} with the parameters of Appendix~\ref{sec:Append_C} and the default Kmeans method to initialize the weights, the means, and the covariance.
        %
        \begin{figure*}[htb]
            \epsscale{2.38}
            \plottwo{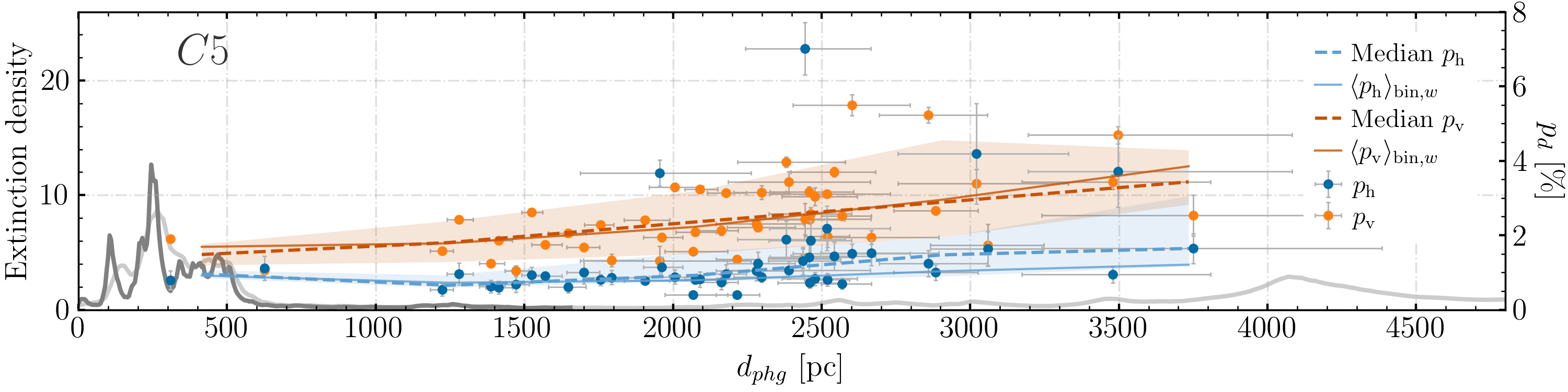}{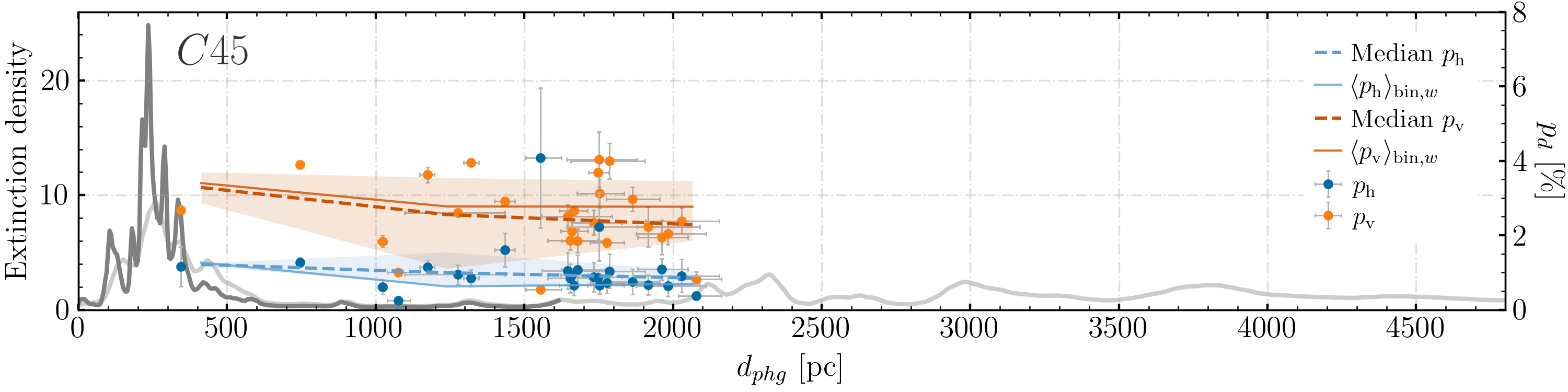}
            \epsscale{1.19}
            \plotone{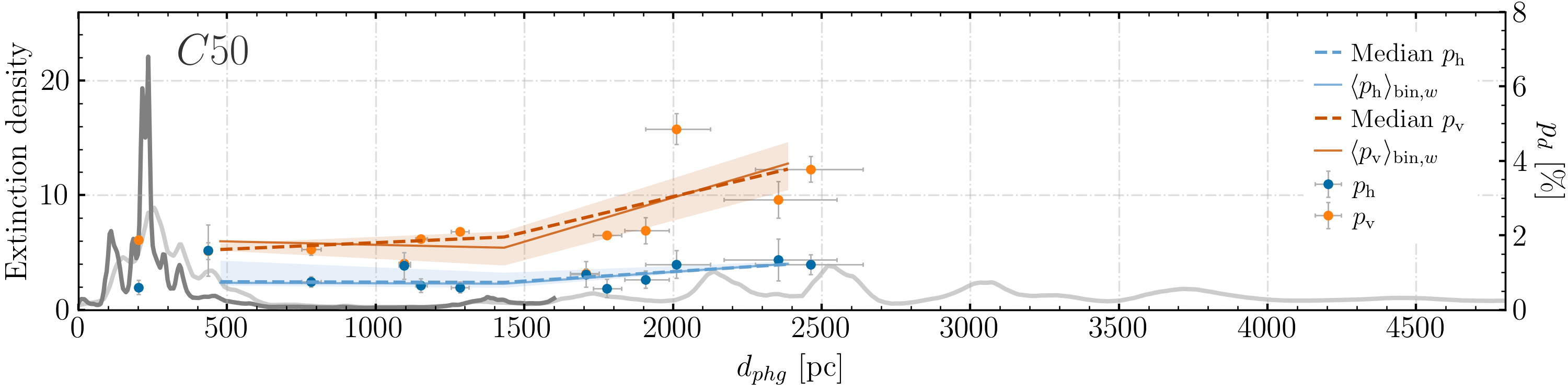}
            \epsscale{0.377}
            \plotone{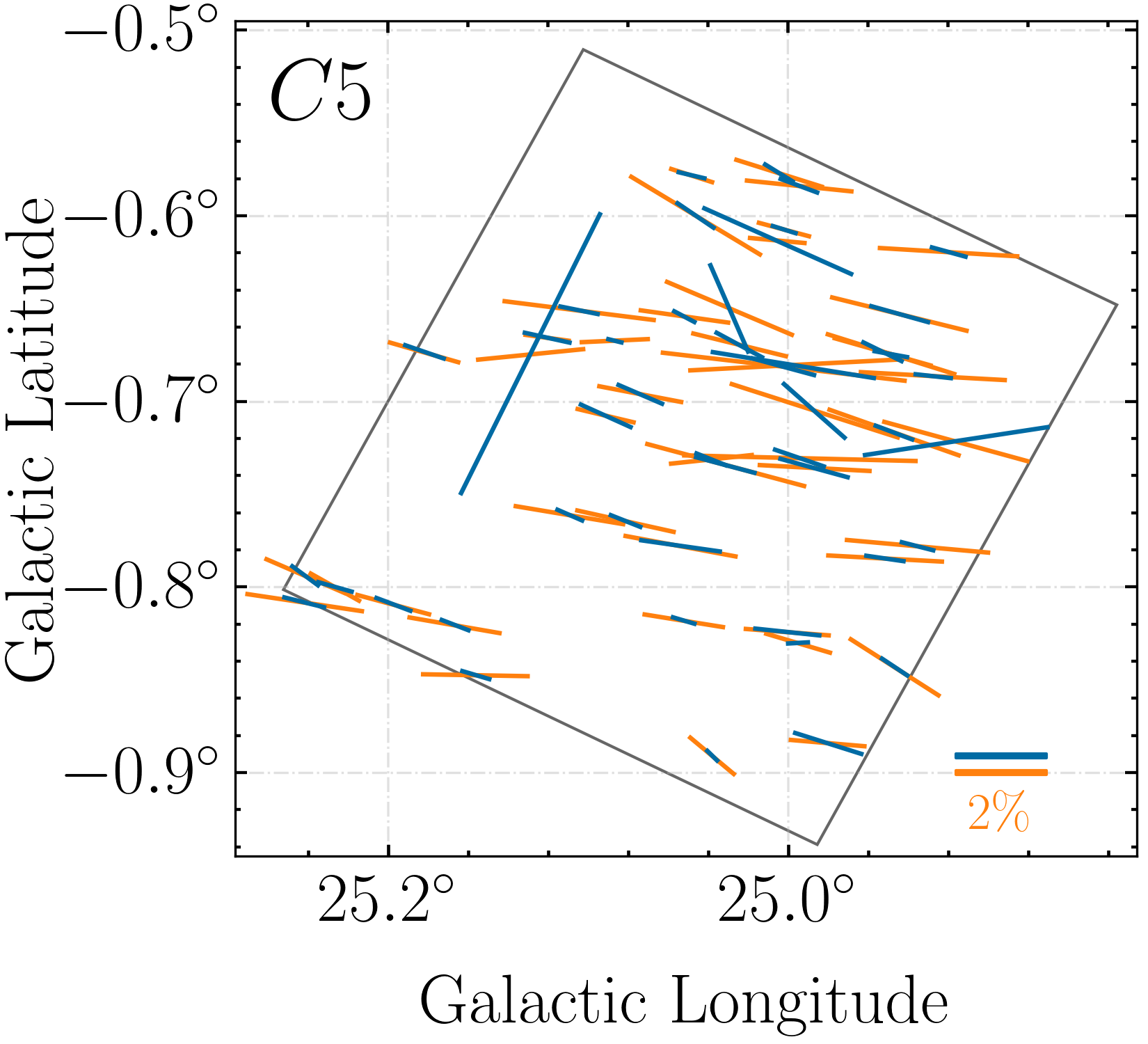}
            \epsscale{0.377}
            \plotone{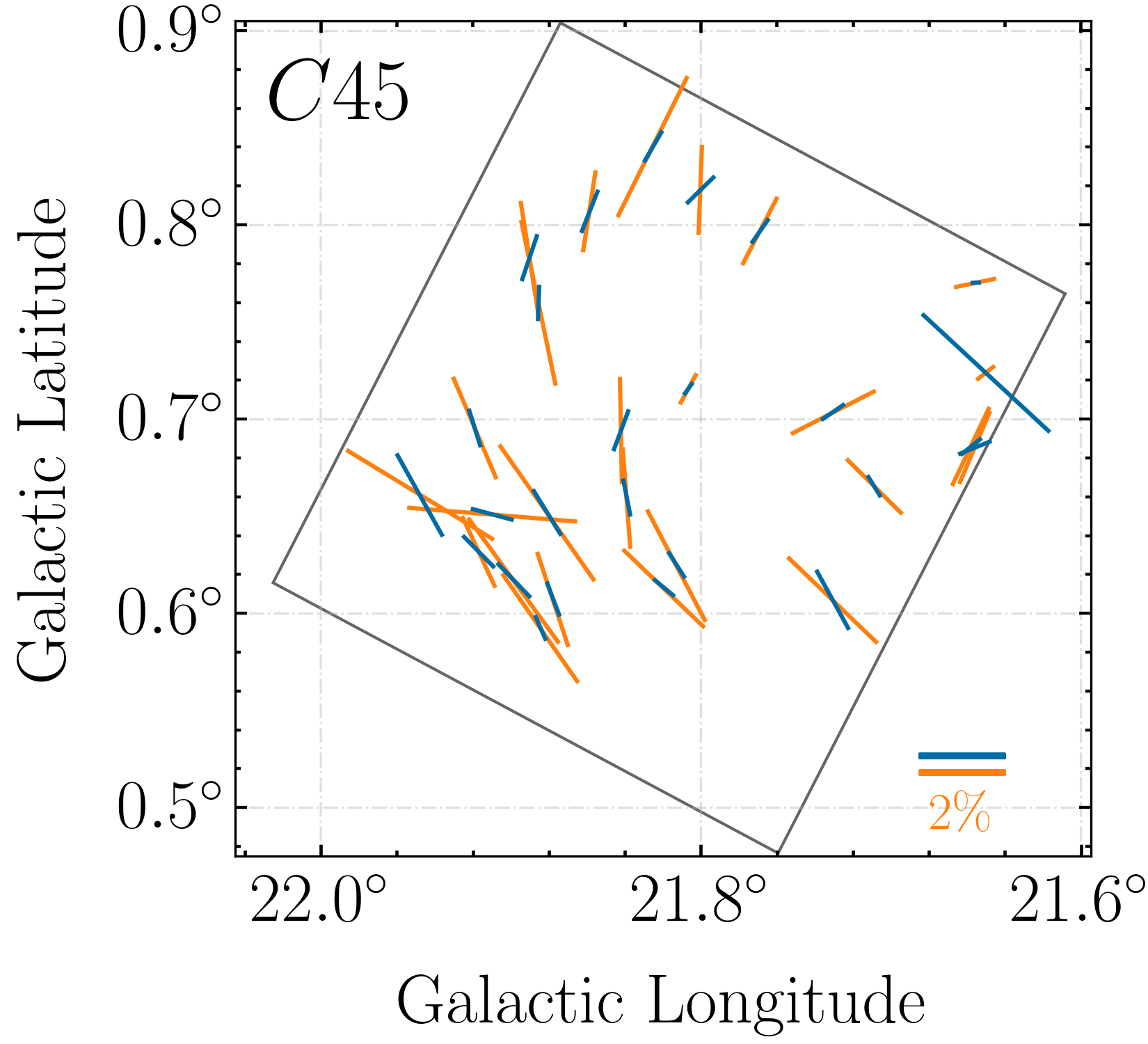}
            \epsscale{0.377}
            \plotone{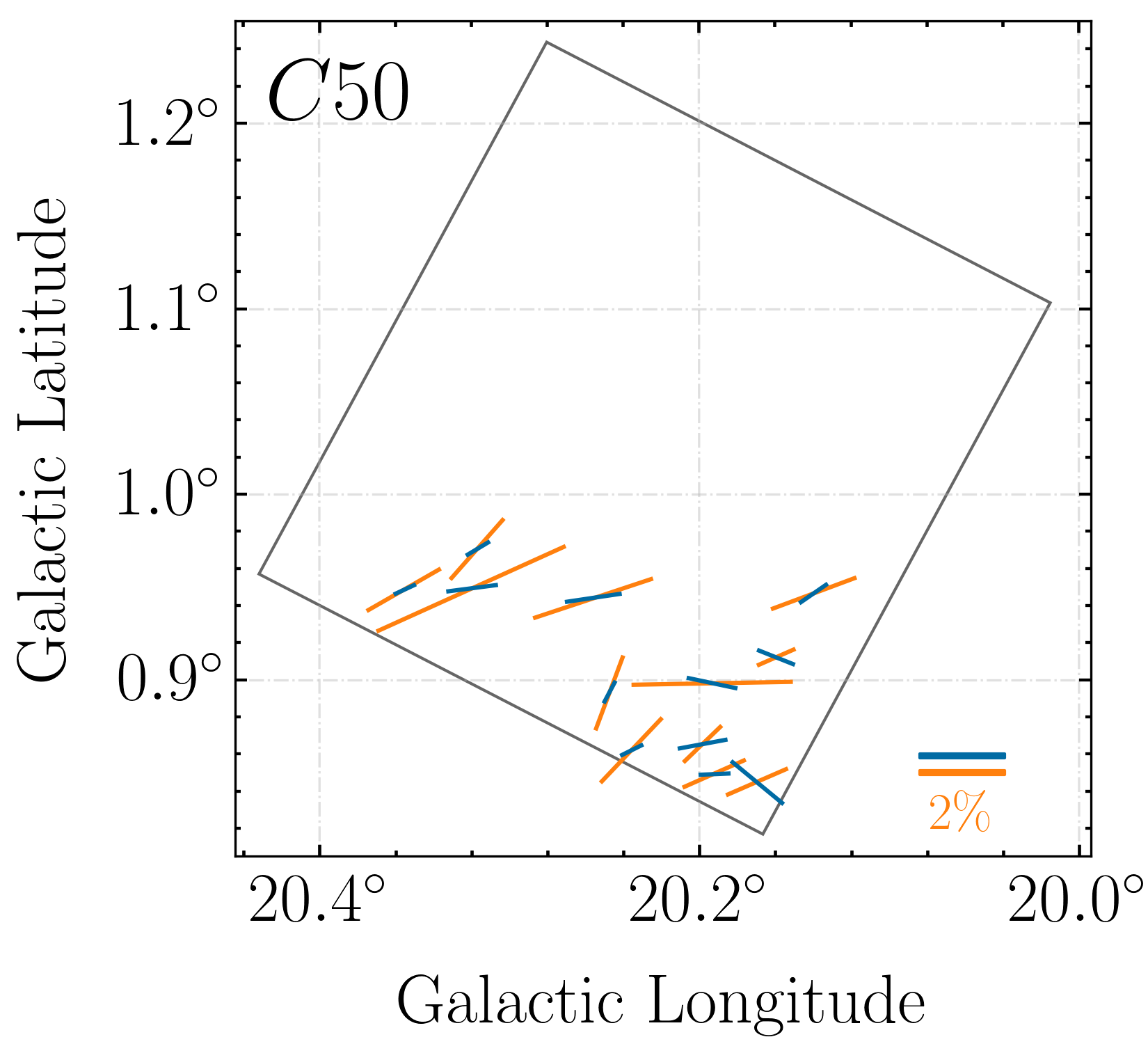}
            \caption{Optical (the orange dots) and NIR (the blue dots) debiased polarization degree as a function of the distance for stars common to both surveys in regions \textit{C5}, \textit{C45}, and \textit{C50} (from top to bottom, respectively). The dashed orange and blue lines show the median polarization degree, and the solid lines show the weighted mean per distance bin. The shaded areas represent the $68\%$ confidence intervals. The polarization (right axis) is compared with optical extinction per parsec ($A_\mathrm{v}$/pc, left axis) from \cite{Vergely_2022}, at $10$~pc (the solid gray curve) and $50$~pc (the solid light-gray curve) resolution. The extinction density units are \mbox{$10^{-3}$~mag pc$^{-1}$}. Bottom row: polarization maps for stars common to IPS-GI (orange) and GPIPS (blue) in the same regions. The gray squares mark the approximate limits of the IPS-GI field of view.
            \label{fig:ips_gpips_Av_dens_dist_pol}}
        \end{figure*}

\section{Results} 
\label{sec:results}

    \subsection{A one-to-one comparison between optical and NIR polarization measurements}  
    \label{subsec:res_V_vs_IR} 

        We present a direct comparison between optical and NIR measurements of stars that are common to both the IPS-GI and GPIPS datasets. To this end, we use the cross-matched polarization datasets (Section~\ref{subsec:data_cross-match}).
        
        %
        \begin{figure*}[htb!]
            \epsscale{0.377}
            \plotone{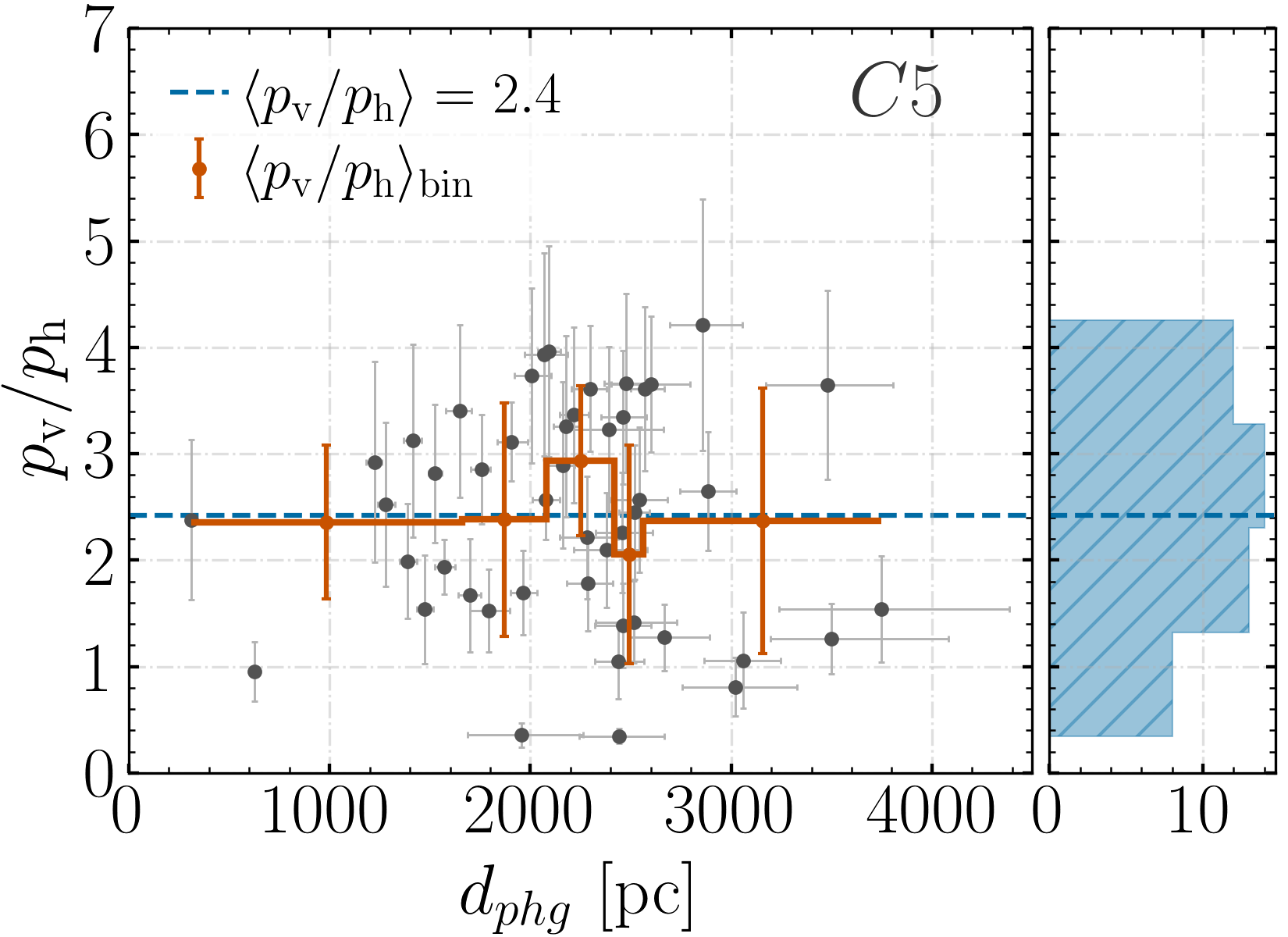}
            \epsscale{0.377}
            \plotone{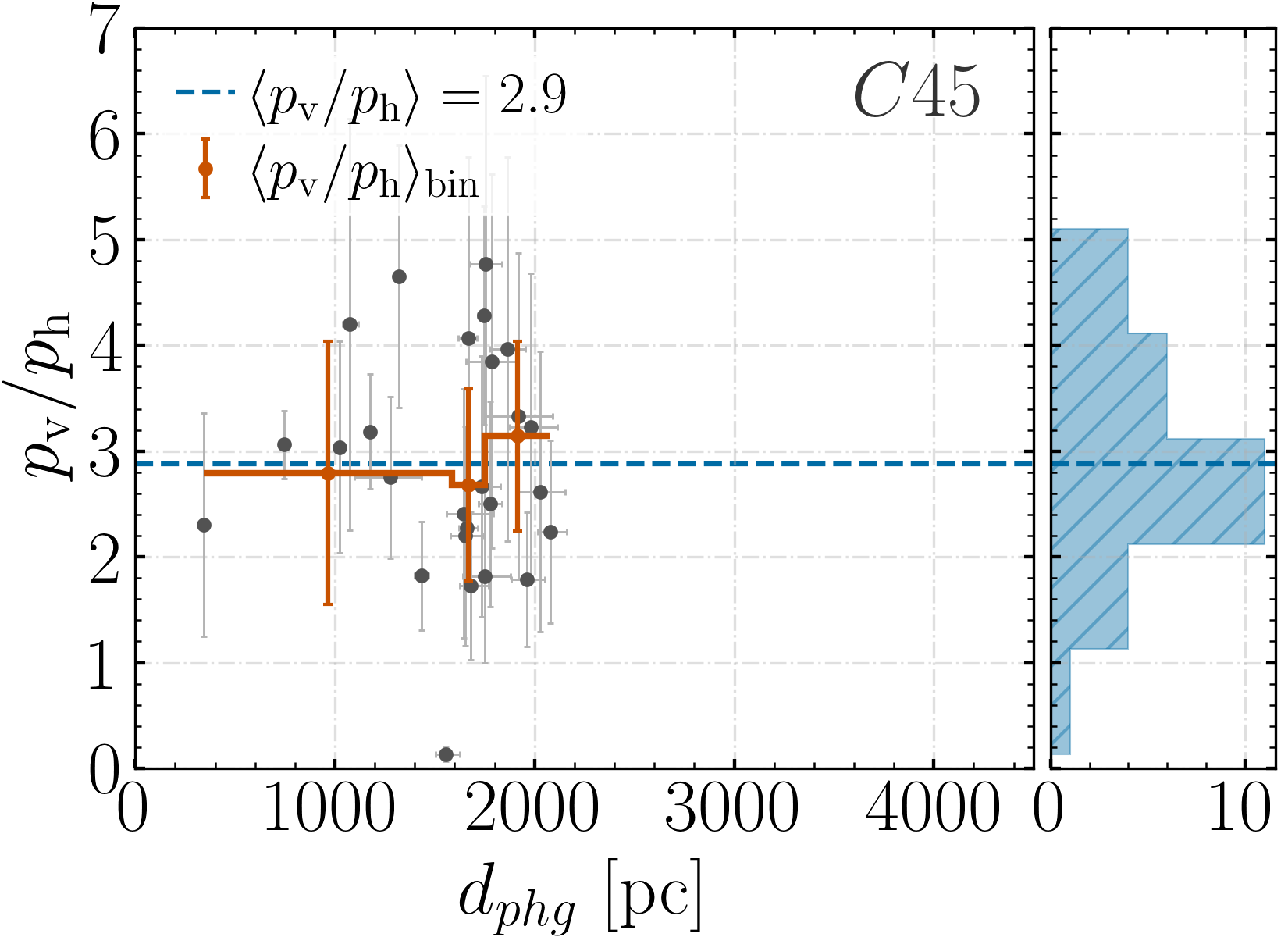}
            \epsscale{0.377}
            \plotone{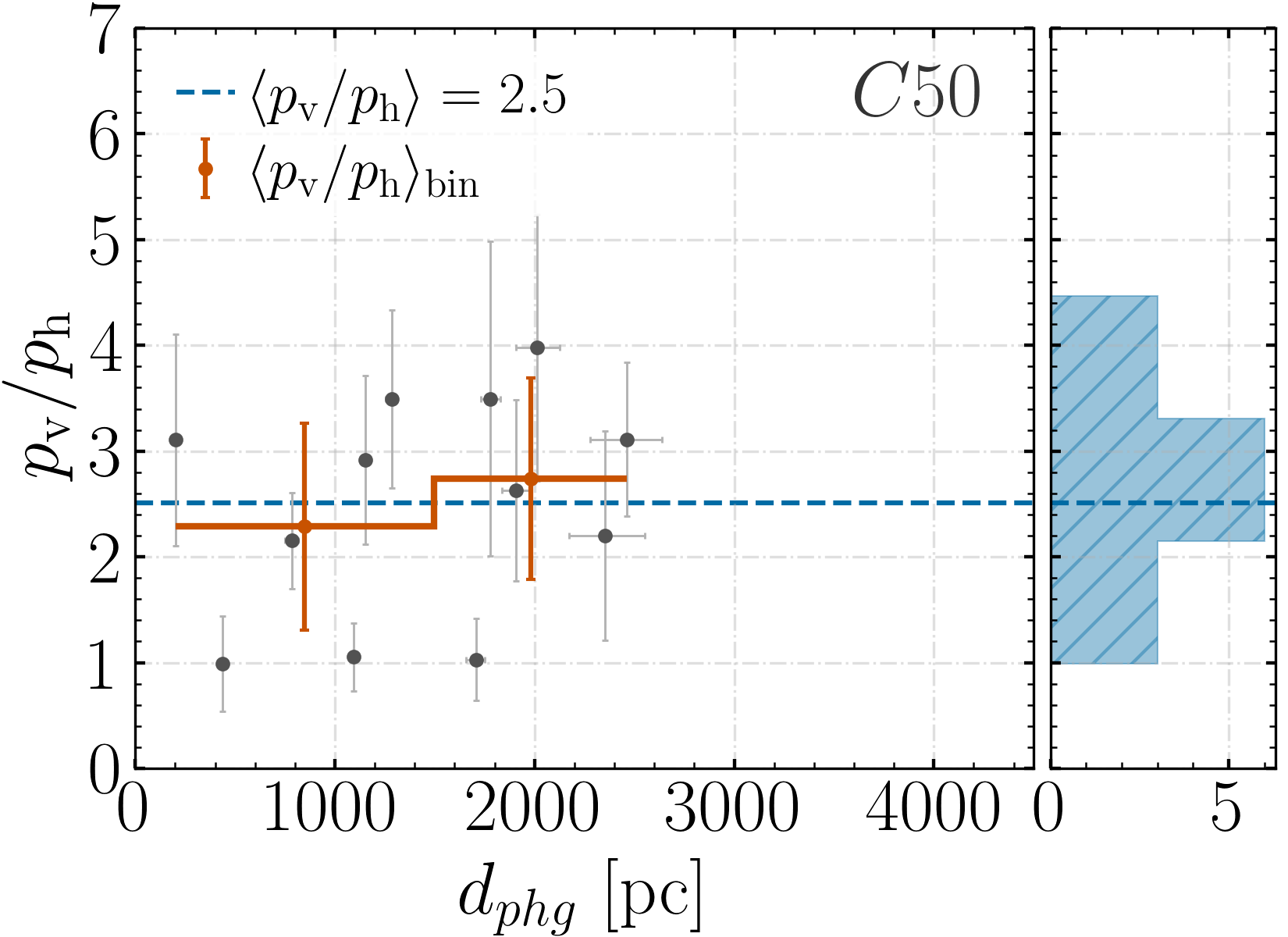}
            \epsscale{0.377}
            \plotone{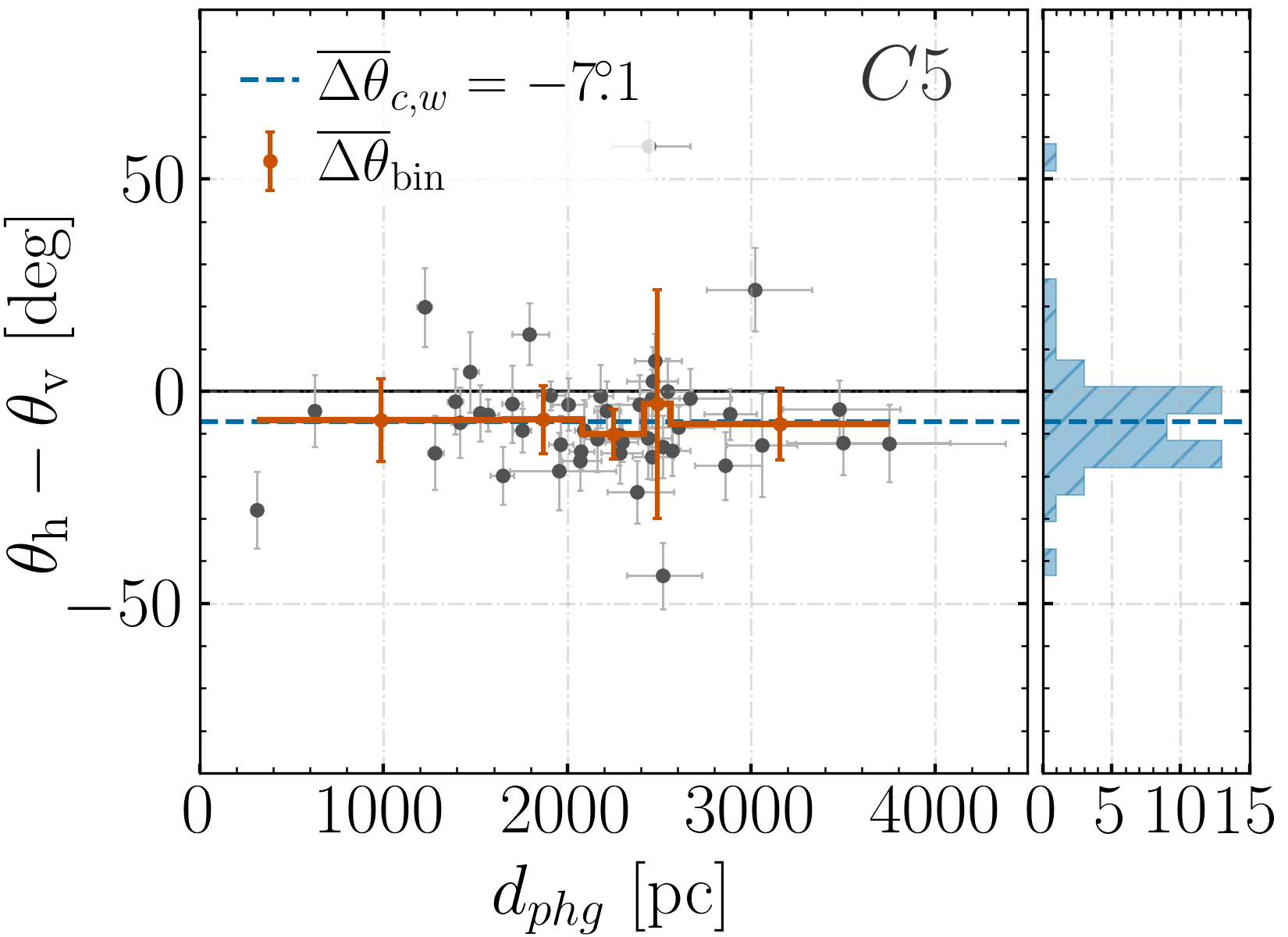}
            \epsscale{0.377}
            \plotone{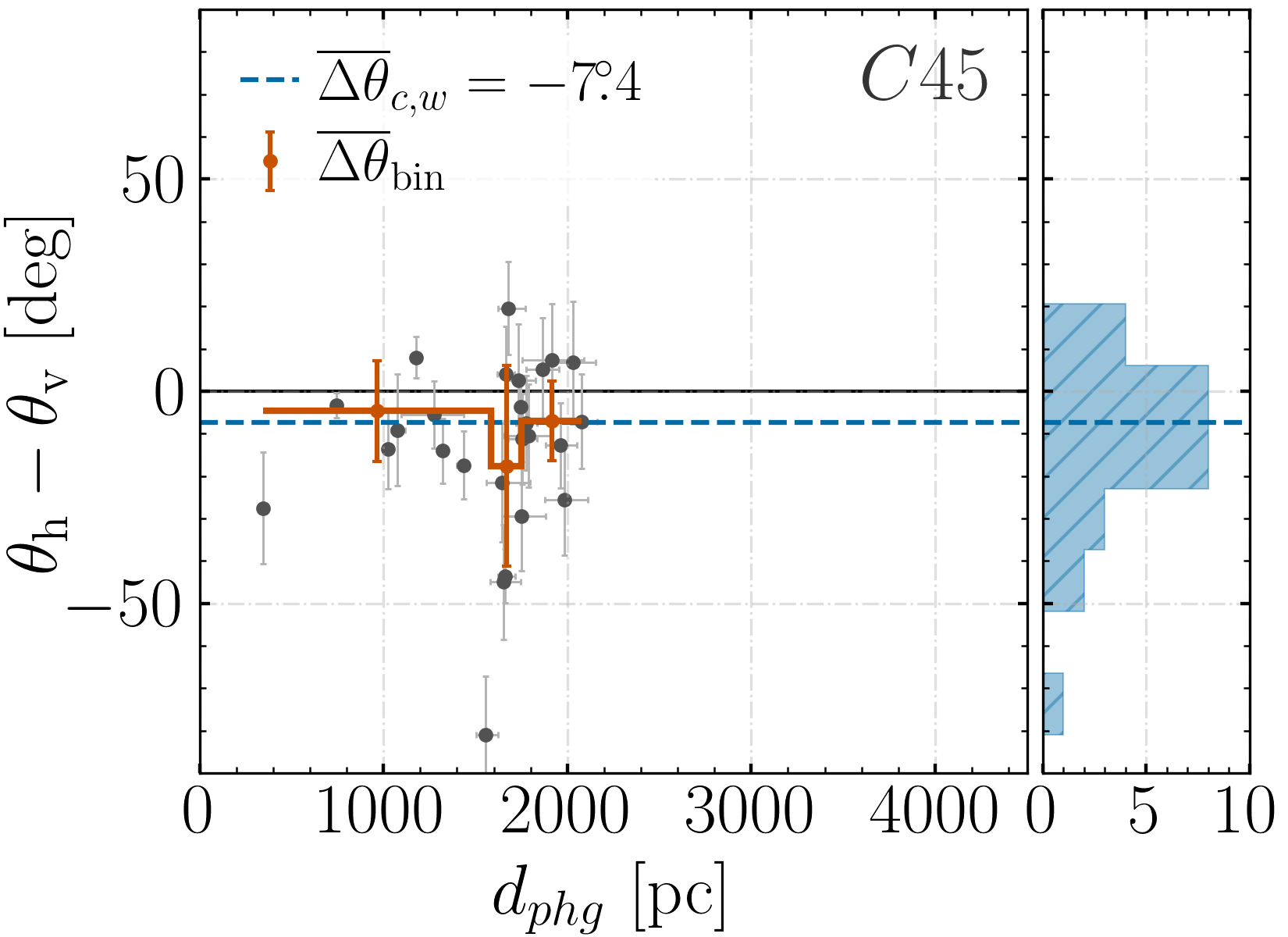}
            \epsscale{0.377}
            \plotone{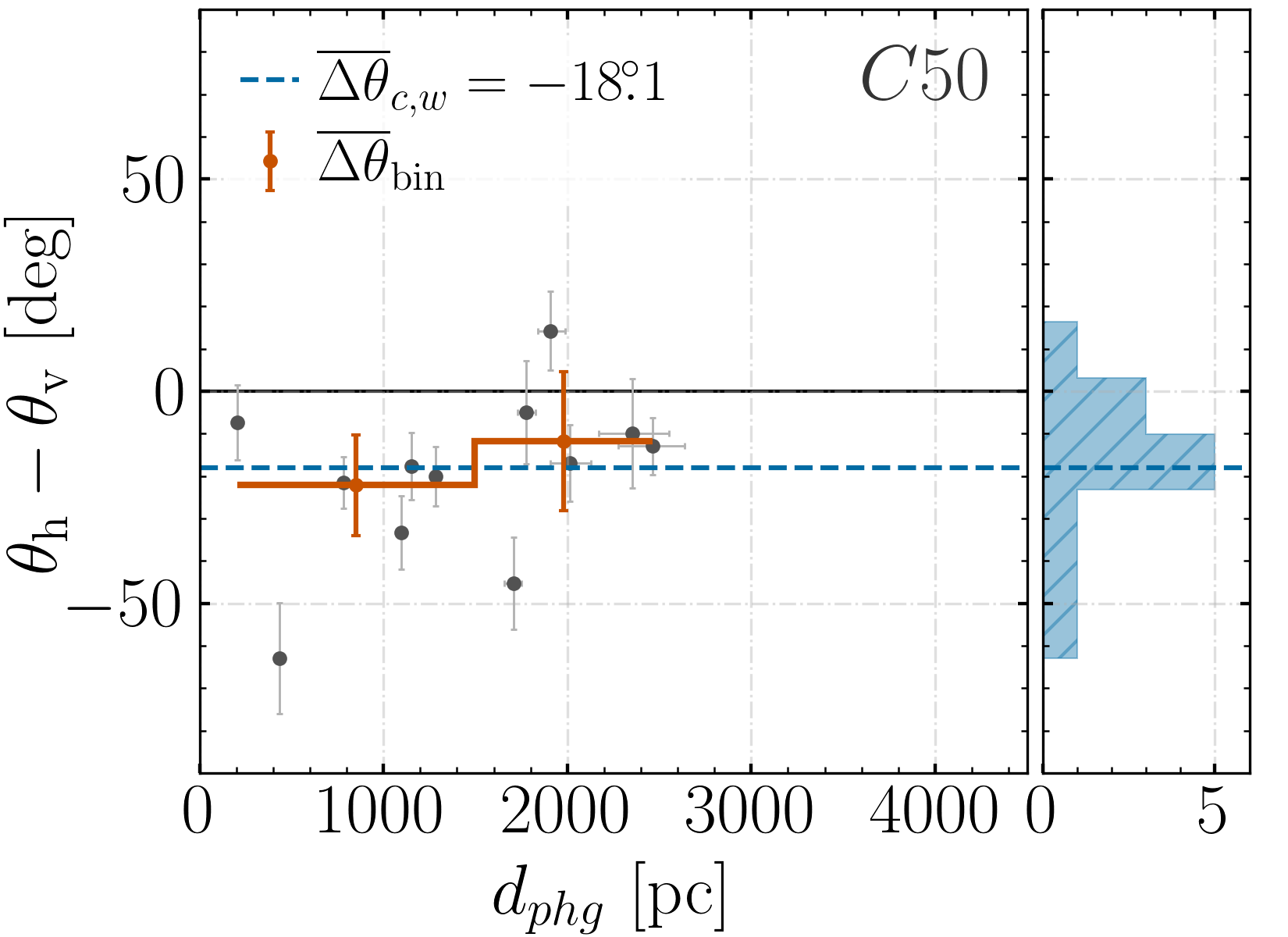}
            \caption{Top row: ratio of the degree of polarization in optical to NIR as a function of distance of stars in common between IPS-GI and GPIPS observations in each region. Bottom row: systematic difference between NIR and optical polarization angles as a function of the distance in each region. The horizontal blue dashed line is the mean value of the distribution shown in the right-side panels. The solid dark-orange lines represent the mean and standard deviation per distance bin. The weighted circular mean and standard deviation were used in the case of the angular difference. 
            \label{fig:ips_gpips_Pratio_and_PAdiff}
            }
        \end{figure*}
        
        \subsubsection{Optical-to-NIR degree of polarization ratio}  
        \label{subsec:res_V_vs_IR_pol}
            
            Figure~\ref{fig:ips_gpips_Av_dens_dist_pol} (top rows) shows the debiased optical and NIR polarization degree compared with the extinction density profiles as a function of distance from \cite{Vergely_2022}. The median and mean polarization degree per distance bin show consistent trends between IPS-GI and GPIPS observations in the three studied regions. \textit{C5} exhibits the longest path lengths, compared to \textit{C45} and \textit{C50} (Figure~\ref{fig:ips_gpips_Av_dens_dist_pol}). Both optical and NIR degrees of polarization steadily increase throughout the entire LOS in \textit{C5} and \textit{C50}. Contrarily, in \textit{C45}, optical and NIR polarization remain roughly constant with distance.

            The top row of Figure~\ref{fig:ips_gpips_Pratio_and_PAdiff} presents the optical-to-NIR debiased polarization ratio ($p_\mathrm{v}/p_\mathrm{h}$) as a function of distance. Each bin contains approximately the same number of measurements, i.e.,~typically around 10 data points per bin. The average ratio of $2{-}3$ observed in the three regions falls within the interstellar polarization expected from the Serkowski curve \citep[][see also (M.~J.~F.~Versteeg et al. 2025, in preparation]{Serkowski_1975} for typical values of $\lambda_\mathrm{max}$ parameter \citep{Whittet_1992}.
            
            In general, variations in $p_\mathrm{v}/p_\mathrm{h}$ could indicate different polarizing dust properties of the ISM structures sampled by both measurements, possibly revealing multiple clouds along the LOS \citep{Patat_2010}. Additionally, the $p_\mathrm{v}/p_\mathrm{h}$ ratio is sensitive to changes in dust extinction $A_V$, as shown by observations \citep[e.g.,][]{Whittet_1992, Whittet_2001, Andersson_Potter_2007}, and grain alignment theory \citep{Lazarian_Hoang_2007}. However, the roughly constant ratios observed in Figure~\ref{fig:ips_gpips_Pratio_and_PAdiff} (top row) suggest uniform polarizing dust properties along our sight lines. A few measurements with small errors and low $p_\mathrm{v}/p_\mathrm{h}$ suggest a deviation from the expected average value in the ISM, but at rather low SNR.

        \subsubsection{Angular alignment}  
        \label{subsec:res_V_vs_IR_PA_align}
        
            Figure~\ref{fig:ips_gpips_Av_dens_dist_pol} (bottom row) shows the polarization maps of stars common to the IPS-GI and GPIPS catalogs. The optical and NIR polarization orientations are generally consistent within the errors. However, a few measurements show significant differences in polarization angles between wavelengths. A systematic polarization angle rotation with wavelength observed in individual clouds likely results from probing different depths \citep[e.g.,][]{Hough_1988}, or from changes in grain size \citep{Codina_Magalhaes_1976}. 
            
            In general, the consistency in the orientation tells us that both observations trace similar polarizing properties of the structures within the regions studied, which is also evidenced through the optical-to-NIR degree of polarization ratio as a function of distance in Figure~\ref{fig:ips_gpips_Pratio_and_PAdiff} (top row). Nevertheless, we also analyzed the systematic difference between the NIR and optical polarization orientations, defined as \mbox{$\Delta\theta=\theta_\mathrm{h}-\theta_\mathrm{v}$} in the range $[-90\degr,90\degr)$. Figure~\ref{fig:ips_gpips_Pratio_and_PAdiff} (bottom row) presents $\Delta \theta$ as a function of distance in \textit{C5}, \textit{C45}, and \textit{C50}. The right-side panels show the distribution of the angular difference along with the weighted circular mean (the dashed blue lines) calculated as,
            \begin{equation}
            \label{eq:circmean_delta_theta}
                \overline{\Delta \theta} = \frac{1}{2} \mathrm{atan2} \left[ \frac{1}{W}\sum_{i=1}^{N}{w_i \mathrm{sin}(2\Delta \theta)}, \frac{1}{W}\sum_{i=1}^{N}{w_i \mathrm{cos}(2\Delta \theta)} \right],
            \end{equation}
            where $w_i$ is the weight of the $i$-th measurement calculated as $1/\sigma_{\Delta\theta, i}^2$, with $\sigma_{\Delta \theta}$ being the error of the angular difference propagated from the observed errors as $(\sigma_{\theta_H}^2 + \sigma_{\theta_V}^2)^{1/2}$. $W$ is the total sum of the weights. 
            
            We found a negative offset in $\overline{\Delta \theta}$ of $-7\fdg1$, $-7\fdg4$, and $-18\fdg1$ in \textit{C5}, \textit{C45}, and \textit{C50}, respectively. We used the reduced $\chi^2$ to test the significance of the offset compared to a regression equal to zero in the angle difference. We used the relation
            \begin{equation}
            \label{Eq:chi2_red}
                \chi^2_{red} = \frac{1}{N-1}\sum^N_{i=1}{\left(\frac{\Delta \theta_i - \overline{\Delta \theta}}{\sigma_{\Delta \theta,i}}\right)^2}~,
            \end{equation}
            where $N$ is the number of measurements and $\overline{\Delta \theta}$ is the average value (Equation~\ref{eq:circmean_delta_theta}) or regression evaluated. Firstly, we calculated the $\chi^2_{red}$ of our best-fits, obtaining values of approximately $5$, $4$, and $3$ in \textit{C5}, \textit{C45}, and \textit{C50}, respectively. Secondly, we calculated the $\chi^2_{red}$ using $\overline{\Delta \theta} = 0$, that is, assuming that the difference in angle is zero, resulting in $6$, $5$, and $8$ for the three regions, respectively. In \textit{C5} and \textit{C45}, the two results are comparable and different from one. Therefore, we cannot conclude that there is a non-zero polarization orientation offset between optical and NIR observations. In \textit{C50}, $\chi^2_{red}$ deviates more, with $\overline{\Delta\theta} = 0$ yielding the poorest regression fit, indicating a possible non-zero angle offset. However, in \textit{C50}, the few common stars between the catalogs are more scattered, and the uncertainties are large. In conclusion, the $\chi^2_{red}$ test indicates that the angular difference observed in the three regions does not deviate significantly from zero. 

            We divided the cross-matched data into distance bins to determine whether there are alignment variations at specific locations along the sightlines. The distance bins are defined as in the top row of Figure~\ref{fig:ips_gpips_Pratio_and_PAdiff}. Then, we calculated the weighted mean per bin with Equation~(\ref{eq:circmean_delta_theta}), as shown by the solid dark-orange line in Figure~\ref{fig:ips_gpips_Pratio_and_PAdiff}, bottom row. The error bars are the weighted standard deviation of each bin. 
            The weighted mean of the systematic difference per distance bin does not show significant deviations from the total mean, represented by the horizontal dashed line in Figure~\ref{fig:ips_gpips_Pratio_and_PAdiff} (bottom row), in any of the regions studied. Therefore, there is no significant correlation between the angle difference and the distance to the stars. Nevertheless, we emphasize that the big errors in the observations and the limited number of stars common between catalogs prevent a more precise analysis. 

    \subsection{Magnetic field tomography with BISP-1}  
    \label{subsec:res_tomo_BISP-1}  
        
\begin{figure*}[p]
            \gridline{
                      \fig{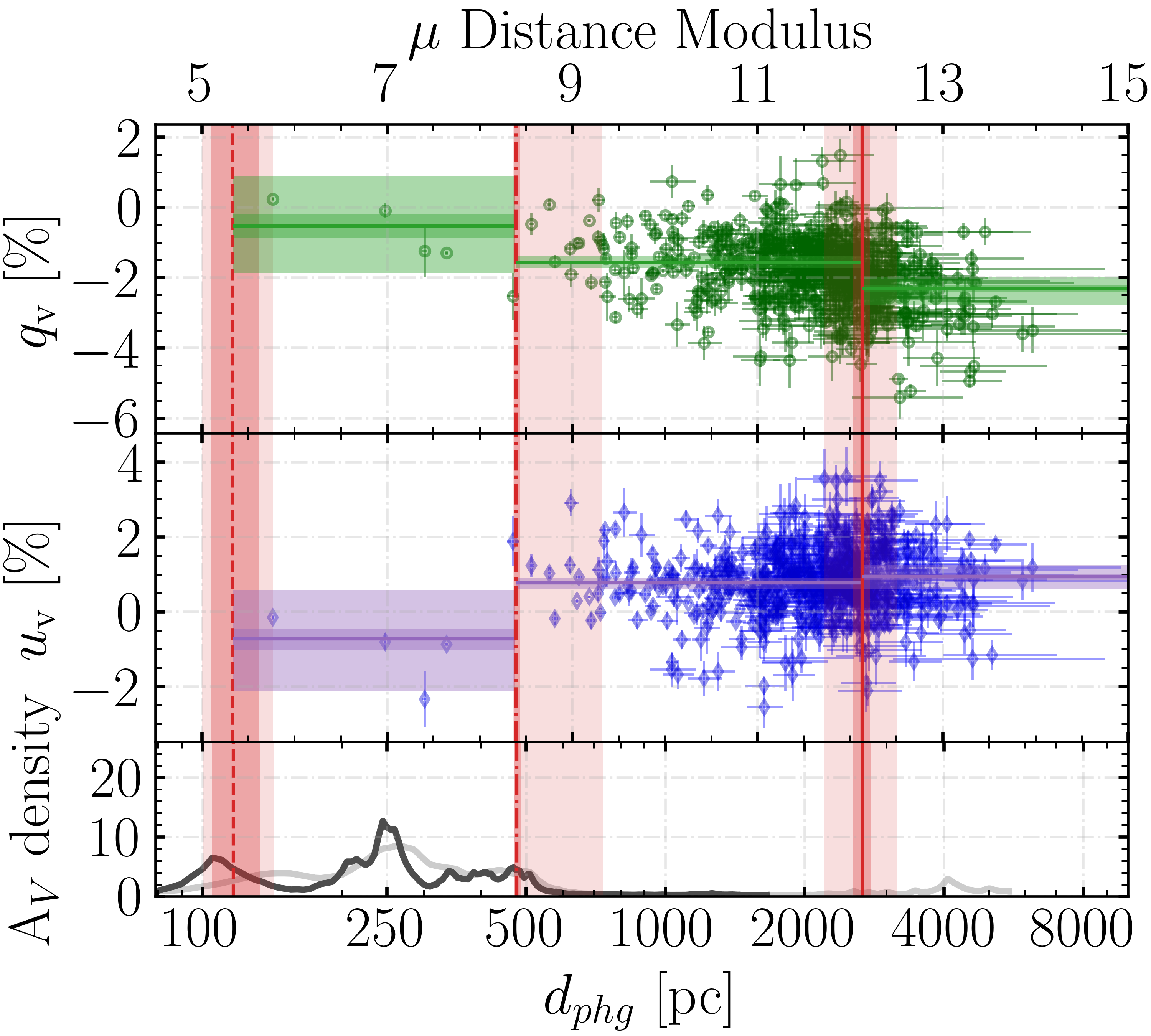}{.32\linewidth}{}
                      \fig{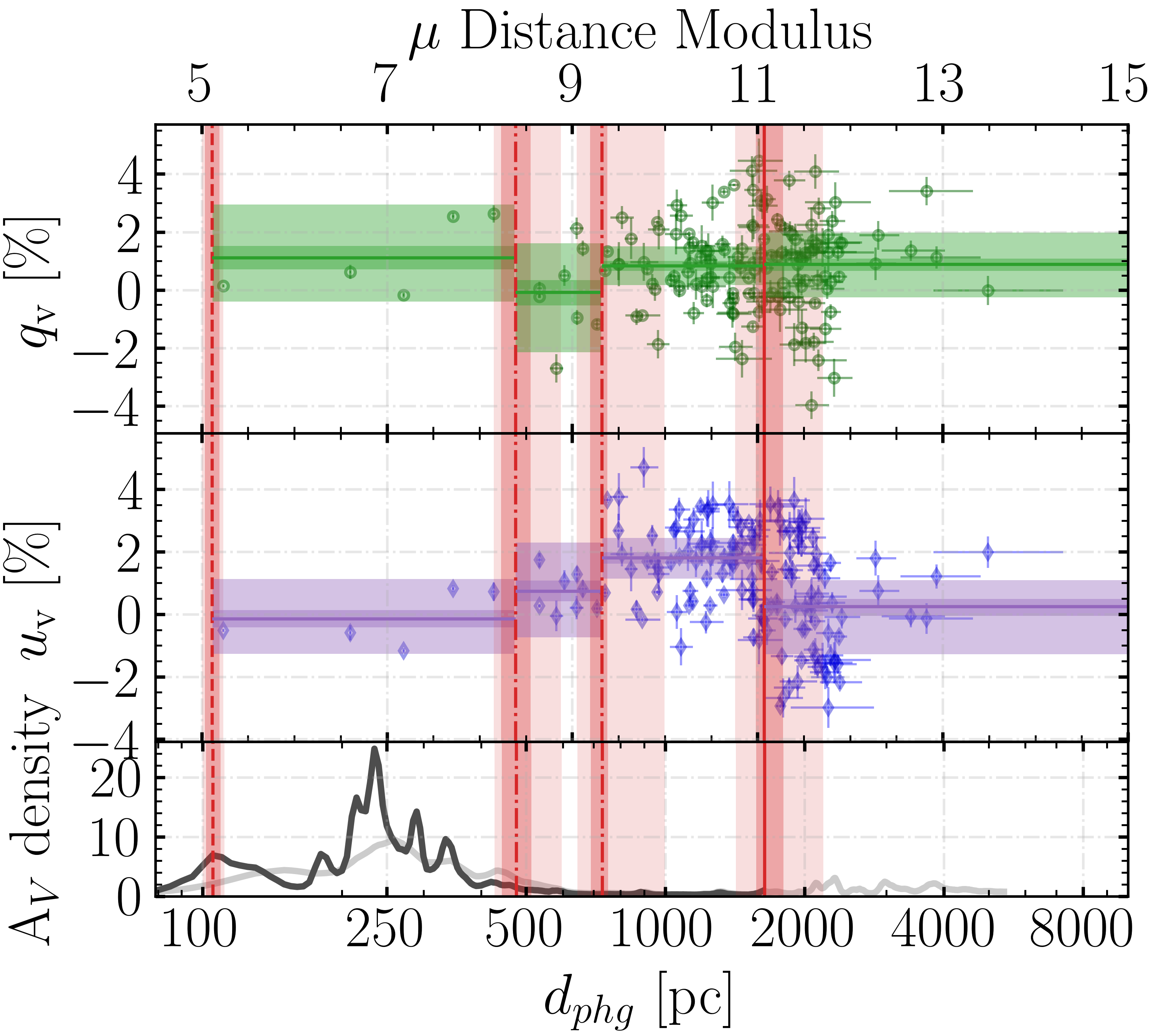}{.32\linewidth}{}
                      \fig{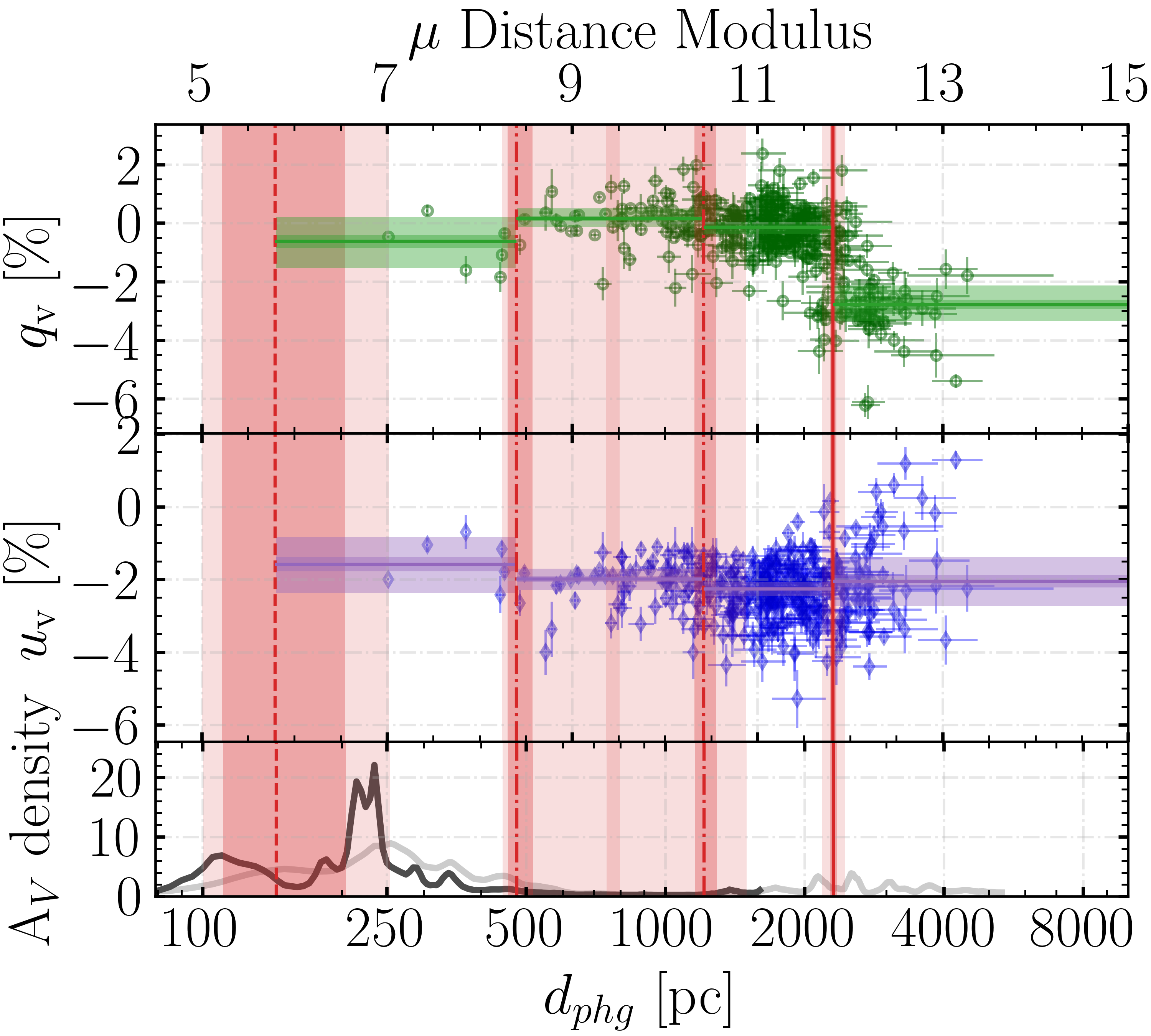}{.32\linewidth}{}
                      }\vspace{-0.9cm}
            \gridline{
                      \fig{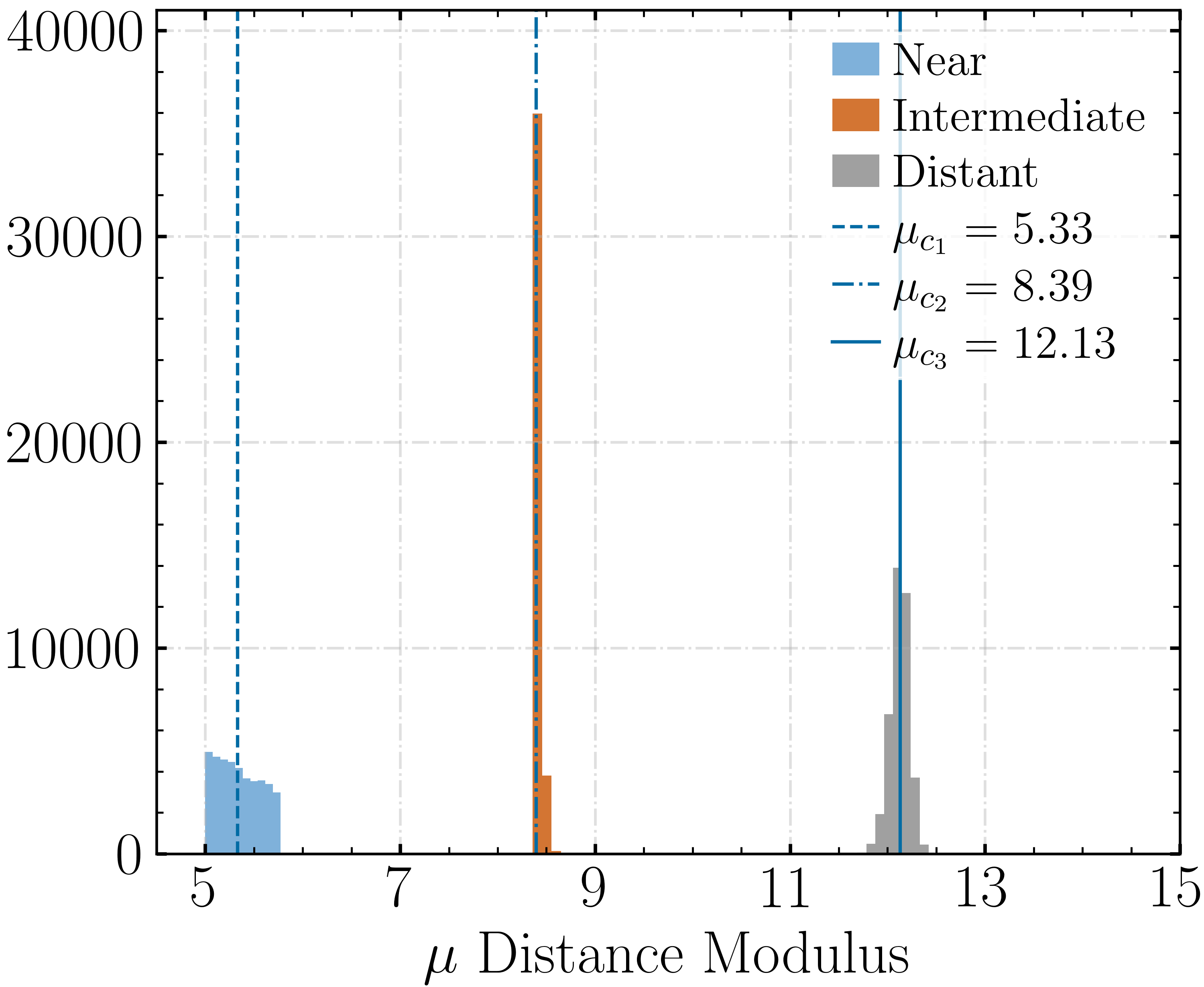}{.32\linewidth}{}
                      \fig{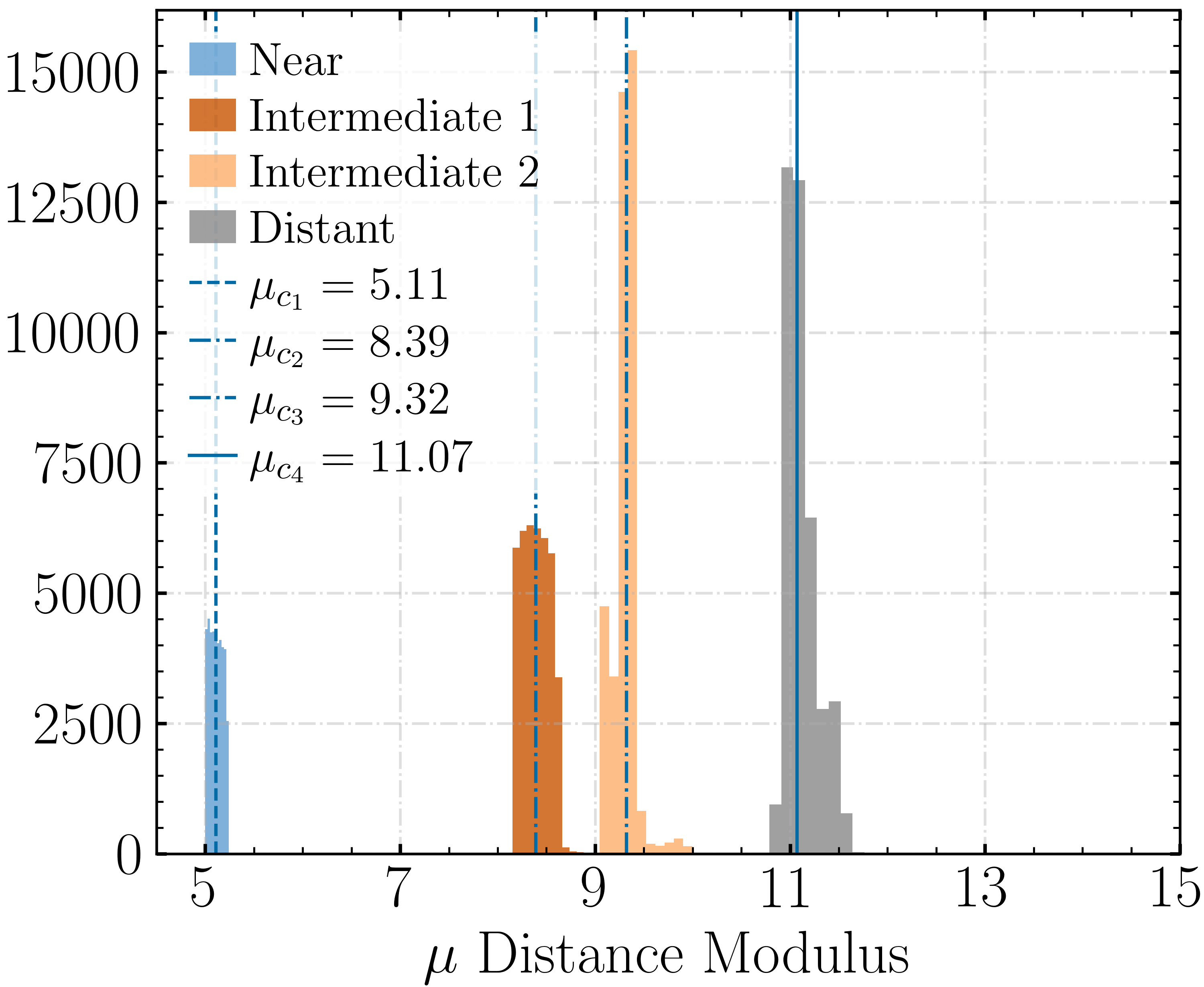}{.32\linewidth}{}
                      \fig{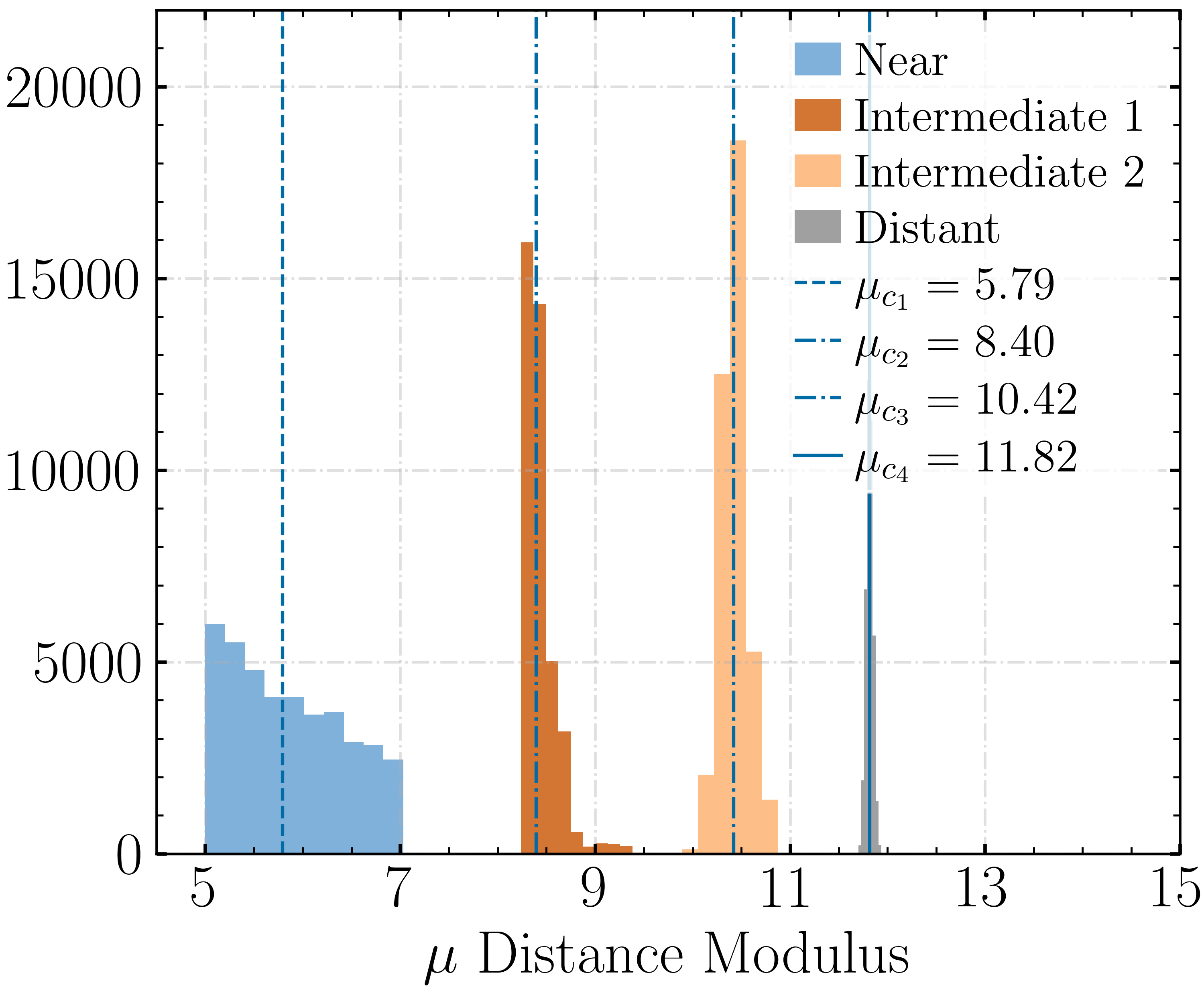}{.32\linewidth}{}
                      }\vspace{-0.9cm}
            \gridline{
                      \fig{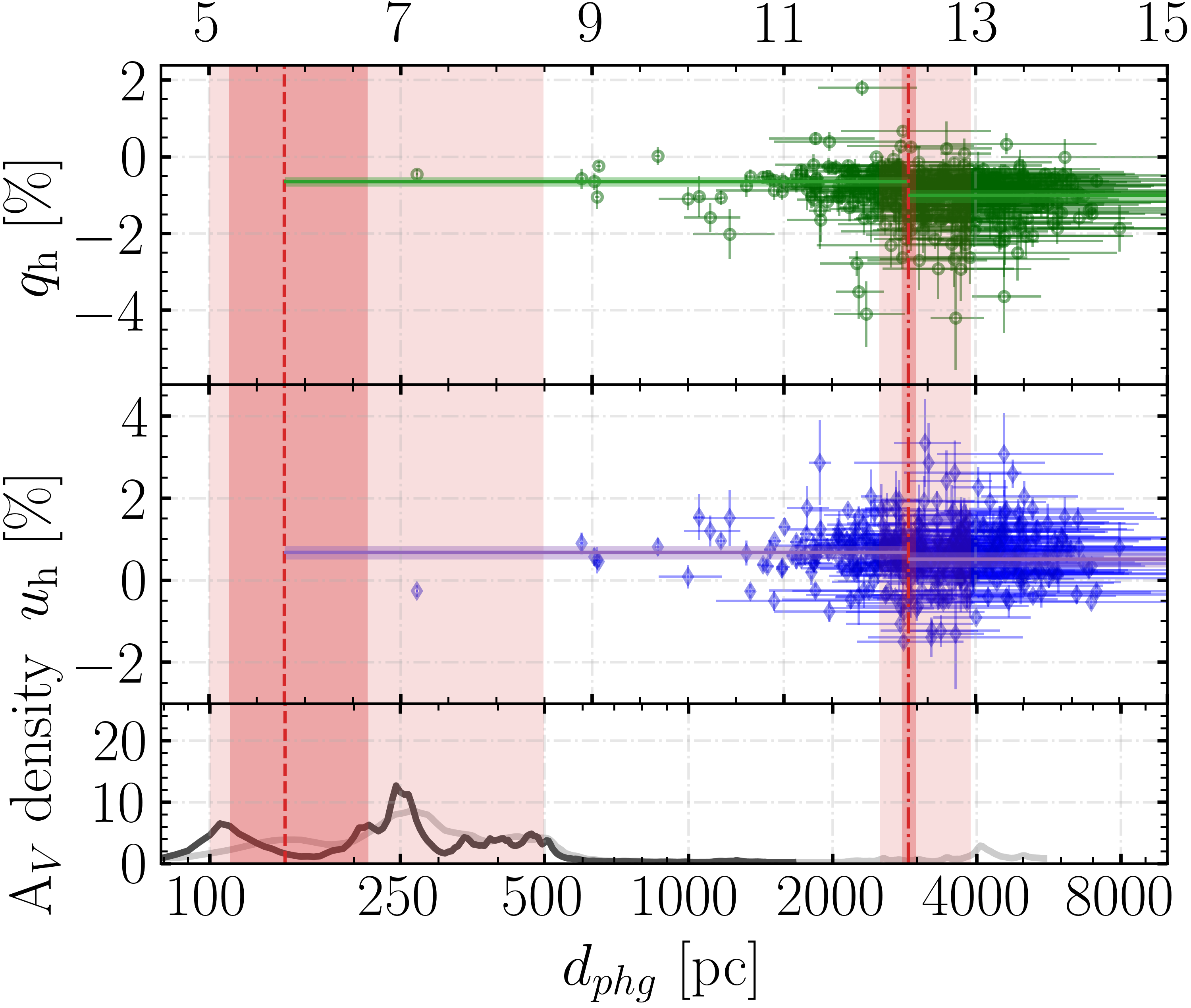}{.32\linewidth}{}
                      \fig{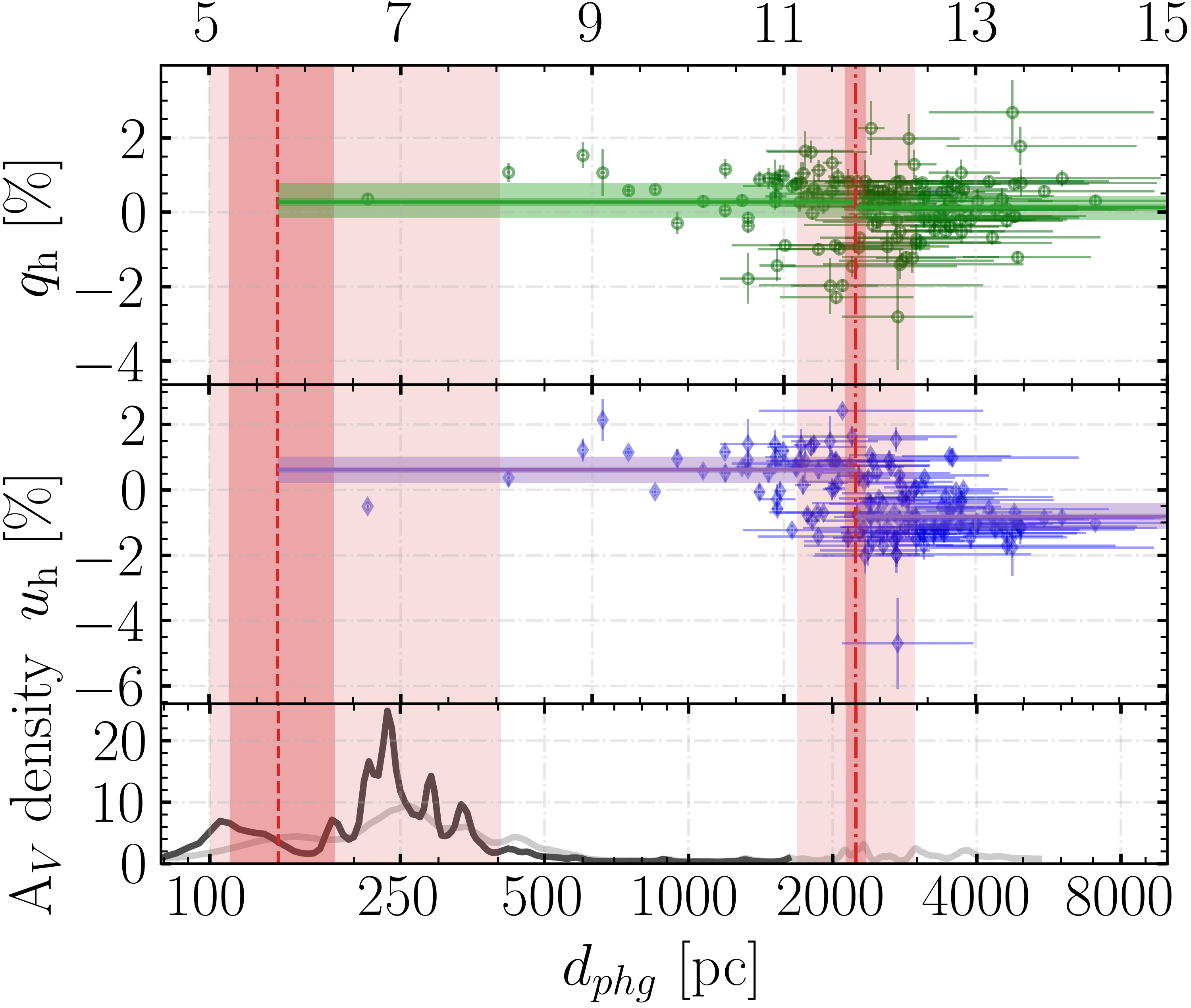}{.32\linewidth}{}
                      \fig{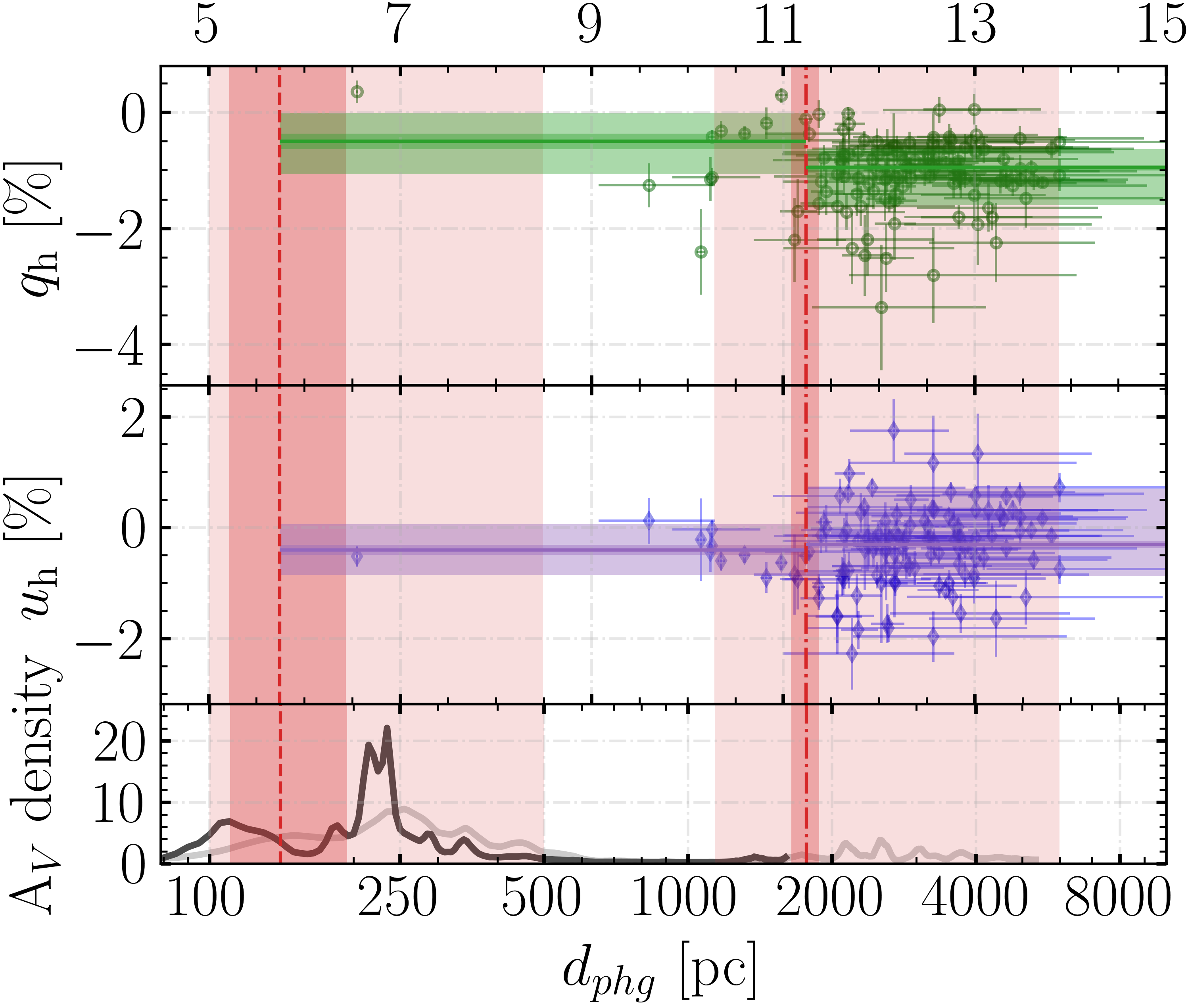}{.32\linewidth}{}
                      }\vspace{-0.9cm}
            \gridline{
                      \fig{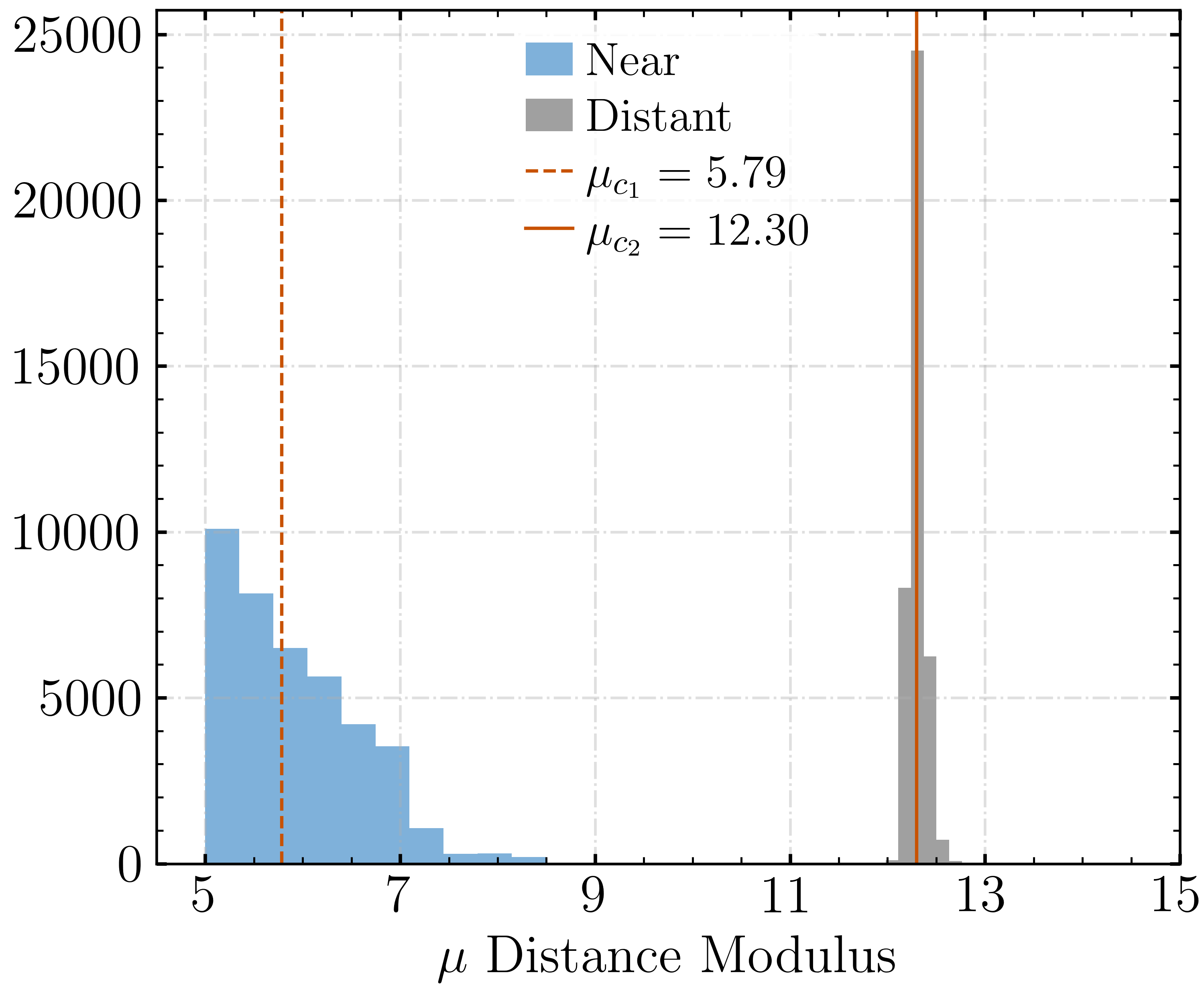}{.32\linewidth}{}
                      \fig{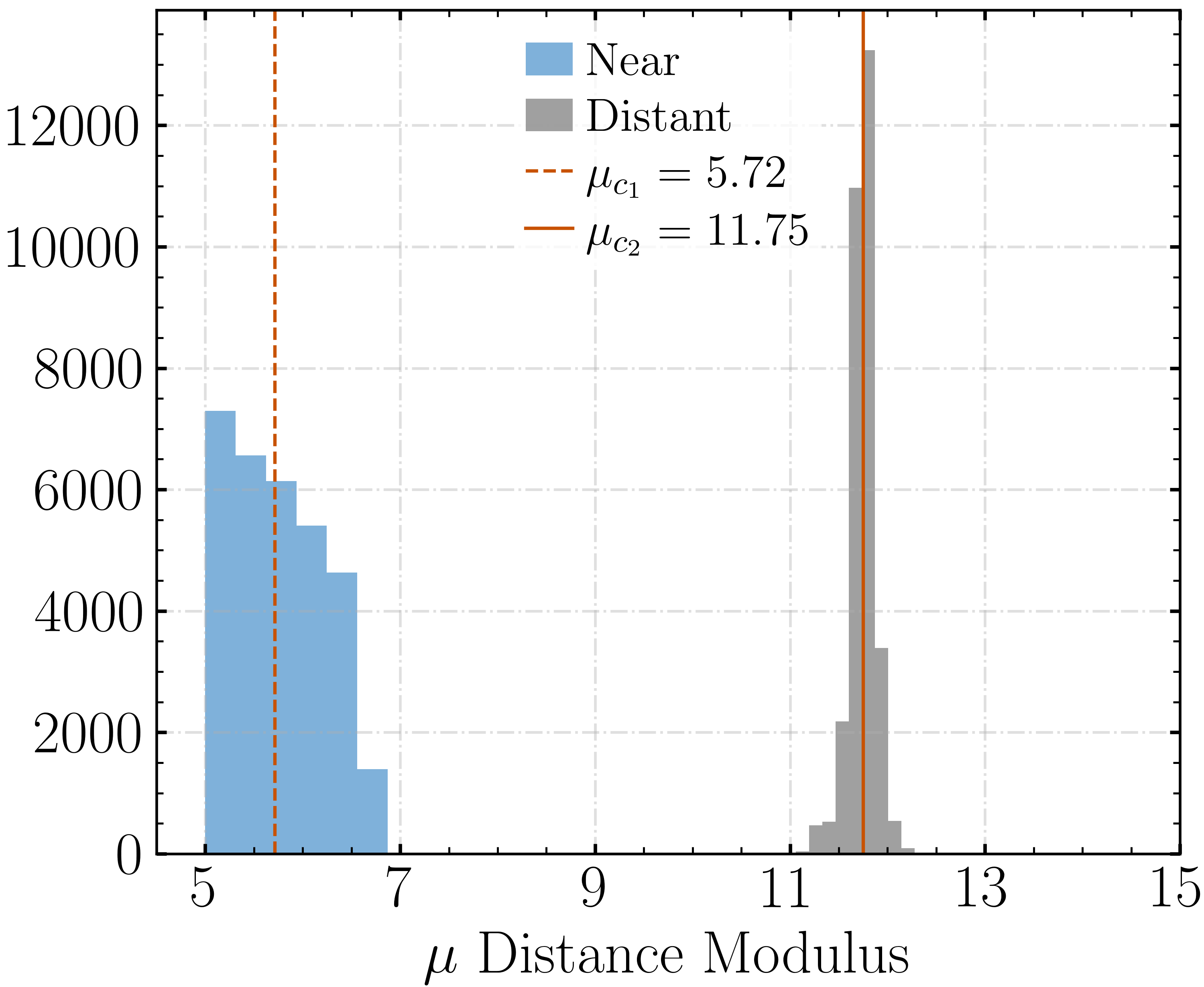}{.32\linewidth}{}
                      \fig{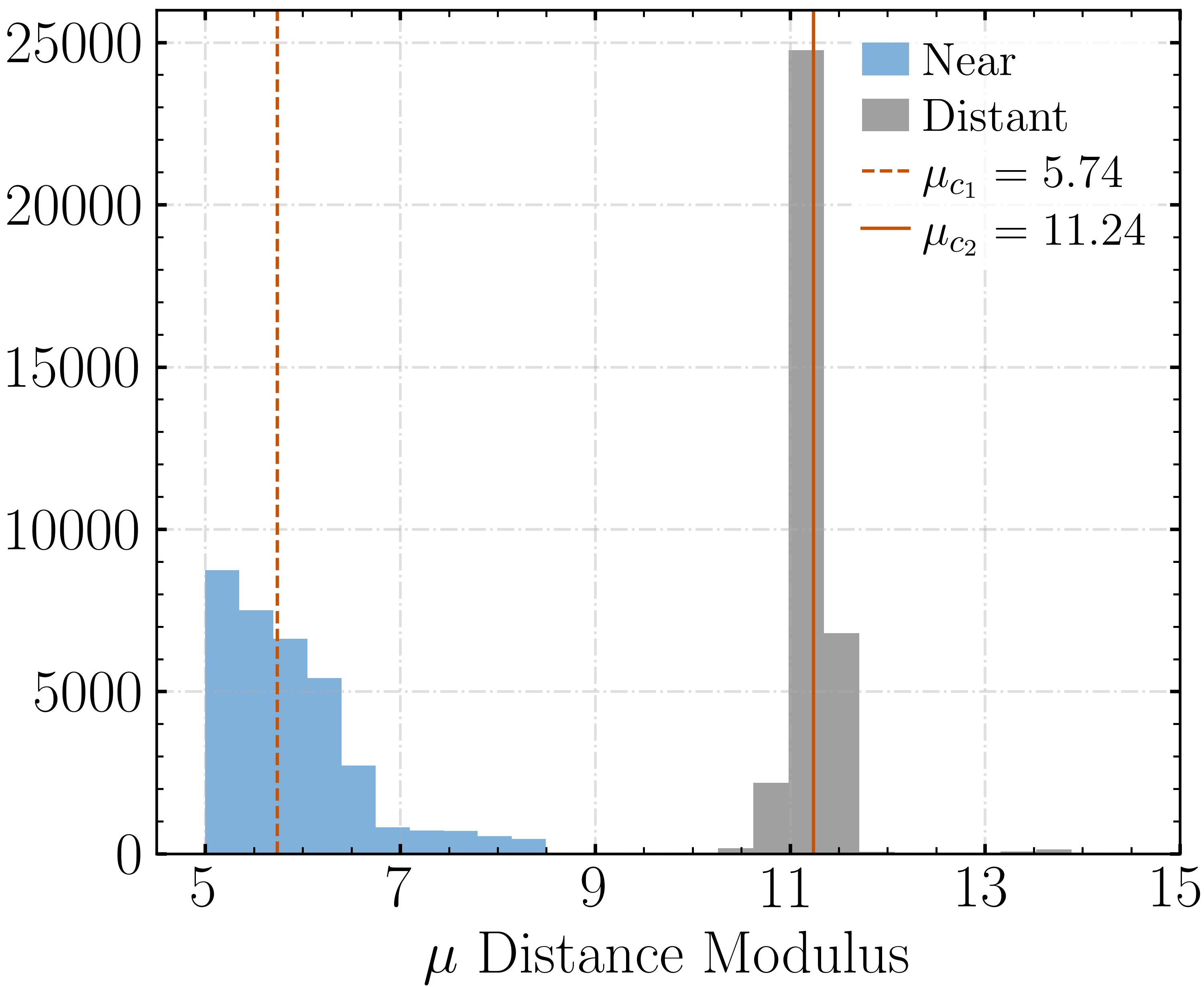}{.32\linewidth}{}
                      }\vspace{-0.8cm}
            \caption{Best solutions of the polarization decomposition using BISP-1 with optical (first two rows) and NIR (last two rows) polarization observations in the regions of \textit{C5} (left), \textit{C45} (middle), and \textit{C50} (right). The first and third rows show the Stokes parameters $q_{(\mathrm{v,h})}$ and $u_{(\mathrm{v,h})}$ as a function of the distance modulus calculated using the inverse parallax. The red vertical lines and shaded areas represent the cloud median distances and their $1\sigma$ and $2\sigma$ confidence intervals. Similarly, the horizontal solid lines and shaded areas represent the LOS-integrated median $q_{(\mathrm{v,h})}$ (green) and $u_{(\mathrm{v,h})}$ (purple) and their $1\sigma$ and $2\sigma$ uncertainties for each group of stars behind the polarizing layers. The bottom panels show the extinction per distance unit ($10^{-3}$~mag~pc$^{-1}$) profiles from \cite{Vergely_2022} at $10$~pc (gray) and $50$~pc (light-gray) resolution; the distance of the profiles is on a logarithmic scale. The second and fourth rows show the posterior distributions of the inverse parallax for each cloud modeled. The vertical dashed lines show the median values.
            \label{fig:BISP-1_tomo_results_IPS}}
        \end{figure*}

        Figure~\ref{fig:BISP-1_tomo_results_IPS} presents the results of the polarization tomography using BISP-1 (Section~\ref{subsec:tomo_BISP1}) with high-quality optical and NIR data. The first and third rows show the polarization observations and the optical extinction density as a function of distance, along with BISP-1's results. The three-cloud and four-cloud models best fit optical observations, while the two-cloud model is the best solution for NIR observations. 
        
        The second and fourth rows of Figure~\ref{fig:BISP-1_tomo_results_IPS} show the posterior distributions of estimated cloud distances. In all cases, the nearest cloud distributions are broad due to the lack of zero-polarization measurements, leading to poorly constrained and prior-dominated solutions. Additionally, \mbox{BISP-1's} minimum requirement of five stars between clouds \citep{Pelgrims_2023} contributes to poor constraints on second-nearest clouds in optical polarization, shifting distance posteriors toward lower values. The priors implemented in the modeling are detailed in Appendix~\ref{sec:Append_B}.  
        
        Table~\ref{tab:tomo_results} summarizes the median LOS-integrated polarization degree, the circular mean magnetic field orientation on the plane-of-sky, and the median cloud distances (calculated as $1/\varpi$ from the median parallax posterior distribution) for each group of stars behind the clouds and each observed region. The reported uncertainties are derived from the spread in the posterior distributions and thus reflect model-based errors; see the notes in Table~\ref{tab:tomo_results} for more details.
        Low-polarization components with large uncertainties have a biased median estimate due to $p/\sigma_p$ ratios (where $\sigma_p$ is the standard deviation of $p$) often falling below five.
        However, since we do not perform any further calculations using the LOS-integrated polarization degree, debiasing is unnecessary in this study. In the following sections, we analyze the tomography results for each region in detail.      
        
        %
        %
        %
        
        %
        \begin{deluxetable*}{clcccc|cc}
        \setlength{\tabcolsep}{9pt}
        \tablecaption{LOS-integrated polarization properties as derived from BISP-1 and GMM modeling. \label{tab:tomo_results}}
        \tablehead{
        \colhead{Region} & \colhead{Cloud} & \multicolumn{4}{c}{BISP-1} & \multicolumn{2}{|c}{GMM} \\
        \cline{3-8}
        \colhead{ } & \colhead{ } & \colhead{$p$} & \colhead{$\theta$} & \colhead{$1/\varpi$} & \colhead{$d_{phg}$} & \colhead{$\overline{p}$} & \colhead{$\overline{\theta}$}   \\
        \colhead{ } & \colhead{ } & \colhead{$(\%)$} & \colhead{$(\degr)$} & \colhead{$(kpc)$} & \colhead{$(kpc)$} & \colhead{$(\%)$} & \colhead{$(\degr)$}
        }
        \colnumbers
        \startdata
        \multicolumn{8}{c}{Optical Polarization}\\[0.15cm] \hline
                     & Near & $1.0^{+0.3}_{-0.3}$ & $118 \pm 10$ & $0.116^{+0.02}_{-0.01}$ & $0.115$ & \multirow{3}{*}{$\Biggl\}2.0\pm0.5$} & \multirow{3}{*}{$78\pm9$} \\[0.15cm]
        \textit{C5}  & Intermediate & $1.7^{+0.05}_{-0.05}$ & $77.0 \pm 0.6$ & $ 0.477^{+0.01}_{-0.01}$ & $0.470$ &  &  \\[0.15cm]
                     & Distant & $2.5^{+0.1}_{-0.1}$ & $79.0 \pm 1.0$ & $2.7^{+0.1}_{-0.1}$ & $2.4$ & $3.3\pm0.8$ & $74\pm10$ \\[0.15cm]\hline
                     & Near & $1.1^{+0.4}_{-0.3}$ & $175 \pm 10$ & $0.105^{+0.004}_{-0.004}$ & $0.104$ & \multirow{4}{*}{$\Biggl\}2.4\pm0.9$} & \multirow{4}{*}{$36\pm24$} \\[0.15cm]
        \multirow{3}{*}{\textit{C45}} & Intermediate 1 & $0.9^{+0.4}_{-0.3}$ & $49 \pm 17$ & $0.476^{+0.04}_{-0.03}$ & $0.469$ &  &  \\[0.15cm]
                     & Intermediate 2 & $2.0^{+0.2}_{-0.2}$ & $33 \pm 3$ & $0.731^{+0.02}_{-0.04}$ & $0.714$ &  &  \\[0.15cm]
                     & Distant & $1.0^{+0.2}_{-0.2}$ & $7 \pm 10$ & $1.64^{+0.2}_{-0.07}$ & $1.55$ & $2.2\pm1.0$ & $164\pm20$ \\[0.15cm]\hline
                     & Near & $1.7^{+0.2}_{-0.2}$ & $124 \pm 4$ & $0.144^{+0.06}_{-0.03}$ & $0.143$ &  \multirow{3}{*}{$\Biggl\}2.1\pm0.6$} & \multirow{3}{*}{$136\pm11$} \\[0.15cm]
        \multirow{3}{*}{\textit{C50}}  & Intermediate 1 & $2.0^{+0.08}_{-0.07}$ & $137.3 \pm 1.2$ & $0.478^{+0.04}_{-0.02}$ & $0.471$ &  &  \\[0.15cm]
                     & Intermediate 2 & $2.3^{+0.05}_{-0.05}$ & $133.2 \pm 0.6$ & $1.21^{+0.08}_{-0.06}$ & $1.16$ & $2.4\pm0.6$ & $134\pm8$ \\[0.15cm]
                     & Distant & $3.5^{+0.1}_{-0.1}$ & $108.2 \pm 1.5$ & $2.31^{+0.04}_{-0.04}$ & $2.10$ & $3.7\pm1.0$ & $109\pm11$ \\[0.15cm]\hline
        \multicolumn{8}{c}{NIR Polarization}\rule{0pt}{0.6cm}\\[0.15cm]\hline
        \multirow{2}{*}{\textit{C5}} & Near & $0.9^{+0.04}_{-0.04}$ & $67.0 \pm 1.2$ & $0.144^{+0.07}_{-0.03}$ & $0.143$ & $1.2\pm0.3$ & $67\pm10$ \\[0.15cm]
                     & Distant & $1.1^{+0.04}_{-0.04}$ & $76.8 \pm 1.2$ & $2.88^{+0.1}_{-0.09}$ & $2.55$ & $1.6\pm0.5$ & $83\pm18$ \\[0.15cm]\hline
                     & Near & $0.7^{+0.1}_{-0.09}$ & $33 \pm 5$ & $0.139^{+0.04}_{-0.03}$ & $0.138$ & $1.0\pm0.4$ & $22\pm24$ \\[0.15cm]
        \textit{C45} & Distant & $0.8^{+0.09}_{-0.1}$ & $139 \pm 3$ & $2.2^{+0.1}_{-0.1}$ & $2.0$ & $1.3\pm0.4$ & $138\pm17$ \\[0.15cm]
                     & -- & -- & -- & -- & -- & $2.2\pm1.0$ & $77\pm18$ \\[0.15cm]\hline
                     & Near & $0.7^{+0.1}_{-0.08}$ & $110 \pm 6$ & $0.141^{+0.05}_{-0.03}$ & $0.140$ & -- & -- \\[0.15cm]
        \textit{C50} & \multirow{2}{*}{Distant} & \multirow{2}{*}{$1.0^{+0.05}_{-0.05}$} & \multirow{2}{*}{$99 \pm 2$} & \multirow{2}{*}{$1.8^{+0.1}_{-0.1}$} & \multirow{2}{*}{$1.7\Biggr\{$} & $1.2\pm0.5$ & $102\pm18$ \\[0.15cm]
                      & &   &   &   &   & $1.7\pm0.6$ & $81\pm11$ \\
        \enddata
        \tablecomments{Columns: (1) Designation of the region; (2) polarizing cloud; (3) median polarization degree; (4) circular mean and standard deviation of the polarization angle; (5) median inverse parallax; (6) corrected photo-geometric distance deduced from the models fitted to the observations (Section~\ref{subsec:Disc_dist_Model}); (7) mean and standard deviation of the polarization degree; (8) circular mean and standard deviation of the polarization angle. The parameters from (3) to (5) were calculated using the posterior distributions of $q_{(\mathrm{v,h})}$, $u_{(\mathrm{v,h})}$, and $\varpi$ obtained from BISP-1 decomposition (Section~\ref{subsec:tomo_BISP1}). The uncertainties of (3) and (5) are estimated from the 84th and 16th percentiles of the posterior distributions. The parameters of (7) and (8) were calculated for each component identified in the GMM clustering analysis (Section~\ref{subsec:tomo_GMM}). The open brackets indicate correspondence with more than one component of the other method.}
        \end{deluxetable*}
        %
        
        %
        \begin{figure*}[ht!]
            \gridline{
                      \fig{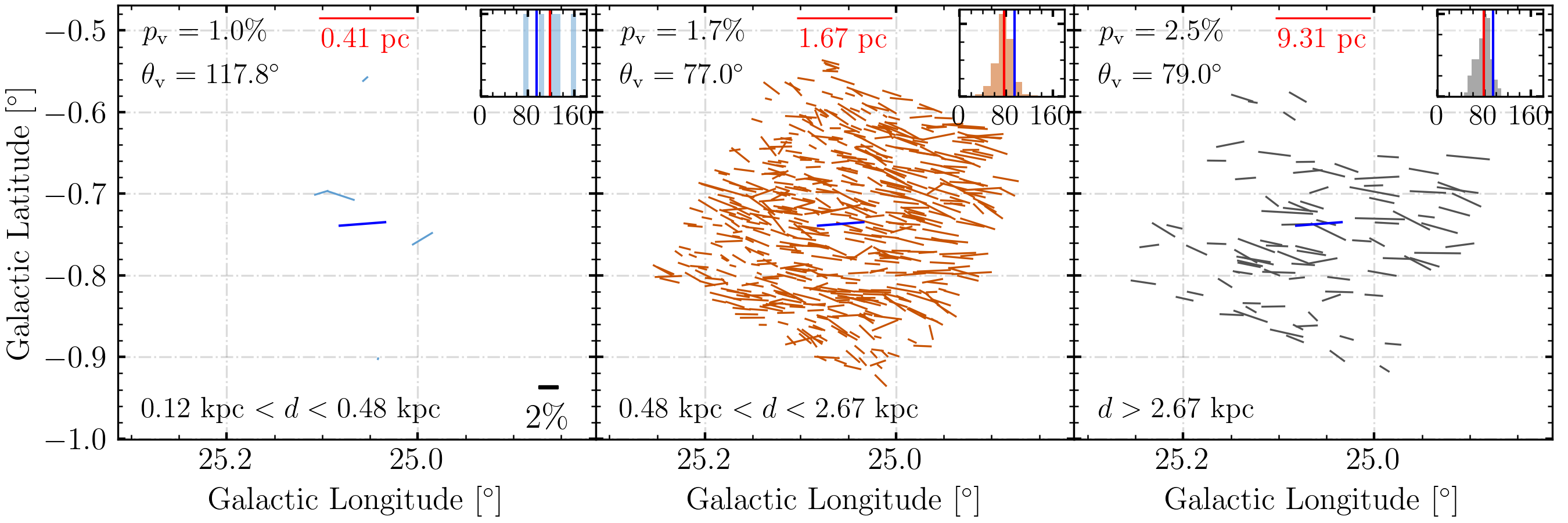}{.8\linewidth}{}
                      }\vspace{-0.9cm}
            \gridline{
                      \fig{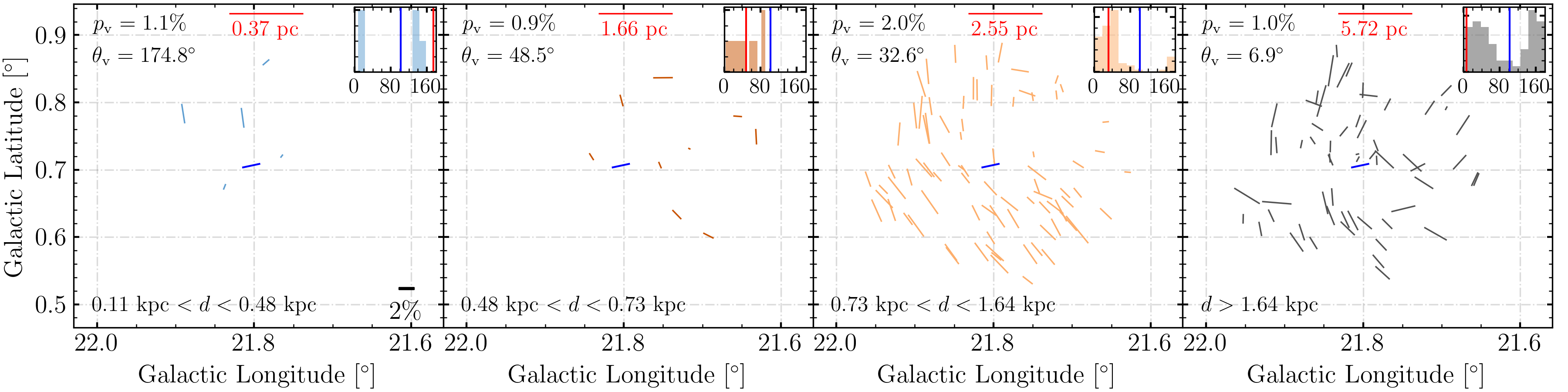}{\linewidth}{}
                      }\vspace{-0.9cm}
            \gridline{
                      \fig{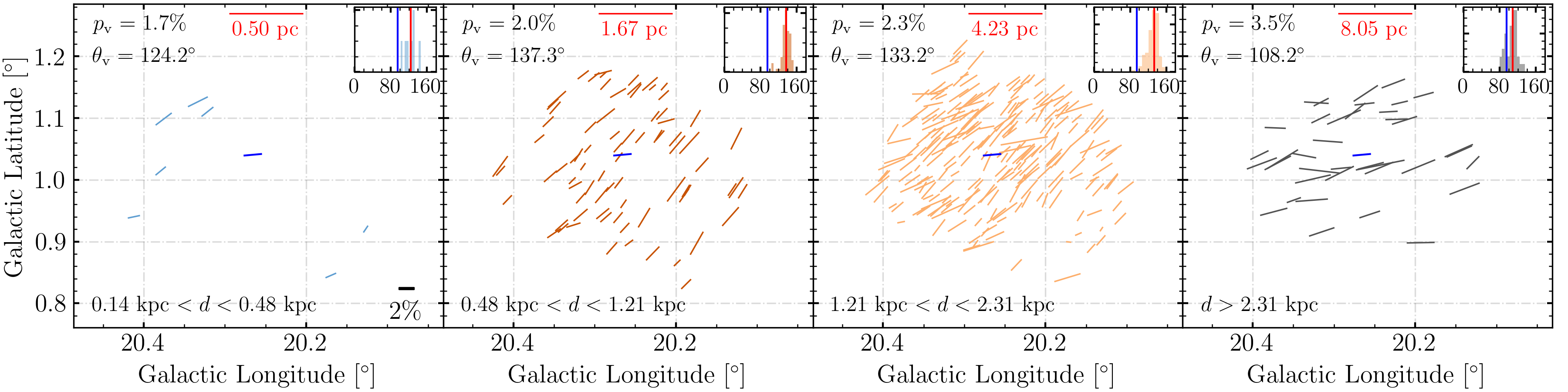}{\linewidth}{}
                      }\vspace{-0.8cm}
            \caption{Polarization maps of star groups located behind each BISP-1 layer in \textit{C5} (top), \textit{C45} (middle), and \textit{C50} (bottom). The insets in the top right show the distribution of the polarization angle observed along with the median value (the red line) and the average magnetic field orientation from Planck (the blue line, also shown as a pseudo-vector in the center of the map). The median polarization properties of the components obtained with BISP-1 are shown in the top left of the panels (also see Table~\ref{tab:tomo_results}). The red ruler shows the actual spatial scale measured for an angular scale of $0\fdg1$ at the distance of the clouds.
            \label{fig:ips_Polvect}}
        \end{figure*}
        %
        \begin{figure*}[ht!]
            \gridline{
                      \fig{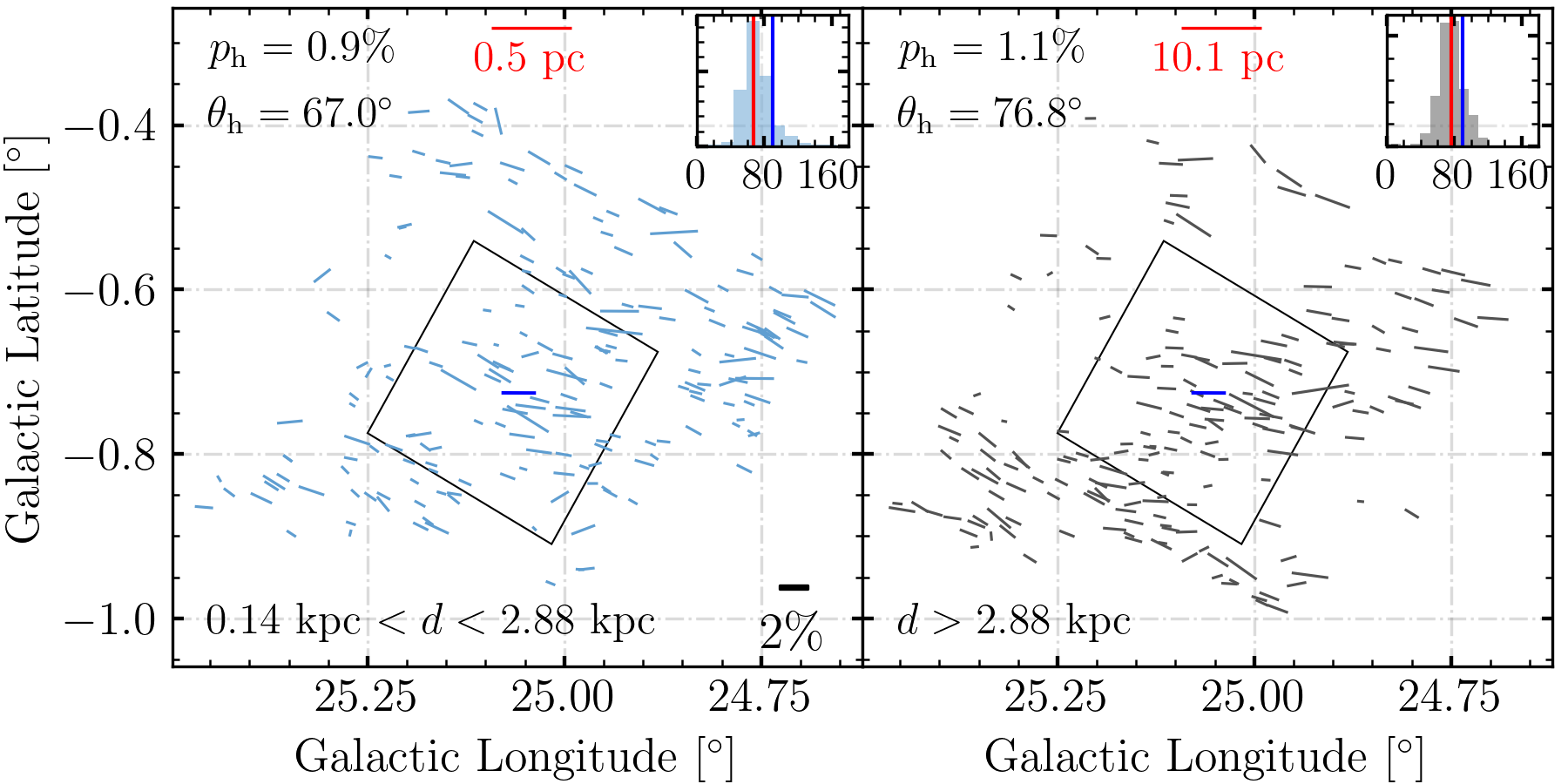}{.55\linewidth}{}
                      }\vspace{-0.9cm}
            \gridline{
                      \fig{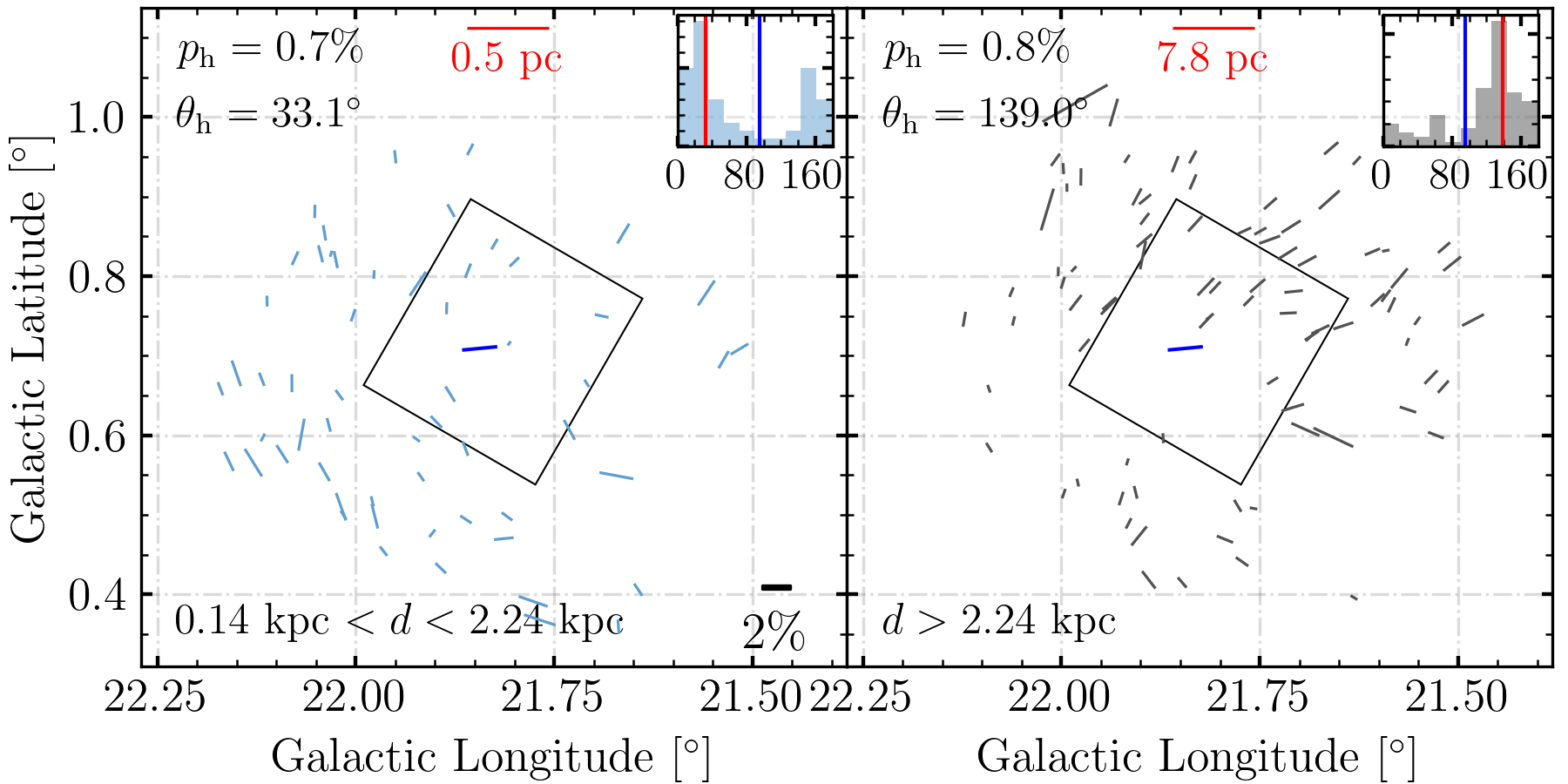}{.55\linewidth}{}
                      }\vspace{-0.9cm}
            \gridline{
                      \fig{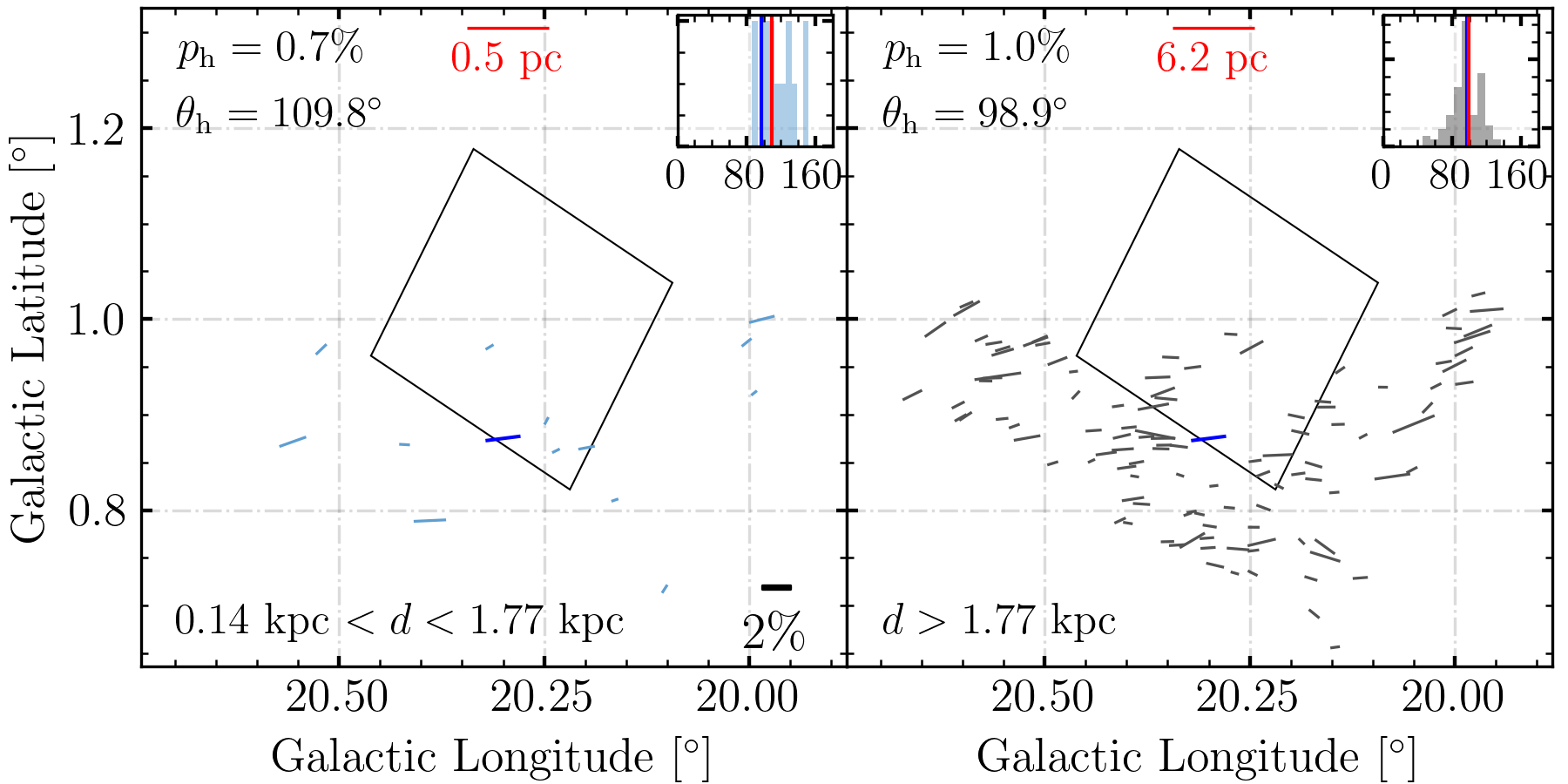}{.55\linewidth}{}
                      }\vspace{-0.8cm}
            \caption{Same as in Figure~\ref{fig:ips_Polvect} but using NIR polarization observations. The black rectangle indicates the approximate location of the IPS-GI field of view.
            \label{fig:gpips_Polvect}}
        \end{figure*}

        \subsubsection{C5}  
        \label{subsec:res_tomo_BISP-1_C5}

            The best optical polarization model obtained in \textit{C5} using BISP-1 includes three polarizing layers at $116^{+20}_{-10}$~pc, $477^{+10}_{-10}$~pc, and $2.7^{+0.1}_{-0.1}$~kpc (\mbox{Figure~\ref{fig:BISP-1_tomo_results_IPS}, left)}. The NIR polarization decomposition model showed two polarizing screens at $144^{+70}_{-30}$~pc and $2.88^{+0.1}_{-0.09}$~kpc, which are consistent within uncertainties with the optical model. 
            The median optical polarization degree increases gradually from $1.0^{+0.3}_{-0.3}\%$ behind the nearest cloud to $1.7^{+0.05}_{-0.05}\%$ and $2.5^{+0.1}_{-0.1}\%$ behind the intermediate and farthest clouds, respectively (Table~\ref{tab:tomo_results}). Meanwhile, the NIR polarization shows a smaller increase from $0.9^{+0.04}_{-0.04}\%$ to $1.1^{+0.04}_{-0.04}\%$. 
            
            The mean LOS-integrated optical polarization orientation (Figure~\ref{fig:ips_Polvect}, top) shifts significantly from $118\degr{\pm}10\degr$ behind the first cloud to $77\fdg0{\pm}0\fdg6$ behind the intermediate cloud and remains stable at $77\degr{-}79\degr$ beyond the distant layer, within a circular standard deviation of at most $1\degr$. NIR polarization follows a similar trend, increasing from $67\fdg0{\pm}1\fdg2$ to $76\fdg8{\pm}1\fdg2$ (Figure~\ref{fig:gpips_Polvect}, top). While both orientations match beyond the distant cloud, they differ significantly after the nearby cloud due to the limited NIR measurements, which likely average the properties of two clouds instead of just the nearest one.
        
        \subsubsection{C45}  
        \label{subsec:res_tomo_BISP-1_C45}

            In \textit{C45}, optical polarization decomposition using BISP-1 identified four polarizing screens at $105^{+4}_{-4}$~pc, $476^{+40}_{-30}$~pc, $731^{+20}_{-40}$~pc, and $1.64^{+0.2}_{-0.07}$~kpc (Figure~\ref{fig:BISP-1_tomo_results_IPS}, middle), while NIR polarization decomposition revealed two clouds at $139^{+40}_{-30}$~pc and $2.2^{+0.1}_{-0.1}$~kpc. The nearby NIR cloud roughly agrees with the optical results, but, as with \textit{C5}, significant uncertainties arise due to the limited number of nearby stars.

            Beyond the near and first-intermediate clouds, the LOS-integrated optical polarization remains around $0.9\%{-}1.1\%$, with uncertainties up to $0.4\%$ (Table~\ref{tab:tomo_results}). It then increases to $2.0^{+0.2}_{-0.2}\%$ past the second-intermediate cloud and decreases to $1.0^{+0.2}_{-0.2}\%$ after the farthest cloud. This trend likely reflects drastic changes in the magnetic field orientation along the LOS and across the observed field of view; for example, see the curved pattern present in each group of stars in Figure~\ref{fig:ips_Polvect}. NIR polarization stays nearly constant at $0.7\%{-}0.8\%$, with an error of $0.1\%$, with no confirmation of the intermediate cloud.

            The optical LOS-integrated polarization angle shifts significantly from $175\degr{\pm}10\degr$ behind the near cloud (Figure~\ref{fig:ips_Polvect}, middle) to $49\degr{\pm}17\degr$ behind the first-intermediate layer, then to $33\degr{\pm}3\degr$ and $7\degr{\pm}10\degr$ beyond the second-intermediate and distant clouds, respectively.
            NIR polarization follows a different trend, changing from $33\degr{\pm}5\degr$ nearby to $139\degr{\pm}3\degr$ farther away (Figure~\ref{fig:gpips_Polvect}, middle). Observations show a polarization angle distribution that broadens with distance behind virtually every cloud (Figures~\ref{fig:ips_Polvect} and~\ref{fig:gpips_Polvect}). The curved patterns, described by background stars, suggest complex, large-scale magnetic field structures that a single bivariate normal distribution with a simple mean orientation cannot fully represent. This complexity also limits the effectiveness of BISP-1 as it accounts only for polarization variations along the line of sight.
            
            %
            \begin{figure*}[ht]
            \gridline{
                      \fig{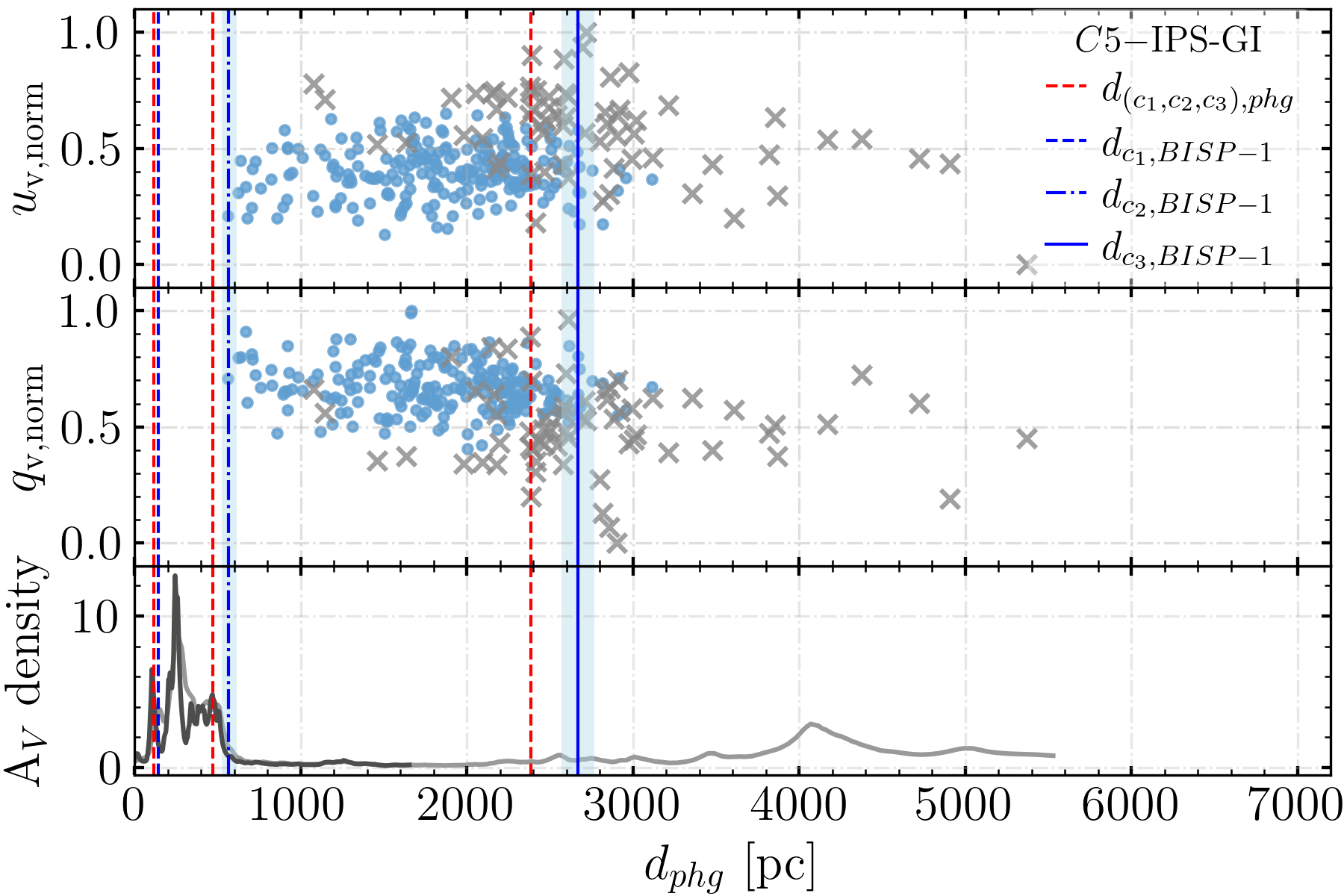}{.49\linewidth}{}
                      \fig{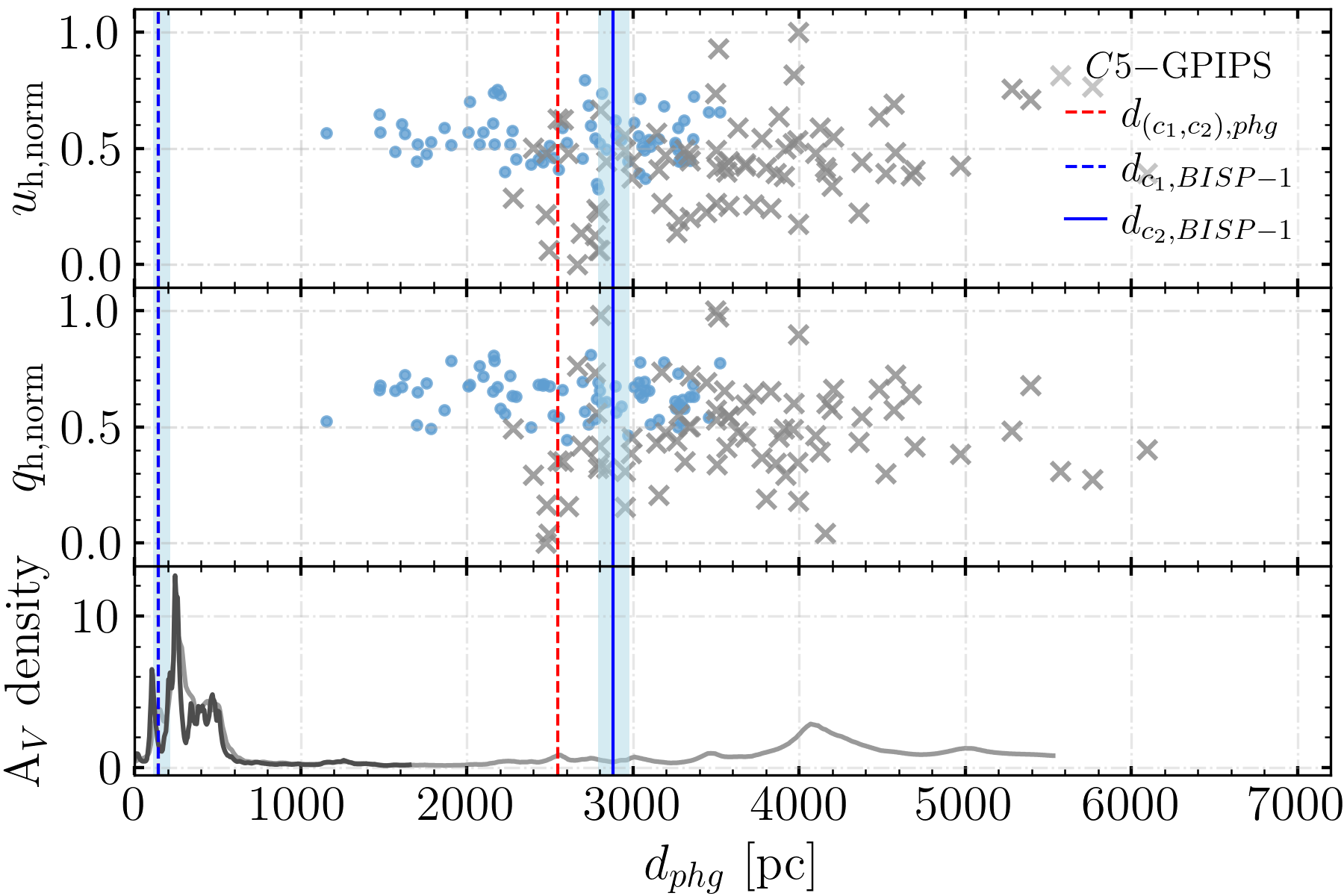}{.49\linewidth}{}
                      }\vspace{-0.9cm}
            \gridline{
                      \fig{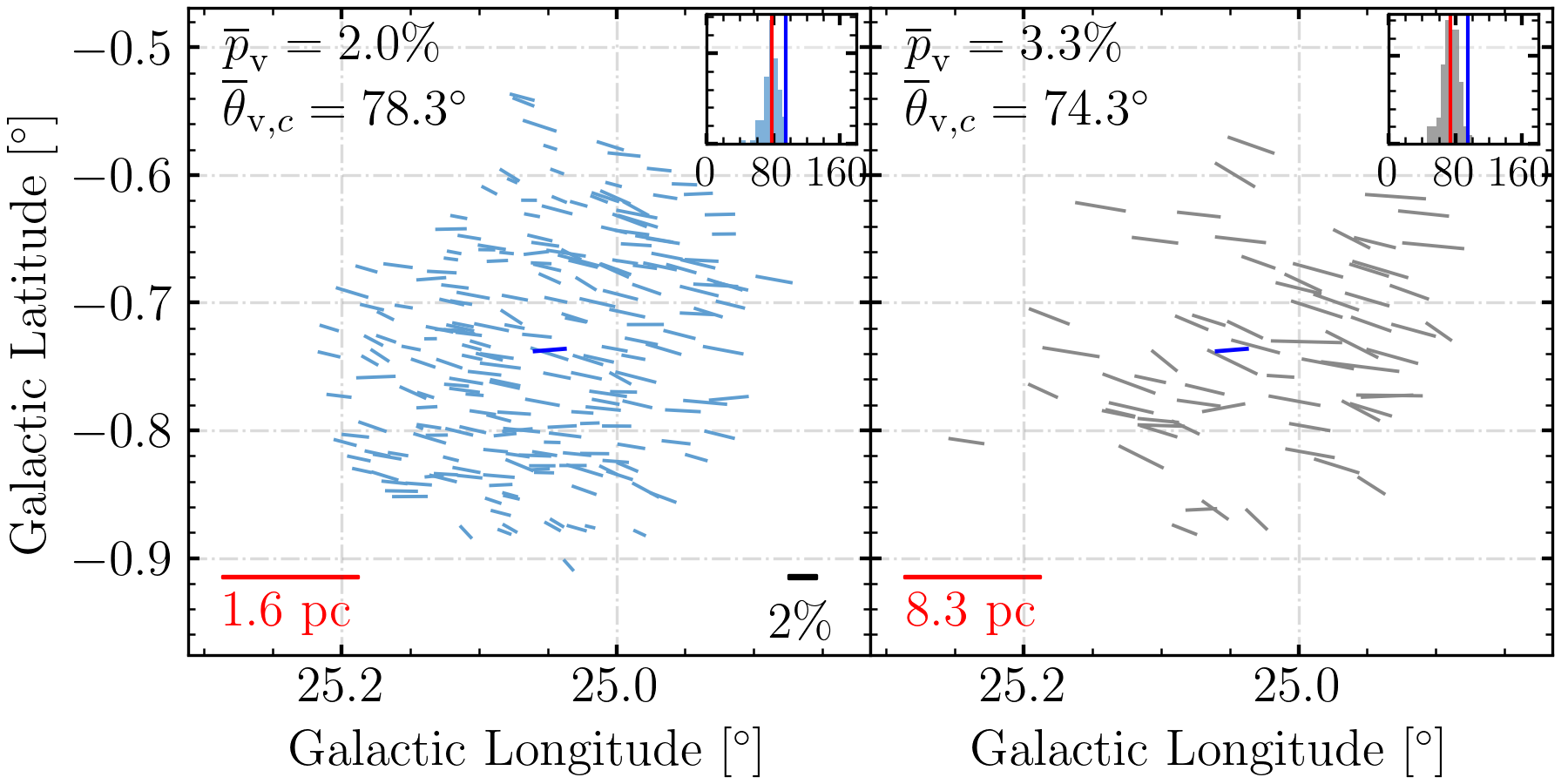}{.49\linewidth}{}
                      \fig{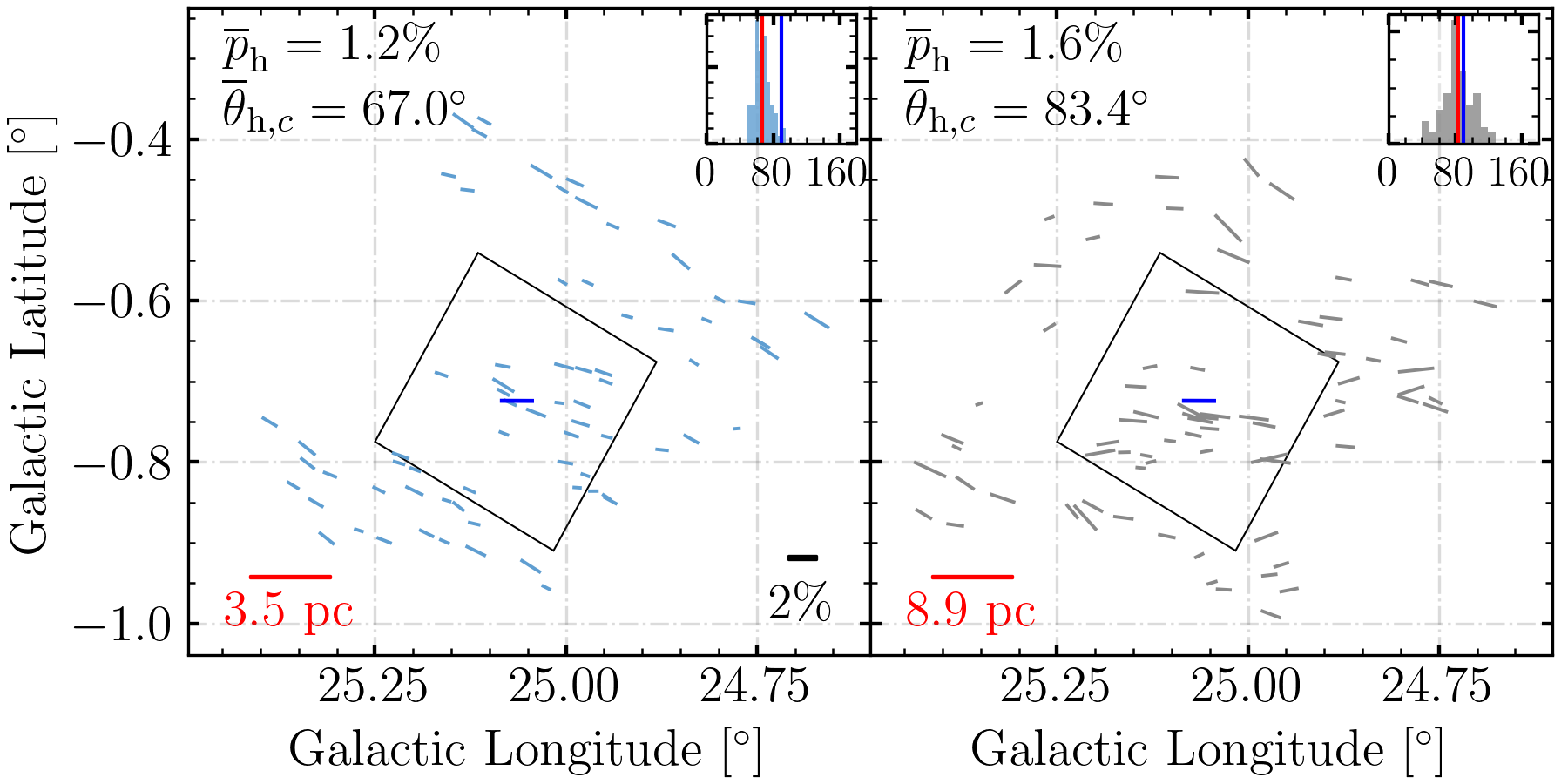}{.49\linewidth}{}
                      }\vspace{-0.8cm}
                \caption{GMM results with optical (left) and NIR (right) polarization in \textit{C5}. Top row: Stokes parameters $q_\mathrm{(v,h)}$ and $u_\mathrm{(v,h)}$ as a function of distance. The different colors and symbols represent the groups identified by the GMM. The extinction density with units of $10^{-3}$~mag~pc$^{-1}$ from \cite{Vergely_2022}, at $10$~pc (dark gray) and $50$~pc (light gray) resolution, is included in the bottom panel for comparison. The blue vertical lines show the location of the polarizing clouds identified with BISP-1; the red dashed vertical lines are the corresponding corrected distance (see Section~\ref{subsec:Disc_dist_Model}). Bottom row: polarization orientation of stars in each cluster. The insets show the polarization angle distributions and the mean values (the red line). The black rectangle in the right panels indicates the approximate location of the IPS-GI field of view. The dark-blue pseudo-vectors represent the average orientation of the magnetic field within the region as observed by Planck.
                \label{fig:GMM_resul_C5}}
            \end{figure*}

        \subsubsection{C50}  
        \label{subsec:res_tomo_BISP-1_C50}

            The four-cloud model is the best result for \textit{C50} using BISP-1 with optical polarization (Figure~\ref{fig:BISP-1_tomo_results_IPS}, right), identifying polarizing screens at $144^{+60}_{-30}$~pc, $478^{+40}_{-20}$~pc, $1.21^{+0.08}_{-0.06}$~kpc, and $2.31^{+0.04}_{-0.04}$~kpc.  NIR polarization decomposition showed two clouds at $141^{+50}_{-30}$~pc and $1.8^{+0.1}_{-0.1}$~kpc, with the near-cloud estimates aligning well with the optical results, but a $600$~pc discrepancy in the distant cloud. This difference may be due to limited NIR data coverage, as GPIPS observations do not cover the entire \textit{C50} field of view (Figure~\ref{fig:FoV_data}), causing optical and NIR to probe different ISM regions.
            
            The $q_{(\mathrm{v,h})}$ and $u_{(\mathrm{v,h})}$ distributions as a function of distance show a significant change in polarization properties of the stars behind the distant layer, which are better-defined in optical polarization. The median LOS-integrated optical polarization degree is higher in \textit{C50} than in other regions, increasing  gradually from $1.7^{+0.2}_{-0.2}\%$ to $3.5^{+0.1}_{-0.1}\%$ with distance (Table~\ref{tab:tomo_results}). In contrast, NIR polarization is lower, ranging from $0.7^{+0.1}_{-0.08}\%$ to $1.0^{+0.05}_{-0.05}\%$. 
            
            The LOS-integrated polarization orientation changes significantly from one group of stars to another in optical and NIR, except between intermediate clouds in optical data, where the main difference is in polarization degree (see the bottom rows of Figures~\ref{fig:ips_Polvect} and~\ref{fig:gpips_Polvect}, respectively). Optical polarization angles change from $124\degr{\pm}4\degr$ behind the near cloud to $137\fdg3{\pm}1\fdg2$ and $133\fdg2{\pm}0\fdg6$ behind the intermediate clouds, and then to $108\fdg2{\pm}1\fdg5$ behind the distant cloud. NIR polarization follows a different trend, shifting from $110\degr{\pm}6\degr$ in the nearby ISM to $99\degr{\pm}2\degr$ farther away. The differences between optical and NIR polarization may result from NIR field of view limitations and the potential averaging of multiple polarizing screens in the NIR near-cloud component. 
            
            %
            \begin{figure*}[ht]
            \gridline{
                      \fig{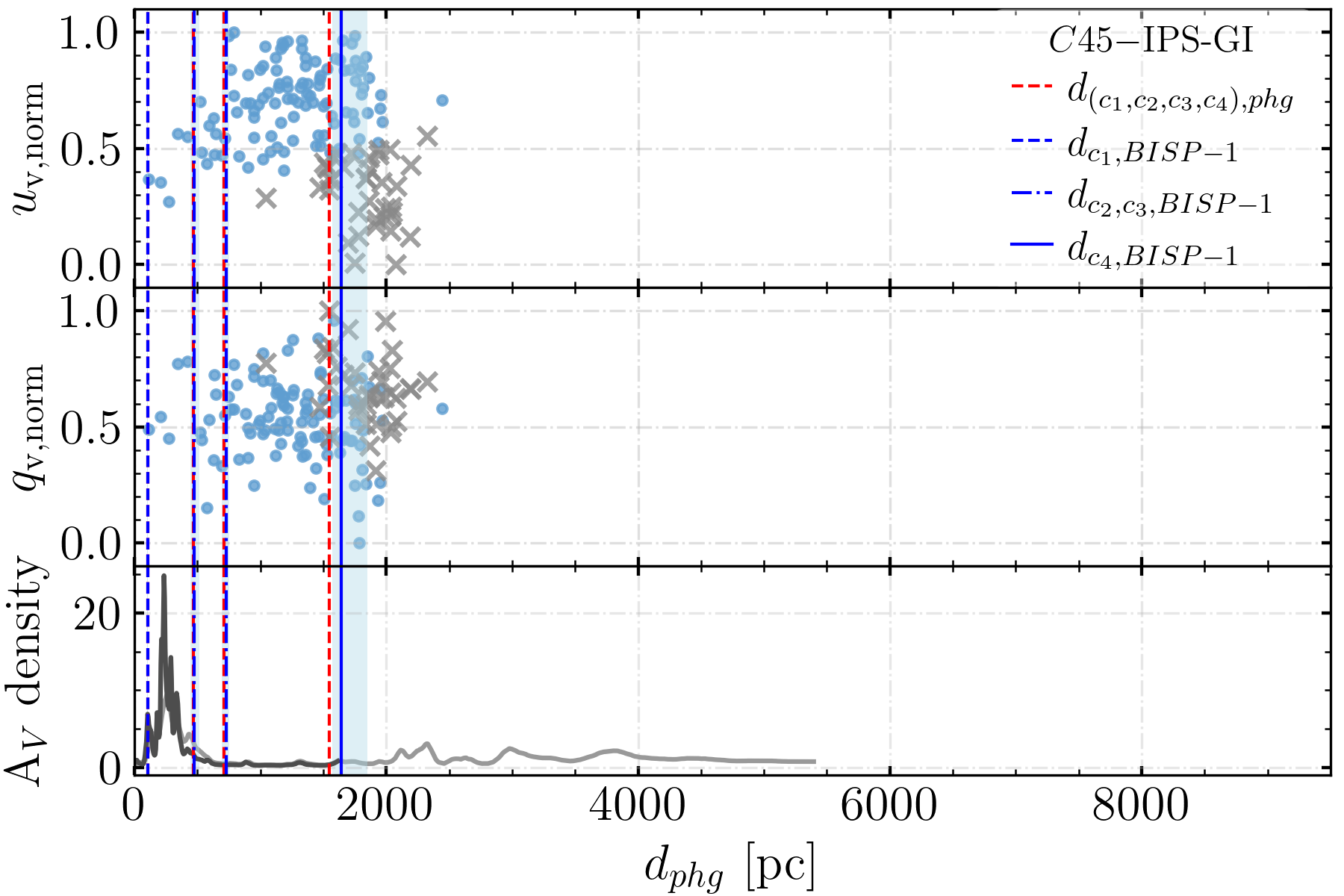}{.49\linewidth}{}
                      \fig{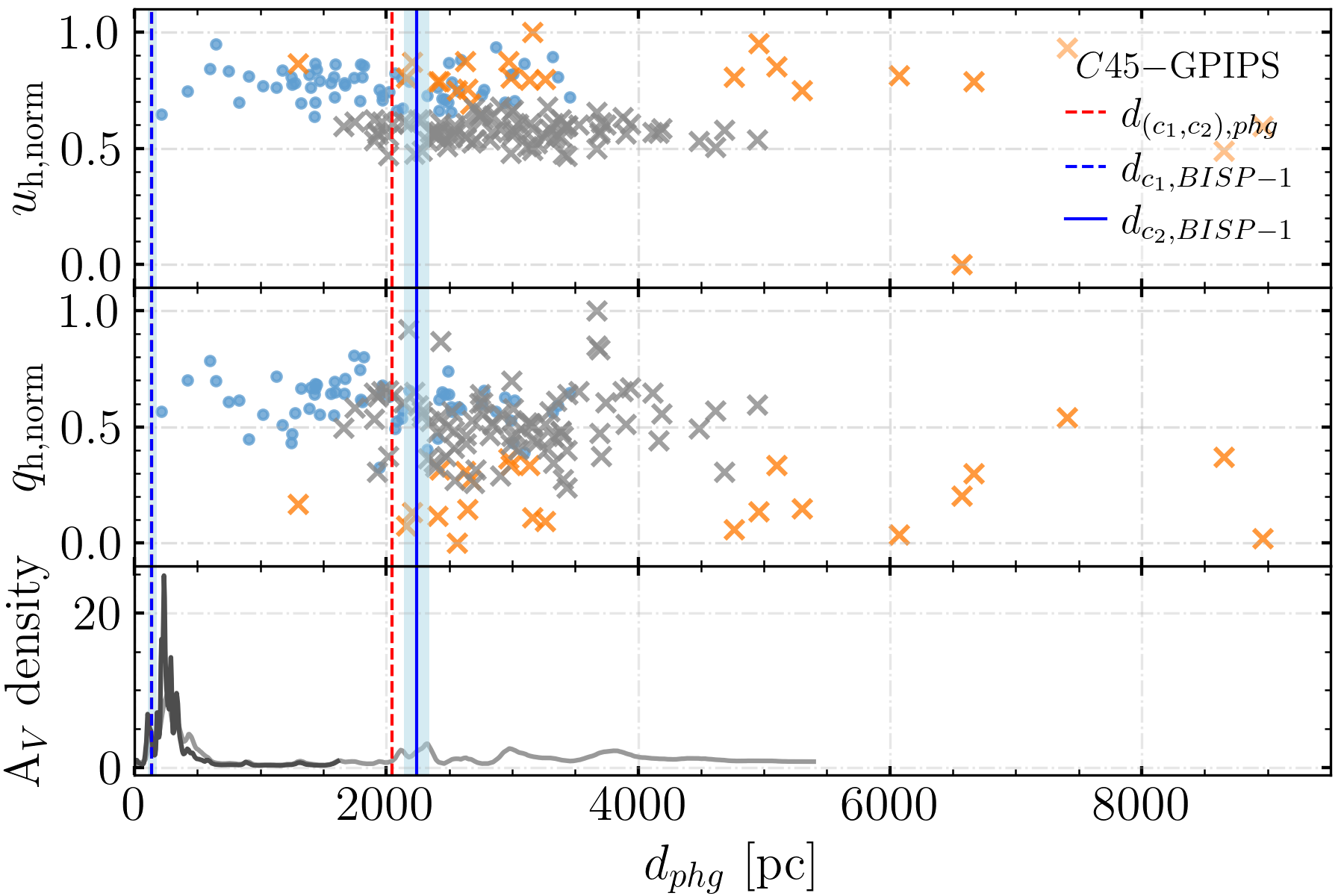}{.49\linewidth}{}
                      }\vspace{-0.9cm}
            \gridline{
                      \fig{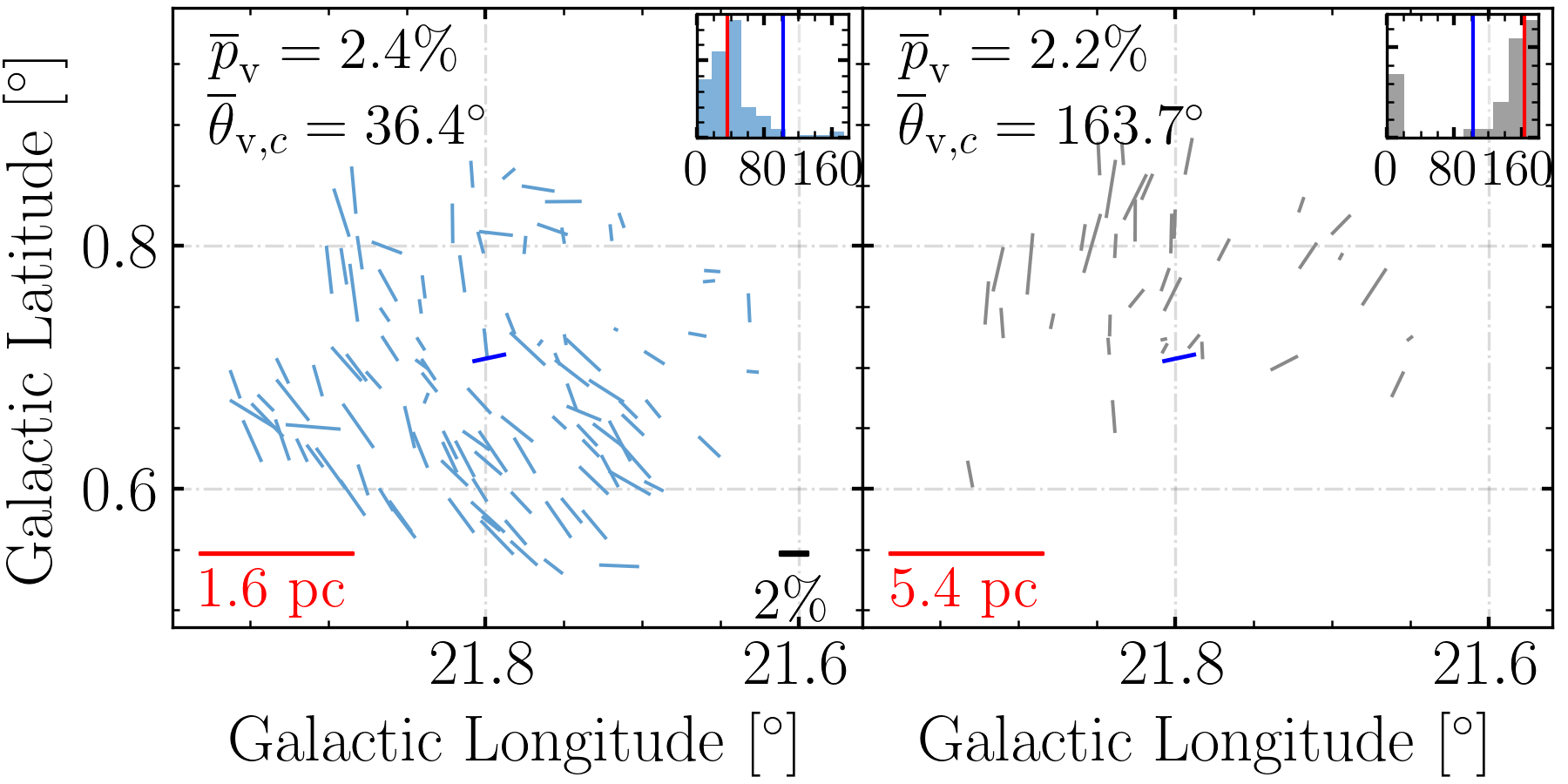}{.39\linewidth}{}
                      \fig{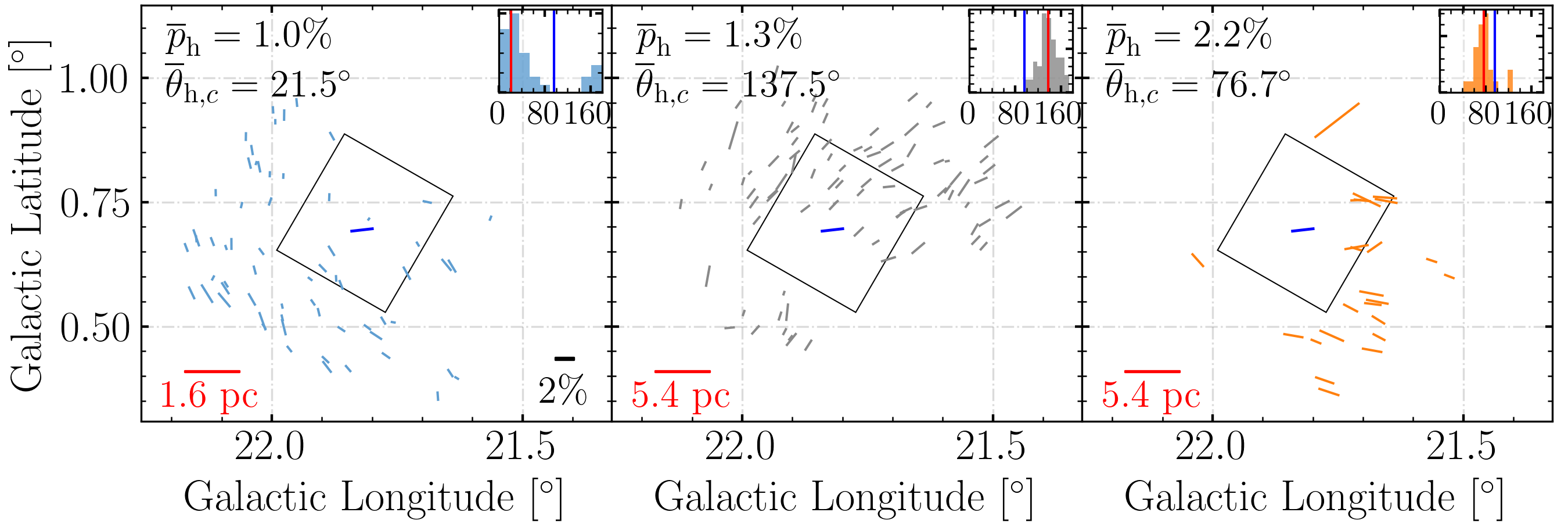}{.59\linewidth}{}
                      }\vspace{-0.8cm}
                \caption{Same description as in Figure~\ref{fig:GMM_resul_C5}, but for \textit{C45}.
                \label{fig:GMM_resul_C45}}
            \end{figure*}
            
            %
            \begin{figure*}[ht]
            \gridline{
                      \fig{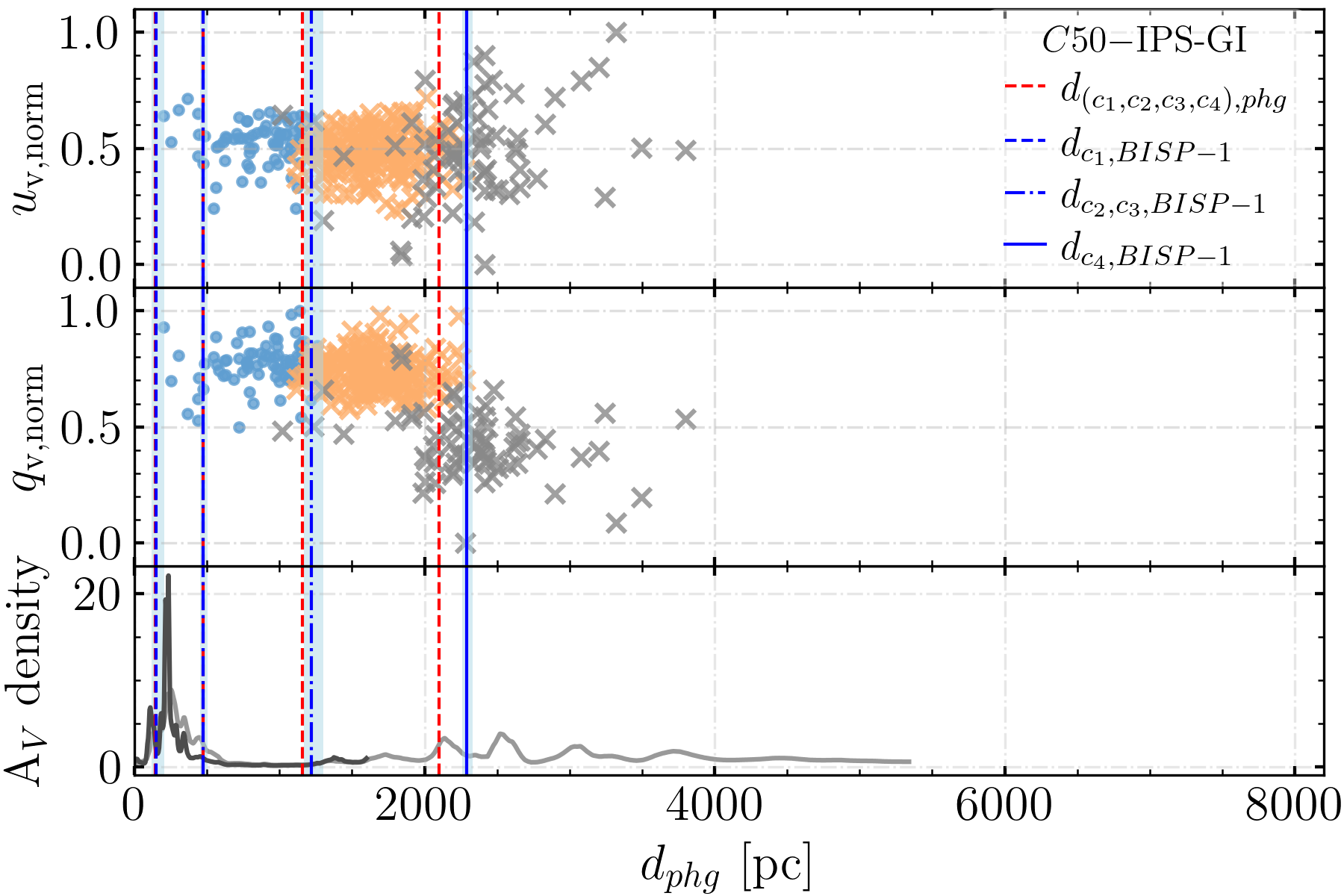}{.49\linewidth}{}
                      \fig{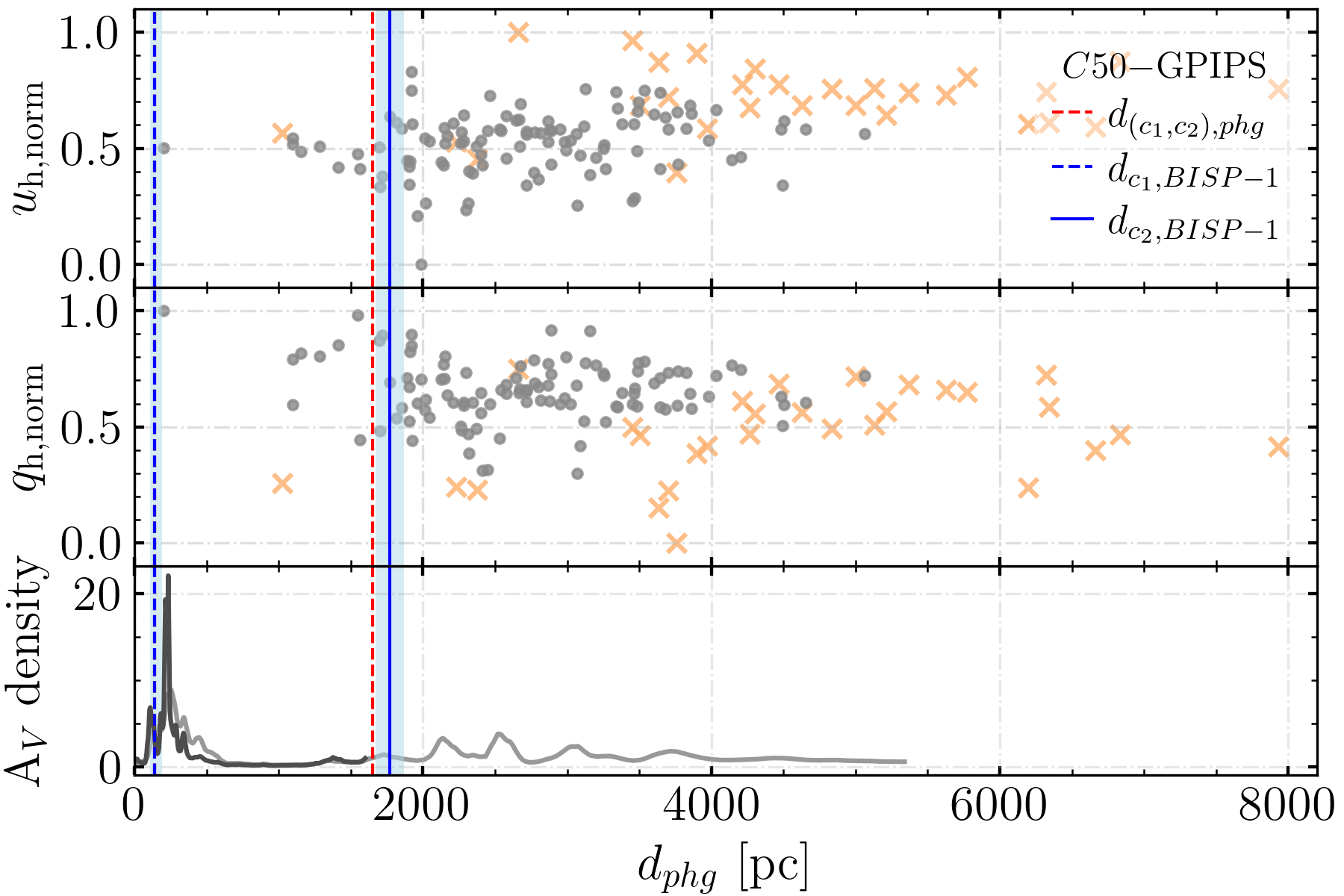}{.49\linewidth}{}
                      }\vspace{-0.9cm}
            \gridline{
                      \fig{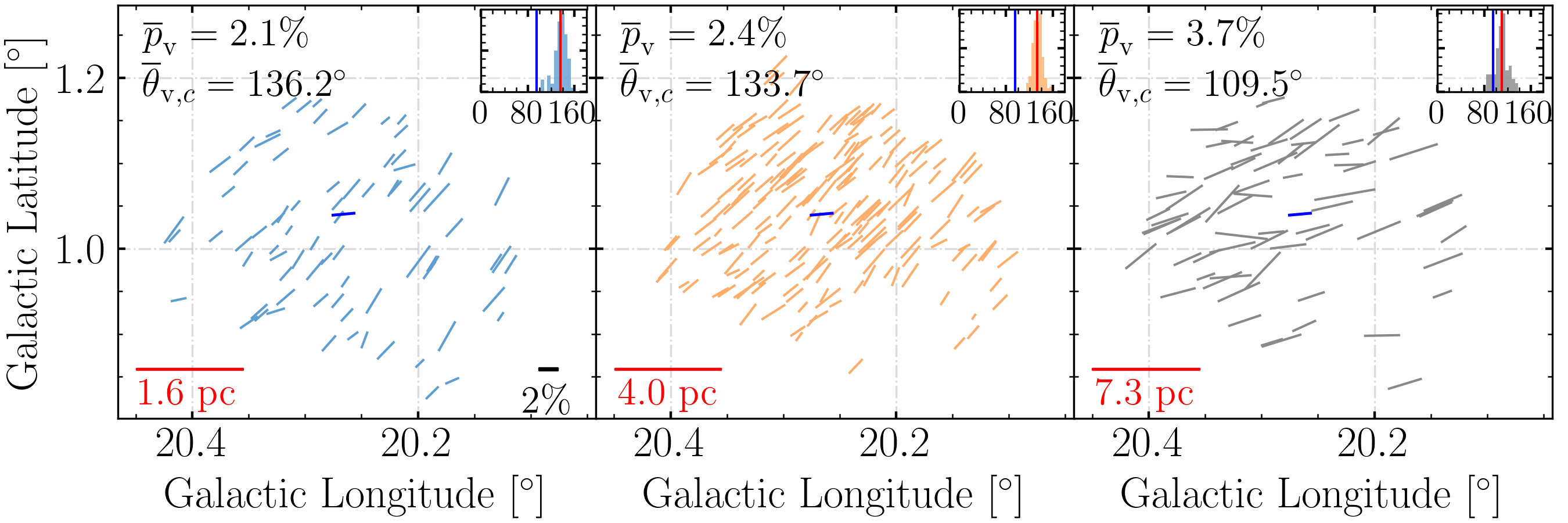}{.59\linewidth}{}
                      \fig{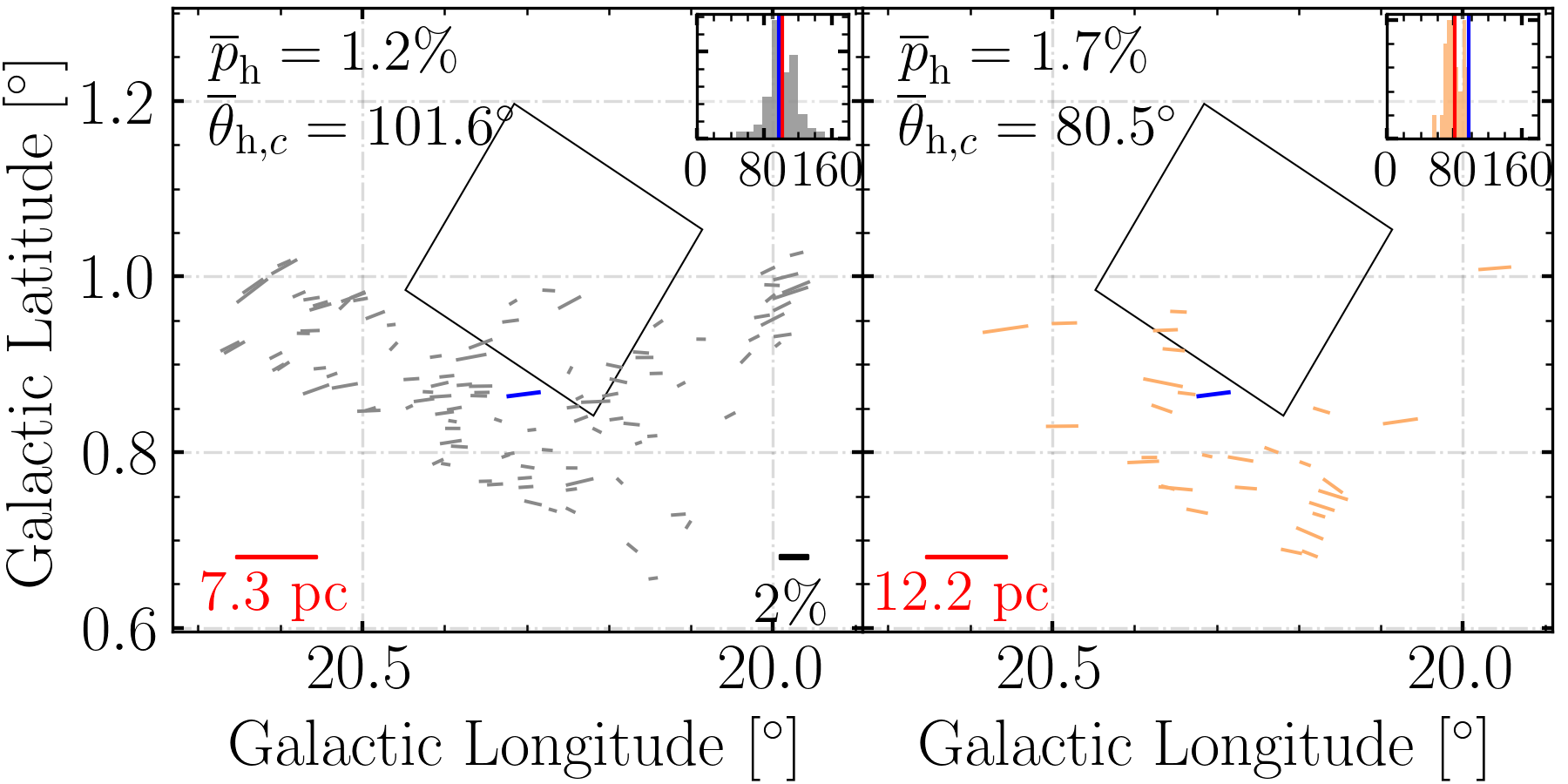}{.39\linewidth}{}
                      }\vspace{-0.8cm}
                \caption{Same description as in Figure~\ref{fig:GMM_resul_C5}, but for \textit{C50}.
                \label{fig:GMM_resul_C50}}
            \end{figure*}

    \subsection{Magnetic field tomography using GMM} 
    \label{subsec:res_tomo_GMM}

        Figures~\ref{fig:GMM_resul_C5} to~\ref{fig:GMM_resul_C50} present the clustering results with optical and NIR polarization in \textit{C5}, \textit{C45}, and \textit{C50}. The solutions showed clear separation between Gaussian components in the $q{-}u$ space, with most measurements having membership probabilities above $\sim80\%$ and well-defined by distance (see Figure~\ref{fig:Append_GMM_result} in Appendix~\ref{sec:Append_C}). 
        We identified at least two components in each region, except for optical polarization in \textit{C50} and NIR polarization in \textit{C45}, where we found three groups that best describe the different polarization properties along the LOS and across the field of view. In the following sections, we present the GMM clustering results for each observed region. The average polarization properties are displayed in Table~\ref{tab:tomo_results}.

        \subsubsection{C5}  \label{subsubsec:res_tomo_GMM_C5}
        
            We identified two main groups in the optical and NIR observations separately, using the GMM method (Figure~\ref{fig:GMM_resul_C5}). The approximate threshold of the farthest group agrees with the distant cloud obtained with BISP-1 using optical and NIR observations (the vertical blue line).
            We increased the SNR cutoff limit to eight in the optical data and five in the NIR, sacrificing information from low-polarization measurements. This reduces the scatter and improves the clustering.

            The mean optical polarization orientation of each group is consistent, within the dispersion, with the counterpart component in the NIR (Figure~\ref{fig:GMM_resul_C5}). Furthermore, although the nearest component of the clustering is expected to represent the combined contribution of the near and intermediate clouds found with BISP-1, the average polarization properties align more closely with those of the intermediate cloud alone (Table~\ref{tab:tomo_results}). 
            
        \subsubsection{C45}  
        \label{subsubsec:res_tomo_GMM_C45}
        
            The clustering of optical polarization in \textit{C45} yielded two components (Figure~\ref{fig:GMM_resul_C45}), which are consistent with the intermediate and distant components of the BISP-1 solution (Figure~\ref{fig:BISP-1_tomo_results_IPS}, middle). However, due to the low number of measurements in the nearby ISM, the clustering algorithm could not differentiate between the near and intermediate components identified by the BISP-1 decomposition, thus treating them as a single group. 
            
            On the other hand, the clustering with NIR polarization showed three groups in \textit{C45} (Figure~\ref{fig:GMM_resul_C45}, right). Two of these groups are closely related to variations of the polarization properties with distance and align with both optical polarization clustering (Figure~\ref{fig:GMM_resul_C45}, left) and the BISP-1 decomposition results (Figure~\ref{fig:BISP-1_tomo_results_IPS}, middle). The third group has different polarization properties across the field of view rather than along the sightline (Figure~\ref{fig:GMM_resul_C45}). Most measurements from this group are outside the IPS-GI field of view and observed distance range. Additionally, these measurements come from the faint red stars removed for the NIR polarization decomposition with BISP-1 (Appendix~\ref{sec:Append_B}).
            
        \subsubsection{C50}  
        \label{subsubsec:res_tomo_GMM_C50}
    
            In \textit{C50}, the optical polarization clustering resulted in three components (Figure~\ref{fig:GMM_resul_C50}, left). All three groups closely align with the two intermediate and the distant components obtained with BISP-1 (Figure~\ref{fig:BISP-1_tomo_results_IPS}, right). However, as in \textit{C45}, the nearest component includes the contribution from two different clouds, the near and intermediate of BISP-1 results. In the case of NIR polarization (Figure~\ref{fig:GMM_resul_C50}, right), there are very few nearby measurements. The majority of the NIR observations are beyond the known threshold of the distant cloud, $1.8$~kpc (Table~\ref{tab:tomo_results}). As a result, the clustering does not recognize the nearest measurements as a separate group. Moreover, it seems to find a component even farther away than what optical polarization can detect. However, this additional polarization layer is not found with BISP-1 using the same observations. We attribute this discrepancy to either GMM failing to separate the components along the LOS due to the lack of nearby stars or to the presence of distant faint red stars, removed for the BISP-1 decomposition (Appendix~\ref{sec:Append_A}), which provide additional information at large distances in \textit{C50}. 
            
\section{Discussion} 
\label{sec:discu}
        
    \subsection{The inverse parallax-distance relation} 
    \label{subsec:Disc_dist_Model}

        The BISP-1 Bayesian algorithm relies on the Gaussian nature of the error in the parallax to evaluate the probability of a star being in front of or behind a polarizing screen \citep{Pelgrims_2023}. However, it is known that the inverse parallax is only a good approximation to the true stellar distance if the parallax SNR is high \citep{Bailer_Jones_2021, GaiaDR3_Fouesneau_StellarParams_2023}. Furthermore, the least-squares astrometric solution obtained from Gaia data processing can result in zero and negative observed parallaxes in the presence of measurement noise \citep{Luri_GaiaDR2_plx_2018}. Since the inverse parallax is constrained to be positive, negative parallaxes are an issue and should be discarded. These conditions are easily fulfilled in high-latitude regions, as \cite{Pelgrims_2024} demonstrated. However, our regions are in the Galactic plane and have high column densities of the order of $10^{22}$~cm$^{-2}$ (Figure~\ref{fig:FoV_data}). Additionally, several stars lie at large distances, resulting in low parallax SNR (Appendix~\ref{sec:Append_A}).
        Consequently, the inverse parallax is not a reliable estimate of the true stellar distance.   
            
        %
        \begin{figure}[t!]
            \epsscale{1.2}
            \plotone{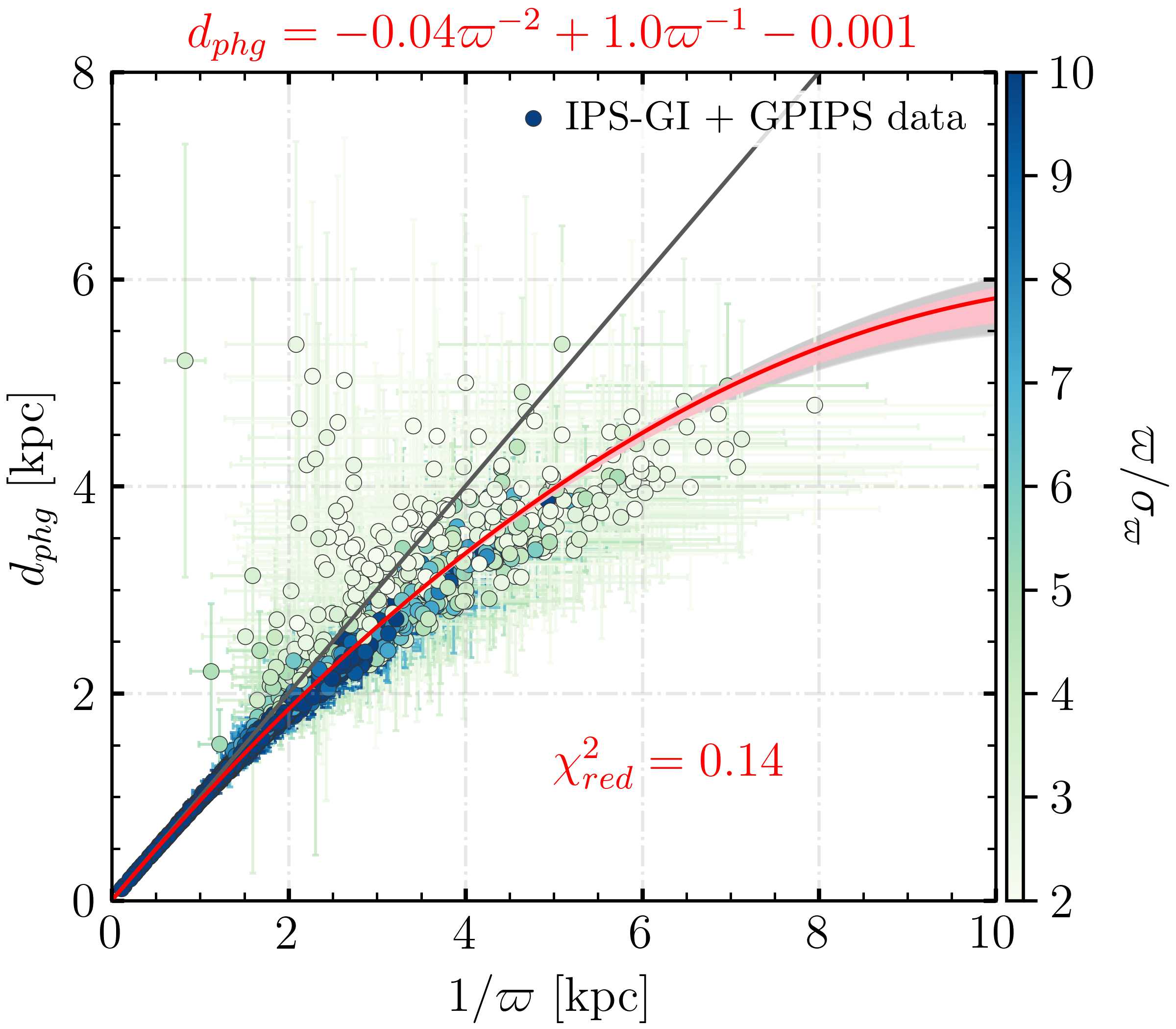}
            \caption{Photo-geometric distances from \cite{Bailer_Jones_2021} as a function of the inverse parallax of the stars observed by IPS-GI and GPIPS in the region of \textit{C5}, \textit{C45}, and \textit{C50}. The solid gray line indicates the expected inverse parallax-distance relation. The shaded pink area represents the best fits of subsamples drawn without replacement from the measured data. The solid red line is the average best fit of the subsamples. The gray shaded area is the $2\sigma$ confidence intervals.
            \label{fig:photogeo_dist_vs_plx_models}}
        \end{figure}
        
        This is confirmed when comparing the photo-geometric distance with the respective inverse parallax within each dataset (Figure~\ref{fig:photogeo_dist_vs_plx_models}). The stellar photo-geometric distances from \cite{Bailer_Jones_2021} are calculated using all Gaia photometry information alongside parallaxes, thereby considering factors such as interstellar extinction. The validation of their solutions against independent distance estimates demonstrated that photo-geometric distances are reliable up to several kiloparsecs \citep{Bailer_Jones_2021}. 
        
        The gray solid line in Figure~\ref{fig:photogeo_dist_vs_plx_models} indicates the linear inverse parallax-distance relation, from which the stars at $d\gtrsim2$~kpc deviate significantly. A two-degree polynomial fit to the data weighted by the uncertainties describes the systematic deviation from the linear relation.
        We used the orthogonal distance regression (ODR) method, which allowed us to include the uncertainties of both variables in the fitting. We used standard error propagation to calculate the inverse parallax errors as $\sigma_{\varpi}/\varpi^2$. We performed $10{,}000$ realizations with subsamples drawn without replacement from the measured data. The size of the subsamples was defined as $k = M-200$, where $M$ is the total number of measurements (${\sim}1{,}800$), including all objects observed in optical and NIR across the three regions. 
        The reduced $\chi^2$ of the average best fit is presented in 
        Figure~\ref{fig:photogeo_dist_vs_plx_models}. The average model allows us to relate the distance obtained from the tomography analysis with BISP-1 (i.e.,~$1/\varpi$) with the photo-geometric distance. 
        The corrected photo-geometric distance values of each cloud are presented in Table~\ref{tab:tomo_results}.

    \subsection{A comparison between polarization decomposition methods}  
    \label{subsec:Disc_tomo_methods_comparison}
        
        The GMM method clusters stars with similar polarization--distance properties, but does not account for uncertainties. The polarization properties of each group are LOS-integrated, and while distance information is used, separation by distance is not enforced.
        In contrast, BISP-1 decomposes the observed polarization to determine the intrinsic properties of individual clouds--only along the LOS--while considering ISM scatter and observational uncertainties. Additionally, BISP-1 provides LOS-integrated polarization properties for stars behind each cloud. In this paper, we directly compare BISP-1's LOS-integrated component properties with the GMM method results with other ISM observations.

        GMM performs well when polarization data have a clear change with distance, as in \textit{C50}, where polarization can be approximated as a bivariate distribution. However, GMM is less effective when the transition between components is very smooth or when the data scatter is large, as seen in \textit{C5}. For instance, increasing the SNR threshold in optical and NIR polarization allowed the identification of at least two components along the LOS in \textit{C5}, but low measurement density in some components can prevent proper group assignment, as in the nearby components. Furthermore, although clustering offers a way to separate the dataset into groups with similar properties and assign a membership probability to each measurement, it cannot determine the distance of the polarizing clouds. 
                
        Nonetheless, GMM provides insights into magnetic field variations along the LOS and also seems to detect spatial variations across the field of view, tackling one limitation of BISP-1, e.g., the third group detected in \textit{C45} with NIR polarization (Figure~\ref{fig:GMM_resul_C45}). As seen in Figure~\ref{fig:ips_Polvect} (bottom), some polarization pseudo-vectors in the second-intermediate cloud have a similar orientation to those of the distant cloud, whereas in Figure~\ref{fig:GMM_resul_C50} (bottom), the components are clearly separated. This discrepancy arises from the sharp cutoff defined by the cloud's distance in BISP-1 decomposition and possibly from the inefficiency in detecting variations in the magnetic field properties across the field of view.
        
        Furthermore, unlike BISP-1, we can use photo-geometric distance instead of the inverse parallax with GMM, which allows us to include the faint red stars in NIR observations (Appendix~\ref{sec:Append_A}). Despite the difficulties of using clustering algorithms and their lack of uncertainty handling, their solution complements the Bayesian inference tomography results. With BISP-1, we can model the number of polarizing clouds and obtain their distances and average polarization properties alongside their uncertainties. Meanwhile, the GMM provides a method to separate the measurements into well-defined groups with different magnetic field properties along the sightline and possibly across the field of view as well. Together, these methods provide a robust picture of the plane-of-sky magnetic field properties.

    \subsection{Complex magnetic field structure in \textit{C45}}  
    \label{subsec:Disc_C45}

        The curved pattern of the plane-of-sky magnetic field orientation is observed in all \textit{C45}'s components along the LOS, more clearly in the BISP-1 decomposition (e.g.,~see Figures~\ref{fig:ips_Polvect} and~\ref{fig:gpips_Polvect}) and perhaps also in Planck's polarized emission (Figure~\ref{fig:FoV_data}). Low polarization with a random orientation is observed towards the Northeast of the \textit{C45} field of view, suggesting a change in the magnetic field orientation, likely perpendicular to the plane of the sky. These patterns remain along the entire LOS, as observed by optical and NIR polarization (e.g.,~see Figures~\ref{fig:ips_Polvect},~\ref{fig:gpips_Polvect}, and~\ref{fig:GMM_resul_C45}). The difference from one group of stars to another is the shift of the location of the low random polarization from the Northeast to the Southwest part of the field of view. 
        
        This means that variations in magnetic field orientations happen primarily across the field of view, in addition to the LOS variations, explaining the depolarization observed. However, BISP-1 cannot accommodate this type of variation and, as a result, is unlikely to accurately determine the polarization properties of the intricate structures in the \textit{C45} region. On the other hand, GMM clustering seems to deal better with variations across the sky, as explained in Section~\ref{subsec:Disc_tomo_methods_comparison}. However, the intricate nature of the GMF in \textit{C45}, showing large-scale variations in the plane-of-sky orientation, could also challenge the performance of the GMM method, as it does not account for observed uncertainties.
            
        Furthermore, a group of long sightlines observed within the \textit{C45} field of view in NIR exhibit a distinct plane-of-sky magnetic field orientation compared to the dominant patterns (Figure~\ref{fig:GMM_resul_C45}, right). These spatially localized measurements were detectable through the clustering method, which performs well when there are significant changes in polarization properties, not just along the LOS but also across the field, a feature that BISP-1 lacks. Moreover, most of these measurements originate from faint red stars, excluded in the BISP-1 analysis but retained during the clustering analysis. It is unclear what might lead to distinct magnetic field orientations in these particular sightlines. We speculate that it could be an elongated dust structure, such as a filament, within a distance not farther than $2$~kpc (Figure~\ref{fig:GMM_resul_C45}, right).
    
    \subsection{Magnetic field properties across the Galactic arms}  
    \label{subsec:Disc_Gal_tomo}
        
        \subsubsection{Nearby ISM or Local arm}  
        \label{subsec:Disc_tomo_GMF_local}

            We found nearby polarizing clouds with both optical and NIR observations. Optical observations better constrain the properties of the nearest clouds around $104$ and $143$~pc (Figure~\ref{fig:BISP-1_tomo_results_IPS} and Table~\ref{tab:tomo_results}). This distance range agrees with the approximate location of the Local Bubble wall \citep{Lallement_2003, Liu_LBW_2017, Pelgrims_2020}. High-resolution 3D dust maps have shown the presence of nearby dense and diffuse dust structures beyond and within the Local Bubble wall in a similar range of distances \citep{Leike_2020, Vergely_2022, Edenhofer_dustmap_2023}. The works of, for example, \cite{Skalidis_Pelgrims_2019} and \cite{Pelgrims_2024} at high Galactic latitudes, \cite{Angarita_2024} at intermediate latitudes, \cite{Andersson_Potter_CoalSack_2005} and \cite{Versteeg_2024} in the Coalsack region, and \cite{Medan_Andersson_2019} in the Local Bubble wall (at high latitudes) demonstrated that a large percentage of starlight polarization is produced in these nearby dust structures.  
        
            %
            \begin{figure}[t!]
                \epsscale{2.35}
                \plottwo{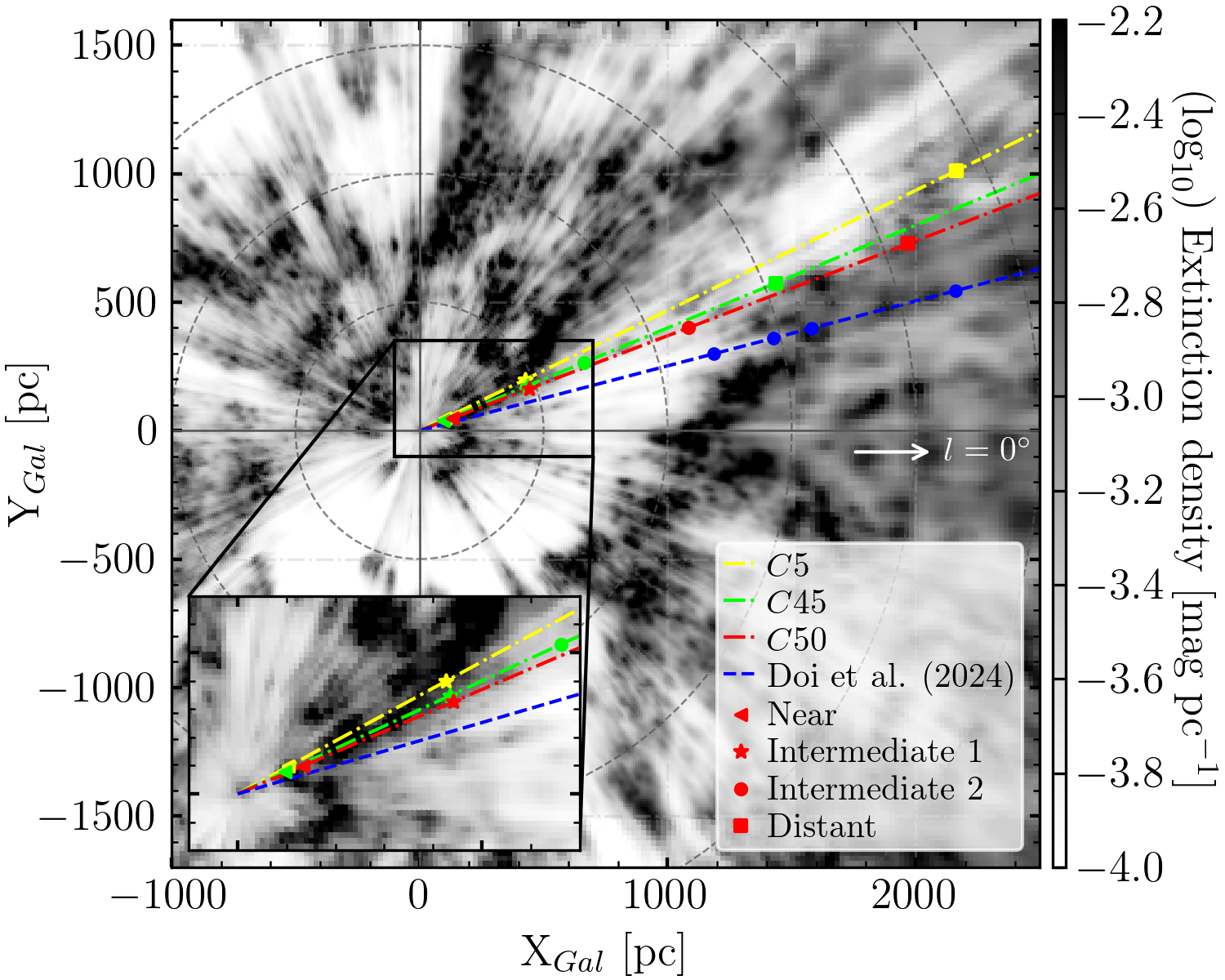}{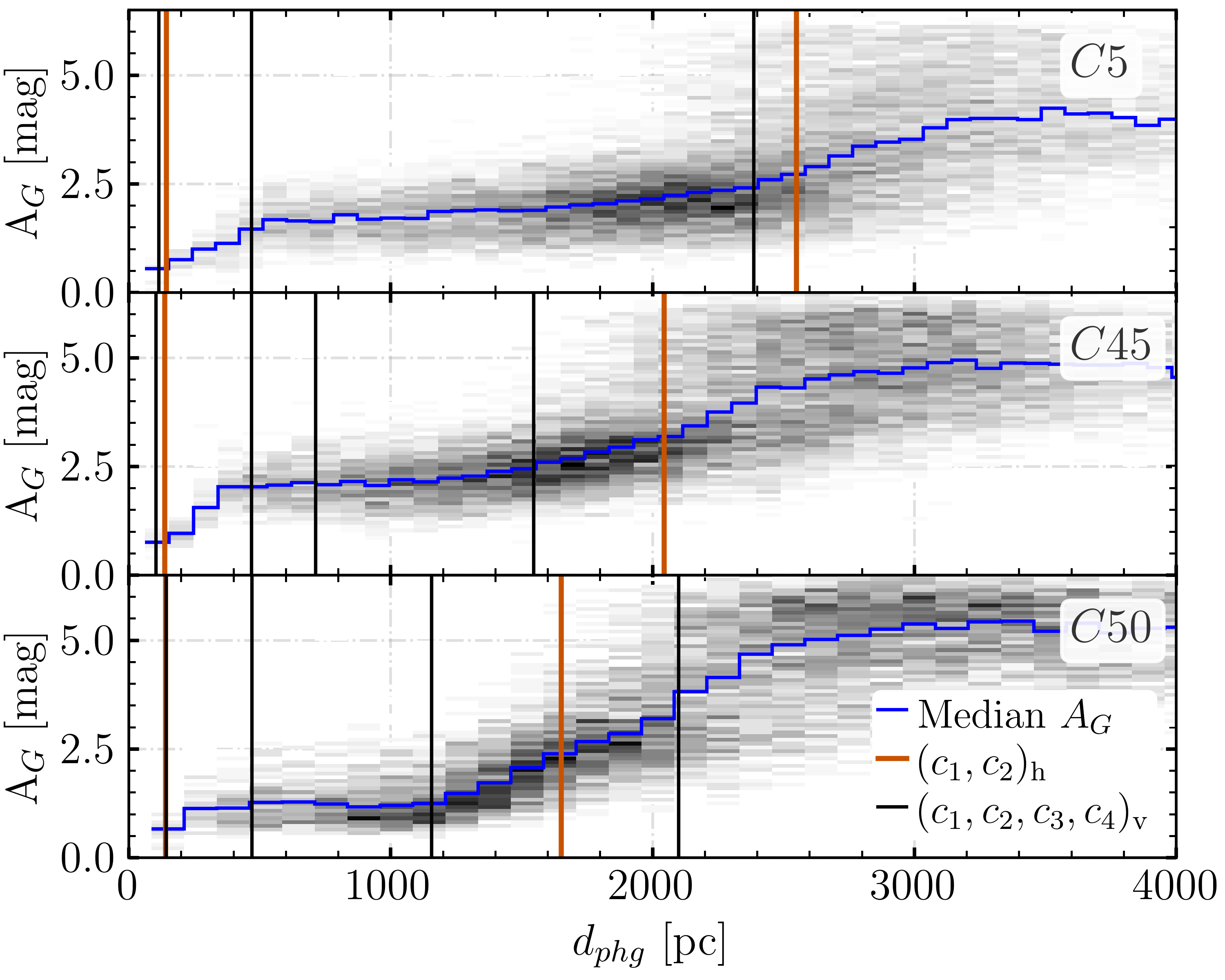}
                \caption{Top: LOSs of \textit{C5} (in yellow), \textit{C45} (in green), and \textit{C50} (in red) drawn over the extinction density XY plane from \cite{Vergely_2022} at $10$~pc ($d<1.5$~kpc) and $25$~pc ($d>1.5$~kpc) resolutions. The markers indicate the adjusted positions of optical polarization clouds. \cite{Doi_2024} results in the Sagittarius arm are displayed in blue for comparison. Bottom panels: \mbox{$G$-band} extinction from \cite{GaiaDR3_Collaboration_2023} as a function of the photo-geometric distance for all Gaia stars within \textit{C5} (top), \textit{C45} (middle), and \textit{C50} (bottom). The solid blue line is the median $A_G$ per distance bin. The vertical lines show the polarizing screens identified with BISP-1 using optical (black) and NIR (orange) observations, corrected for the bias in the inverse parallax-distance relation (Section~\ref{subsec:Disc_dist_Model}).
                \label{fig:LOS_3D_dustmap}}
            \end{figure}
            
            The intermediate polarizing screens observed with optical polarization are located between $470$ and $714$~pc. Figure~\ref{fig:LOS_3D_dustmap} (bottom) shows the LOS-integrated \mbox{$G$-band} extinction as a function of distance, indicating that these intermediate clouds correspond to significant changes in stellar dust extinction. According to high-resolution 3D dust maps (Figure~\ref{fig:LOS_3D_dustmap}, top), this distance range falls within the Local arm. More specifically, the intermediate clouds may be linked to the Southern structure of the Aquila Rift.

            In NIR polarization, the intermediate clouds (in the Local arm) are not detected in either method. This is likely because of the very low number of stars at distances at or in front of the Local arm, which makes the distance determination of (nearby or intermediate) clouds very uncertain. NIR polarization from nearby stars (i.e.,~those crossing the Local Bubble wall structures only) could be very low, causing fewer detections with low SNR.

        \subsubsection{Distant ISM or Sagittarius Arm}  
        \label{subsec:Disc_tomo_GMF_far}

            The second-intermediate cloud found at $1.16$~kpc (Table~\ref{tab:tomo_results}) in \textit{C50} (Figure~\ref{fig:BISP-1_tomo_results_IPS}, right) is expected to be beyond the Local arm, probably representing the transition between the interarm region and the Sagittarius arm according to 3D dust maps (Figure~\ref{fig:LOS_3D_dustmap}, top, and Figure~\ref{fig:ips_gpips_Av_dens_dist_pol}, third row). Furthermore, the distances to the farthest clouds are consistently traced by optical and NIR polarization between $1.5$ and $2.5$~kpc, far beyond the Local arm and within the Sagittarius arm. These distance ranges align with the magnetic field tomography analysis by \cite{Doi_2024} in a region close to \textit{C50} (Figure~\ref{fig:LOS_3D_dustmap}, top). 
            
            The location of the second-intermediate cloud in \textit{C50} matches that identified by \cite{Doi_2024} as their second component at $1.23$~kpc in the Sagittarius arm. Moreover, the mean magnetic field orientation of $134\fdg5$ in their foreground component, likely including contributions from various structures, aligns with the average orientation of $130\degr$ between the clouds of the Local arm in \textit{C50} (Table~\ref{tab:tomo_results}).     
    
            The distance between the intermediate and distant clouds is larger in \textit{C5} than in the other regions. This could indicate that \textit{C5} has longer path lengths in the interarm region. This is in agreement with the extinction density profiles of Figure~\ref{fig:ips_gpips_Av_dens_dist_pol} and the stellar $G$-band extinction of Figure~\ref{fig:LOS_3D_dustmap} (bottom) as a function of the distance. The $G$-band extinction also aligns well with the distant clouds, although sampling issues in the NIR observations toward \textit{C50} seem to affect the accuracy of the distant cloud’s inferred location.

    \subsection{Magnetic field orientation comparison between starlight polarization and polarized thermal dust emission}  
    \label{subsec:Disc_GMF_PA_planck}
        %
        \begin{figure*}[t!]
            \epsscale{0.53}
            \plotone{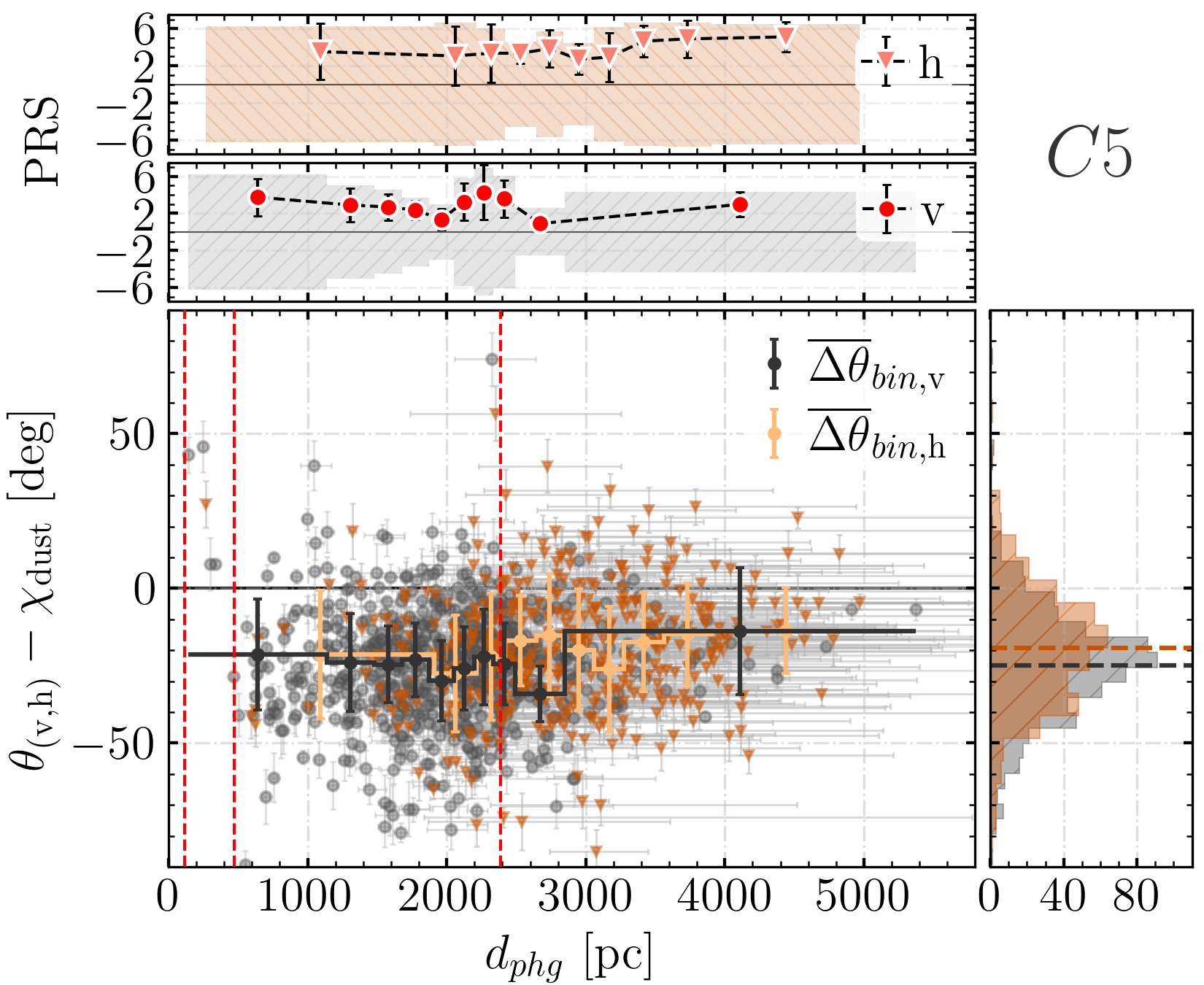}
            \epsscale{0.53}
            \plotone{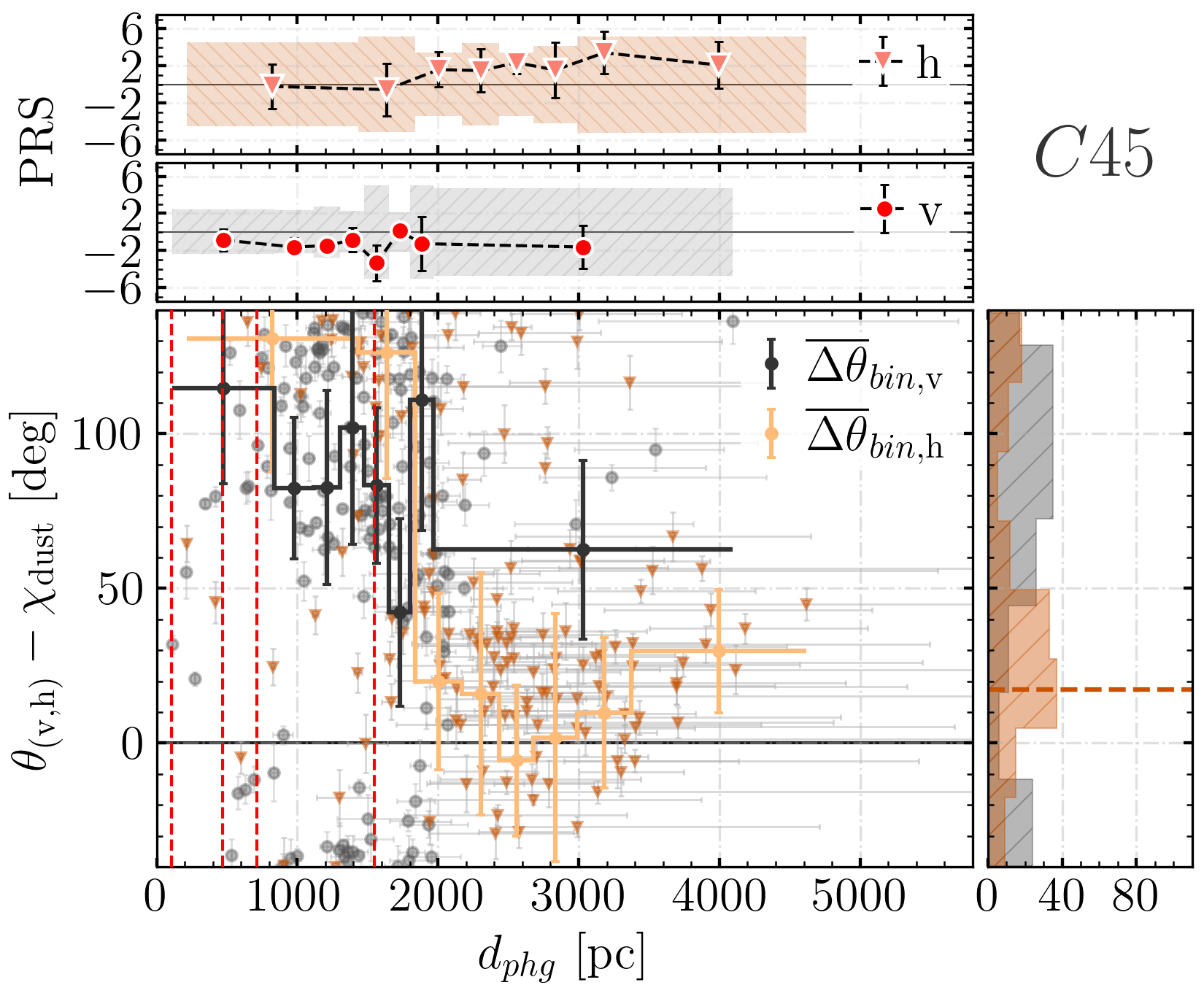}
            \epsscale{0.53}            
            \plotone{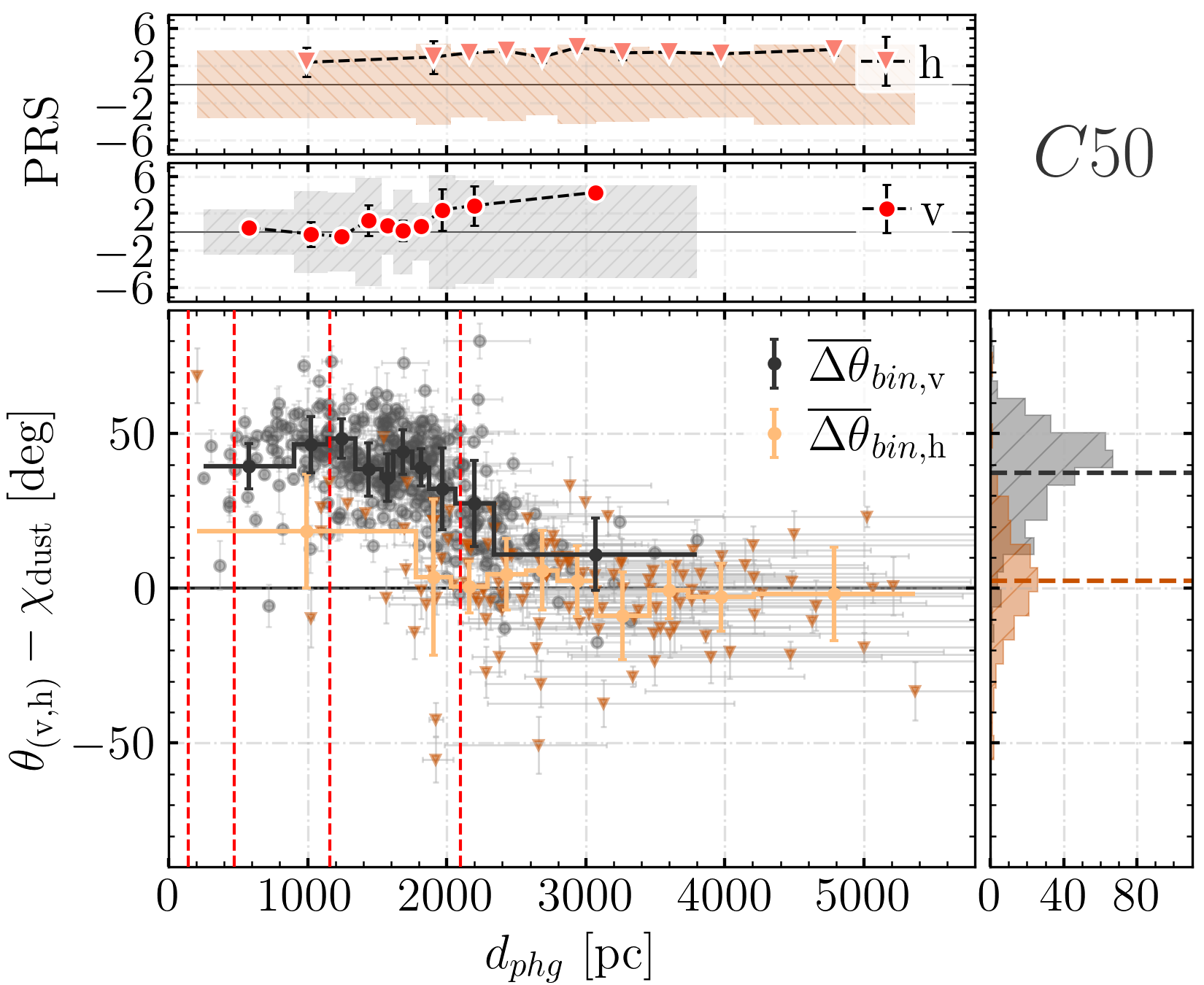}
            \caption{Magnetic field orientation difference between starlight polarization and polarized thermal dust emission from \cite{Planck-Collaboration_2018_20} as a function of distance in \textit{C5} (top left), \textit{C45} (top right), and \textit{C50} (bottom). Due to the $\pi$-ambiguity, the angular difference range in \textit{C45} was shifted by $50\degr$ for visualization.
            For each region, the bottom panels show the weighted circular mean difference per distance bin using NIR (the solid light-orange) and optical (the solid black) observations. The error bars are the weighted circular standard deviation on each distance bin. The vertical red dashed lines show the corrected distance of the polarizing clouds found with BISP-1. The right-side panels show the angular difference distributions with the circular means (the horizontal dashed lines). The two top panels show the PRS (Equation~\ref{eq:PRS}) per distance bin along with the weighted error. The gray and orange shaded areas represent the values between the maximum and minimum PRS per distance bin (see explanation in Section~\ref{subsec:Disc_GMF_PA_planck}).
            \label{fig:C50_PAdiff_ips_vs_planck}}
        \end{figure*}

        Planck's 353 GHz polarized thermal dust emission may be depolarized beyond a certain distance because the larger beam averages over large-scale magnetic field variations \citep{Planck-Collaboration_2018_20}. This phenomenon, known as polarization horizon \citep{Uyaniker_2003}, limits the observed path length to less than the full Galactic disk. Comparing Planck data with optical/NIR stellar polarization helps locate this horizon.

        Figure~\ref{fig:FoV_data} shows the plane-of-sky GMF orientation from starlight polarization (white and black pseudo-vectors, Section~\ref{sec:observ_data}) and polarized thermal dust emission at 353 GHz from Planck (white texture), rotated by $90\degr$. Overall, the magnetic field orientations from optical, NIR, and thermal dust observations partially align in the three regions, although some discrepancies exist between the optical/NIR and Planck data. Notably, Planck's average magnetic field orientation (see blue pseudo-vectors of Figures~\ref{fig:ips_Polvect} and~\ref{fig:gpips_Polvect}) aligns better with the longest starlight polarization sightlines, which is expected given that Planck captures longer path lengths in the Galaxy.
        
        For a quantitative comparison, we interpolated Planck’s values at each star’s position and computed the difference between the magnetic field orientations inferred from starlight polarization and polarized thermal dust emission ($\theta_\mathrm{v,h}-\chi_{\mathrm{dust}}$, Figure~\ref{fig:C50_PAdiff_ips_vs_planck}). 
        We calculated $\chi_{\mathrm{dust}}$ using Planck's High-Frequency Instrument (HFI) processed maps at $353$~GHz from \cite{Planck-Collaboration_2020} as 
        \begin{equation}
            \label{eq:PA_planck}
            \chi_{\mathrm{dust}} = \frac{1}{2} \mathrm{arctan2} (-U_{353},Q_{353}) + 90\degr,
        \end{equation}
        where $Q_{353}$ and $U_{353}$ are the Stokes parameters in HEALPix’s COSMO convention \citep{Planck-Collaboration_2020}. The negative sign converts $\chi_{\mathrm{dust}}$ to the IAU (North-to-East) convention. Uncertainties, $\sigma_{\chi_{\mathrm{dust}}}$, are calculated following Equation~(B.5) of \cite{Planck-Collaboration_XIX_2015}. We used bilinear interpolation\footnote{\url{https://astropy-healpix.readthedocs.io/en/latest/interpolation.html}} on the HEALPix Stokes parameters and covariance maps to obtain the values at each star’s location.
        
        In regions \textit{C45} and \textit{C50}, the plane-of-sky GMF orientations differ significantly at near and intermediate distances (vertical lines in Figure~\ref{fig:C50_PAdiff_ips_vs_planck} mark the polarizing clouds identified with BISP-1). However, the angular difference decreases to nearly zero beyond the distant clouds, as the orientations become increasingly aligned in the farthest stars, which are primarily those observed in the NIR (the orange triangles in Figure~\ref{fig:C50_PAdiff_ips_vs_planck}). 
        Conversely, in \textit{C5}, the angular difference is about $-20\degr$ at all distances. This could be caused by dust at distances greater than what starlight polarization can probe.

        We tested the significance of the relative alignment using the Projected Rayleigh Statistic (PRS, \citealt{Durand_Greenwood_1958}), a metric often used to measure the alignment between linear polarization observations \citep[e.g.,][and references therein]{Jow_2018, Panopoulou_2021}. The PRS statistic can discern between parallel ($\text{PRS} > 0$) or perpendicular ($\text{PRS} < 0$) alignments. We calculated the PRS and its weighted variance as in \cite{Jow_2018},
        \begin{equation}
            \label{eq:PRS}
            \mathrm{PRS} =  \frac{1}{\sqrt{\sum_i^N{w_i^2/2}}} \sum_i^N{w_i\mathrm{cos (2\Delta\theta_i)}}~,
        \end{equation}
        where  $\Delta\theta_i = \theta_{\mathrm{(v,h)}, i} - \chi_{\mathrm{dust}, i}$ is the angular difference of the $i$-th value defined in the range $[-90\degr,90\degr)$ and $w_i$ is the weight defined as $1/\sigma_{\Delta\theta_{i}}^2$, with $\sigma_{\Delta\theta_{i}} = \sqrt{\sigma_{\theta_{\mathrm{(v,h), i}}}^2 + \sigma_{\chi_{\mathrm{dust}, i}}^2}$. The PRS depends on the number of measurements, $N$, in each distance bin. Therefore, we binned the data evenly and took care of the $\pi$-ambiguity. 
        
        The two top panels of Figure~\ref{fig:C50_PAdiff_ips_vs_planck} show the PRS values as a function of distance for each region and wavelength observed. The maximum and minimum values from Equation~(\ref{eq:PRS}) correspond to the perfect parallel (i.e.,~\mbox{$|\Delta\theta| = 0^\circ$}) and perpendicular (i.e.,~\mbox{$|\Delta\theta| = 90^\circ$}) alignments, respectively, based solely on measurement errors. The shaded areas represent possible value ranges due to measurement errors.
        In \textit{C5}, the PRS remains positive and fairly constant with distance but is not significantly higher than zero due to large uncertainties, suggesting weak parallel alignment of the plane-of-sky magnetic field orientations. In \textit{C45} and \textit{C50}, the PRS is negative in the near and intermediate polarizing clouds, mainly traced by optical polarization, indicating a misalignment in the structures of the Local Bubble wall and the Local arm. Beyond $2$~kpc, towards the Sagittarius arm, parallel alignment becomes more significant, especially in \textit{C50}, where the values are maximum and the error bars are very small. 
        
        In summary, NIR observations in at least two regions are close to the maximum parallel alignment with Planck's magnetic field orientation, while the optical data often show little alignment, especially in \textit{C45} and \textit{C50}. This indicates that the plane-of-sky magnetic field direction changes significantly along the LOS, at least in the Local Bubble wall and the Local arm.

\section{Summary} 
\label{sec:summary}

    We presented a tomography analysis of the LOS-integrated plane-of-sky magnetic field component throughout the Galactic spiral arms based on optical and NIR starlight polarization observations in the Galactic plane in the regions named \textit{C5}, \textit{C45}, and \textit{C50}. These LOSs probe the ISM structures in directions close to the Aquila Rift and the Sagittarius arm. The high-quality measurements of both optical and NIR observations consistently exhibit similar polarizing properties of the interstellar structures common to both observations, regardless of the number of these structures and their location along the LOS. Furthermore, optical and NIR polarization data proved to be complementary, with optical polarization better tracing the nearby ISM, while NIR polarization primarily captures the distant ISM. 

    We determined the LOS-integrated polarization properties of the stars behind different magnetized dust clouds using two methods: the polarization decomposition tool BISP-1 and the GMM clustering tool. Both methods produced consistent results, particularly in the distant ISM (i.e.,~the Sagittarius arm). Moreover, the information from each method is complementary: BISP-1 accurately locates the polarizing clouds and accounts for ISM scatter, while GMM clustering identifies large-scale magnetic field variations across the sky in addition to LOS variations and more effectively separates the star groups behind each cloud.

    The polarization decomposition revealed three to four polarizing clouds along the LOS with optical observations and at least two with NIR data. The lack of NIR measurements with high SNR in the nearby ISM prevented us from finding the intermediate clouds observed with optical polarization. Nevertheless, the distant stars observed in the NIR provided insights into the properties of the GMF and the ISM at greater distances than optical polarization. The near clouds between $104$ and $143$~pc are consistent with dust structures within the Local Bubble wall. 
    The intermediate clouds around $469$ and $714$~pc are within the Local arm. The second-intermediate cloud in \textit{C50} at $1.16$~kpc is beyond the Local arm, as demonstrated by 3D dust maps from \cite{Vergely_2022}, and it is probably in the interarm region or the near edge of the Sagittarius arm. Finally, the distant clouds between $1.5$ and $2.5$~kpc in the Sagittarius arm are consistent with the results from \cite{Doi_2024}. We emphasize that these values were obtained using solely starlight polarization and stellar distance information.

    The starlight polarization tomography demonstrated large variations along the LOS of the plane-of-sky magnetic field properties in \textit{C45} and \textit{C50}, from the Local Bubble wall to the Sagittarius arm. This was also confirmed by comparing starlight polarization and polarized thermal dust emission from Planck. 
    The small difference in alignment observed in \textit{C5} along the entire LOS may be due to more dust observed by Planck farther away.
    Additionally, the magnetic field in \textit{C45} is complex and difficult to understand fully from the tomography analysis alone. Both methods consistently identified clouds along the LOS, but the curved magnetic field orientation and signs of depolarization between clouds suggest a complicated GMF structure across the Galactic arms, which both methods likely struggle to describe accurately.
    In conclusion, the above results show, first, the potential of using multiwavelength starlight polarization observations for Galactic tomography and, second, the need for complementary tomography tools or the development of a new method that accounts for the intricate properties of the interstellar magnetic fields. 

\begin{acknowledgments}
    We would like to thank the anonymous referee for their valuable comments and suggestions, which have greatly improved the quality of this work.
    Over the years, IPS data have been gathered by a number of dedicated observers, to whom the authors are very grateful: Flaviane C. F. Benedito, Alex Carciofi, Cassia Fernandez, Tib\'erio Ferrari, Livia S. C. A. Ferreira, Viviana S. Gabriel, Aiara Lobo-Gomes, Luciana de Matos, Rocio Melgarejo, Antonio Pereyra, Nadili Ribeiro, Marcelo Rubinho, Daiane B. Seriacopi, Fernando Silva, Rodolfo Valentim, and Aline Vidotto.
    Y.A., M.H., and M.J.F.V. acknowledge funding from the European Research Council (ERC) under the European Union’s Horizon 2020 research and innovation program (grant agreement No 772663).
    V.P. acknowledges funding from a Marie Curie Action of the European Union (grant agreement No. 101107047).
    C.V.R. acknowledges support from {\it Conselho Nacional de Desenvolvimento Científico e Tecnológico} - CNPq (Brazil, grant~310930/2021-9) and the Brazilian Ministry of Science, Technology and Innovation (MCTI) and the Brazilian Space Agency (AEB) by the support from PO 20VB.0009.
    A.M.M.'s work and optical/NIR polarimetry at IAG have been supported over the years by several grants from S\~ao Paulo state funding agency FAPESP, especially 01/12589-1 and 10/19694-4. A.M.M. has also been partially supported by the Brazilian agency CNPq (grant 310506/2015-8). A.M.M. graduate students have been provided with grants over the years from the Brazilian agency CAPES.
    This research has used data, tools, and materials developed as part of the \mbox{EXPLORE} project that has received funding from the European Union’s Horizon 2020 research and innovation program under grant agreement No 101004214.
    Finally, this work has made use of data from the European Space Agency (ESA) mission Gaia (\url{https://www.cosmos.esa.int/gaia}), processed by the Gaia Data Processing and Analysis Consortium (DPAC, \url{https://www.cosmos.esa.int/web/gaia/dpac/consortium}). Funding for the DPAC has been provided by national institutions, in particular, the institutions participating in the Gaia Multilateral Agreement.
\end{acknowledgments}





%

\facility{LNA:BC0.6m}

\software{Astropy \citep{Astropy_Collaboration_2013,Astropy_Collaboration_2018},
          Matplotlib \citep{Hunter_Matplotlib_2007}, NumPy \citep{Harris_numpy_2020}, SciPy \citep{Virtanen_SciPy_2020}, Scikit-learn \citep{Pedregosa_Scikitlearn_2011}}.


\appendix

\section{Parallax of faint red stars}
\label{sec:Append_A}  
    
    %
    \begin{figure}[hbt]
        \epsscale{2.4}
        \plottwo{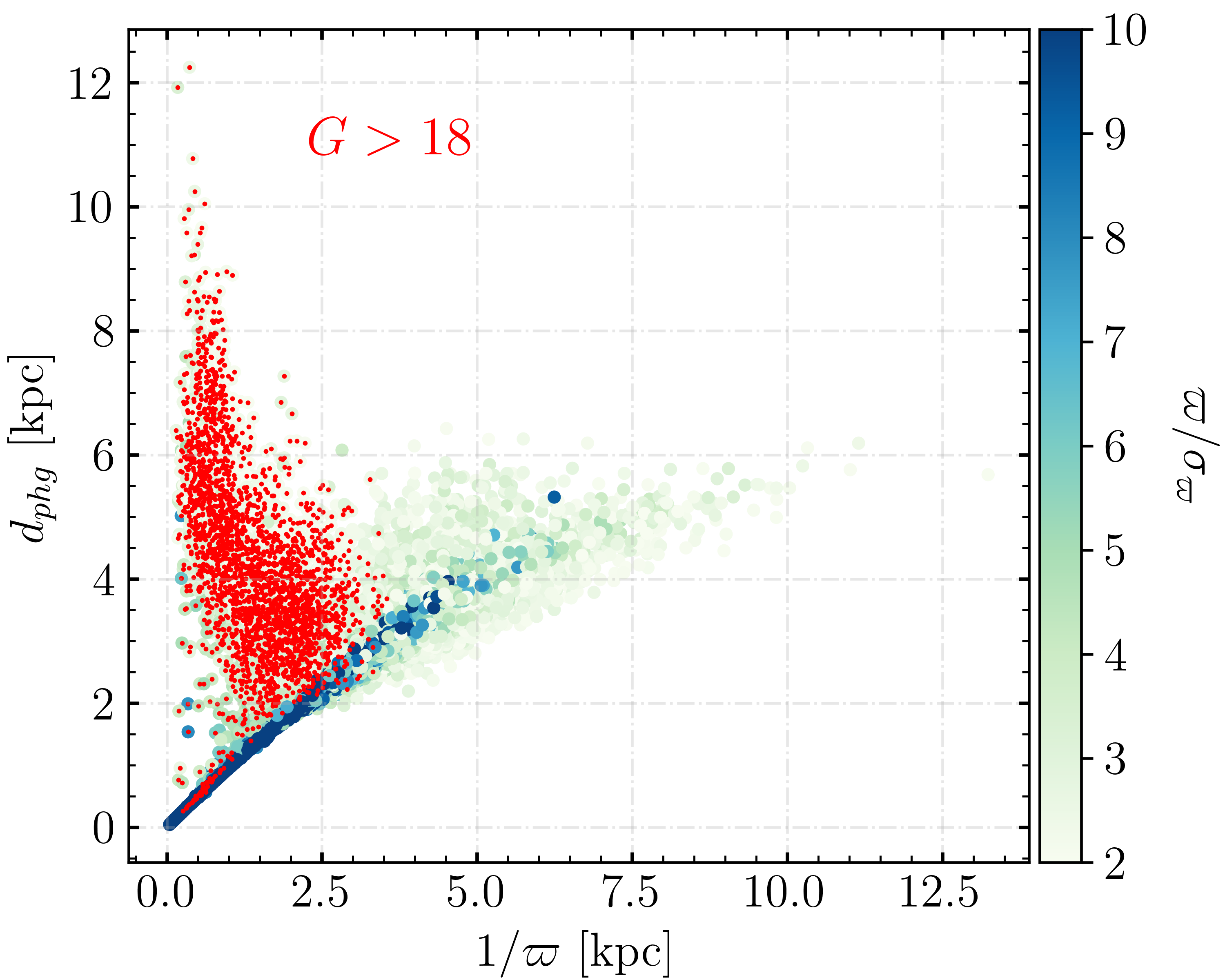}{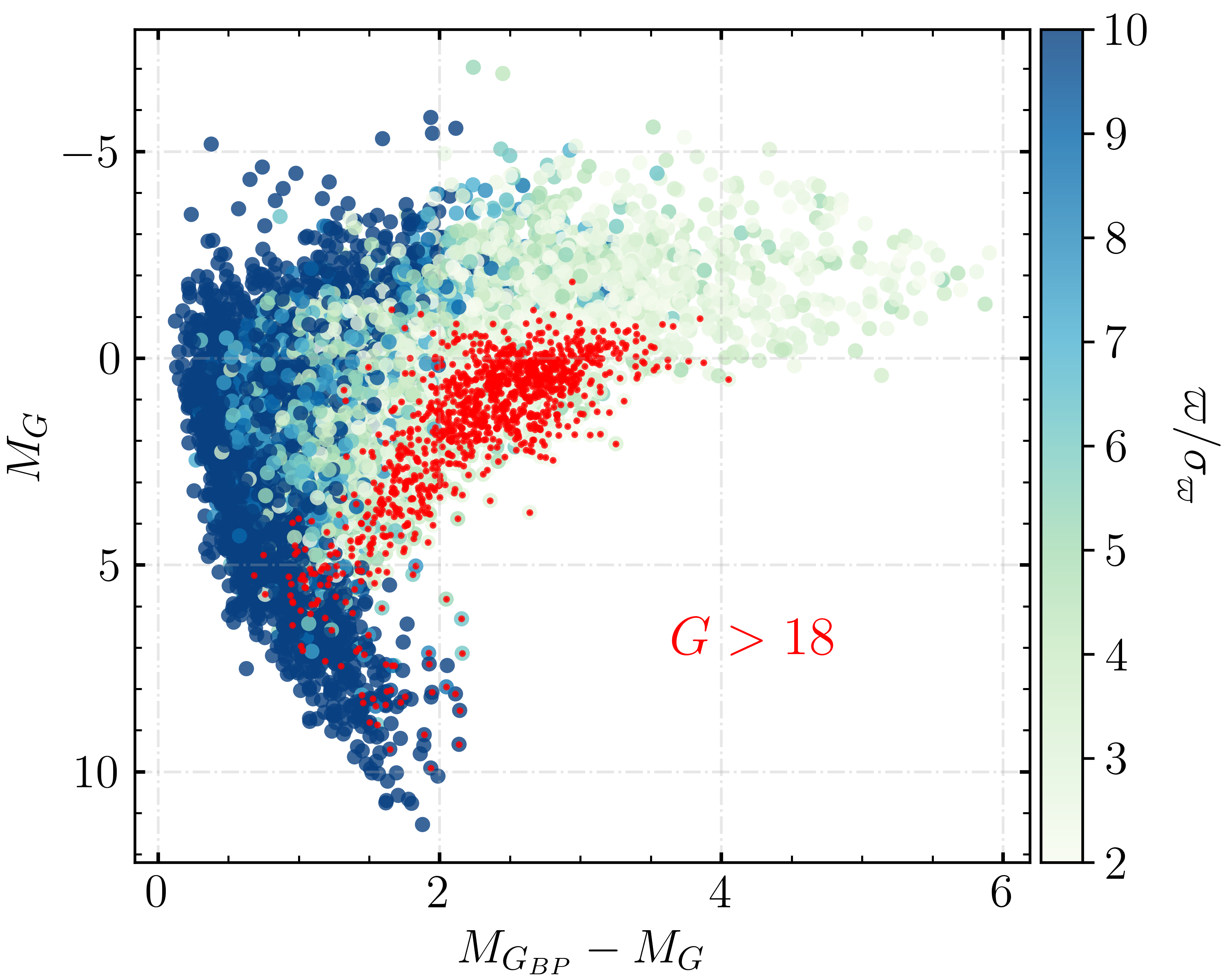}
        \caption{Top: comparison between the inverse parallax and the photo-geometric distance from \cite{Bailer_Jones_2021}. Bottom: color-magnitude diagram from Gaia-EDR3 photometry \citep{Gaia_Collaboration_2021b}. Both panels present all high-quality Gaia measurements (i.e.,~$\varpi/\sigma_{\varpi}>2$, RUWE $<1.4$, and $\varpi>0$) in the region \textit{C5}, colored by the parallax SNR saturated at 10. The red points represent faint stars with $G>18$.
        \label{fig:append_dist_plx_cm_diagram}}
    \end{figure}

    The inverse parallax-distance relation of all high-quality Gaia observations in the regions studied revealed that the parallaxes of faint sources are biased. For example, Figure~\ref{fig:append_dist_plx_cm_diagram} presents the photo-geometric distance \citep{Bailer_Jones_2021} as a function of the inverse parallax using all high-quality Gaia observations in the region of \textit{C5}, colored by the parallax SNR. The red points show stars with spurious parallaxes and characterized by magnitudes above 18. Distant, faint stars are often assigned a large parallax, so they will erroneously appear as nearby sources (Figure~\ref{fig:append_dist_plx_cm_diagram}, top). These sources are typically red stars (Figure~\ref{fig:append_dist_plx_cm_diagram}, bottom). This is also observed with all high-quality Gaia data in the regions \textit{C45} and \textit{C50}.

    It is known that Gaia's uncertainty in parallax, position, and proper motion increases significantly for faint stars \citep{Lindegren_Gaiaedr3_astrometric_2021}. Moreover, biased parallax measurements are a known issue reported in faint distant sources, such as quasars, and the solution to this issue is non-trivial \citep{Lindegren_Gaiaedr3_Plx_bias_2021}. Since BISP-1 relies on the parallax to perform the polarization decomposition along the LOS, spurious parallaxes from faint red sources represent an issue. To avoid this issue, we remove the faint stars with $G>18$ when performing the polarization tomography with the GPIPS dataset. This issue is not observed in the IPS-GI dataset.

\section{Initial parameters of BISP-1 decomposition}
\label{sec:Append_B}

    We present the priors used in the polarization decomposition with BISP-1 in Table~\ref{tab:BISP_priors}. We followed the recipe of \cite{Pelgrims_2023}, primarily keeping the default values for most parameters. Subsequently, we adjusted the distance and the diagonal elements of the intrinsic-scatter covariance matrix priors for each region in the sky. The lower limit of the distance prior for the nearest clouds was set at $100$~pc following the approximate distance to the Local Bubble wall observed in the extinction density profiles of Figure~\ref{fig:ips_gpips_Av_dens_dist_pol} \citep[see also][]{Lallement_2003, Liu_LBW_2017, Pelgrims_2020}. The upper limit of the distance prior in the farthest clouds is set to the maximum observed distance by default. In all cases, we verified that the $C_{qu}$ elements of the intrinsic-scatter covariance matrix fulfill its definition in the range ($-\sqrt{C_{qq}C_{uu}}, \sqrt{C_{qq}C_{uu}}$).
                
    \begin{deluxetable*}{clccccccc}
        \tablecaption{Limits of the uniform priors used in BISP-1 polarization decomposition. \label{tab:BISP_priors}}
        \tablehead{
        \colhead{Region} & \colhead{Cloud} & \colhead{$\varpi$} & \colhead{$1/\varpi$} & \colhead{$q$} & \colhead{$u$} & \colhead{$C_{qq}$} & \colhead{$C_{uu}$} & \colhead{$C_{qu}$}   \\
        \colhead{ } & \colhead{ } & \colhead{(mas)} & \colhead{(pc)} & \colhead{(\%)} & \colhead{(\%)} & \colhead{($\%^2$)} & \colhead{($\%^2$)} & \colhead{($\%^2$)}
        }
        \colnumbers
        \startdata
         \multicolumn{9}{c}{Optical Polarization}\\[0.15cm] \hline
                     & Near & $[2.00, 10.0]$ & $[500.0, 100.0]$ & $[-5.4, 5.4]$ & $[-5.4, 5.4]$ & $[0, 4]$ & $[0, 4]$ & $(-4, 4)$ \\[0.15cm]
         \textit{C5} & Intermediate & $[1.25, 3.33]$ & $[800.0, 300.0]$ & $[-5.4, 5.4]$ & $[-5.4, 5.4]$ & $[0, 4]$ & $[0, 4]$ & $(-4, 4)$ \\[0.15cm]
                     & Distant & $[0.161, 1.43]$ & $[6219.2, 700.0]$ & $[-5.4, 5.4]$ & $[-5.4, 5.4]$ & $[0, 1]$ & $[0, 1]$ & $(-1, 1)$ \\[0.15cm]\hline
                      & Near & $[3.33, 10.0]$ & $[300.0, 100.0]$ & $[-4.7, 4.7]$ & $[-4.7, 4.7]$ & $[0, 2]$ & $[0, 2]$ & $(-2, 2)$ \\[0.15cm]
         \textit{C45} & Intermediate 1 & $[1.67, 3.33]$ & $[600.0, 300.0]$ & $[-4.7, 4.7]$ & $[-4.7, 4.7]$ & $[0, 2]$ & $[0, 2]$ & $(-2, 2)$ \\[0.15cm]
                      & Intermediate 2 & $[1.00, 2.50]$ & $[1000.0, 400.0]$ & $[-4.7, 4.7]$ & $[-4.7, 4.7]$ & $[0, 2]$ & $[0, 2]$ & $(-2, 2)$ \\[0.15cm]
                      & Distant & $[0.20, 0.83]$ & $[4984.4, 1200.0]$ & $[-4.7, 4.7]$ & $[-4.7, 4.7]$ & $[0, 3]$ & $[0, 3]$ & $(-3, 3)$ \\[0.15cm]\hline
                      & Near & $[2.00, 10.0]$ & $[500.0, 100.0]$ & $[-6.2, 6.2]$ & $[-6.2, 6.2]$ & $[0, 1]$ & $[0, 1]$ & $(-1, 1)$ \\[0.15cm]
         \textit{C50} & Intermediate 1 & $[1.25, 3.33]$ & $[800.0, 300.0]$ & $[-6.2, 6.2]$ & $[-6.2, 6.2]$ & $[0, 1]$ & $[0, 1]$ & $(-1, 1)$ \\[0.15cm]
                      & Intermediate 2 & $[0.667,1.43]$ & $[1500.0, 700.0]$ & $[-6.2, 6.2]$ & $[-6.2, 6.2]$ & $[0, 1]$ & $[0, 1]$ & $(-1, 1)$ \\[0.15cm]
                      & Distant & $[0.222,1.00]$ & $[4495.7, 1000.0]$ & $[-6.2, 6.2]$ & $[-6.2, 6.2]$ & $[0, 2]$ & $[0, 2]$ & $(-2, 2)$ \\[0.15cm]\hline
         \multicolumn{9}{c}{NIR Polarization}\rule{0pt}{0.6cm}\\[0.15cm]\hline
         \textit{C5} & Near & $[2.00, 10.0]$ & $[500.0, 100.0]$ & $[-4.2, 4.2]$ & $[-4.2, 4.2]$ & $[0, 1]$ & $[0, 1$] & $(-1, 1)$ \\[0.15cm]
                     & Distant & $[0.126, 3.33]$ & $[7951.6, 300.0]$ & $[-4.2, 4.2]$ & $[-4.2, 4.2]$ & $[0, 1]$ & $[0, 1]$ & $(-1, 1)$ \\[0.15cm]\hline
         \textit{C45} & Near & $[2.00, 10.0]$ & $[500.0, 100.0]$ & $[-4.7, 4.7]$ & $[-4.7, 4.7]$ & $[0, 1]$ & $[0, 1]$ & $(-1, 1)$ \\[0.15cm]
                      & Distant & $[0.141, 3.33]$ & $[7080.6, 300.0]$ & $[-4.7, 4.7]$ & $[-4.7, 4.7]$ & $[0, 1]$ & $[0, 1]$ & $(-1, 1)$ \\[0.15cm]\hline
         \textit{C50} & Near & $[2.00, 10.0]$ & $[500.0, 100.0]$ & $[-3.4, 3.4]$ & $[-3.4, 3.4]$ & $[0, 1]$ & $[0, 1]$ & $(-1, 1)$ \\[0.15cm]
                      & Distant & $[0.167, 3.33]$ & $[5988.8, 300.0]$ & $[-3.4, 3.4]$ & $[-3.4, 3.4]$ & $[0, 1]$ & $[0, 1]$ & $(-1, 1)$ \\
        \enddata
        \tablecomments{Columns: (1) Designation of the region; (2) polarizing cloud; (3) prior limits of the parallax; (4) the corresponding inverse parallax of the values in (3); (5) and (6) are the prior limits of the Stokes parameters $q_{(\mathrm{v,h})}$ and $u_{(\mathrm{v,h})}$, respectively; (7), (8), and (9) are the elements of the intrinsic-scatter covariance matrix $C_{qq}$, $C_{uu}$, and $C_{qu}$, respectively.}
    \end{deluxetable*}

\section{Initial parameters and validation of the GMM clustering}
\label{sec:Append_C} 

    The GMM package\footnote{\url{https://scikit-learn.org/stable/modules/mixture.html}} in the scikit-learn Python library  \citep{Pedregosa_Scikitlearn_2011} provides four fitting models based on the choice of the covariance matrix:
    \begin{enumerate}
        \item[\textit{i})] The full covariance type assigns a covariance matrix to each Gaussian component, meaning that each component can have its own shape and orientation in all dimensions. 
        \item[\textit{ii})] A tied covariance indicates that all components have the same covariance, forcing all components to have the same shape and orientation. 
        \item[\textit{iii})] The diagonal covariance type means that each Gaussian component has a diagonal covariance matrix, allowing components to have different variances (i.e.,~shapes) and the same orientation along the coordinate axes. 
        \item[\textit{iv})] The spherical covariance means that all elements of the diagonal covariance matrix are the same, assigning a single variance to each Gaussian component, which results in identical shapes and orientations along the axes.
    \end{enumerate}
    A Gaussian distribution is defined by its covariance matrix and mean. The covariance determines the shape and orientation of the density distributions in all dimensions, while the mean specifies their location in space.

    We tested each type of covariance model with different numbers of Gaussian components between one and nine.
    Table~\ref{tab:GMM_params} presents the number of Gaussian components and the covariance model that minimized the BIC in the GMM method for each dataset. These parameters are subsequently used in the GMM clustering described in Section~\ref{subsec:res_tomo_GMM}.
    For NIR polarization in \textit{C45}, we chose the second-best option for the initial parameters (Table~\ref{tab:GMM_params}). This decision was made because the additional group consisted only of the four farthest measurements, which are too few to characterize any properties of the magnetic field. The same stars beyond $2.8$~kpc were removed from the analysis with optical observation to improve the clustering, which can be affected by the contrasting long distances of these few measurements.

\begin{figure*}[ht]
        \gridline{
                  \fig{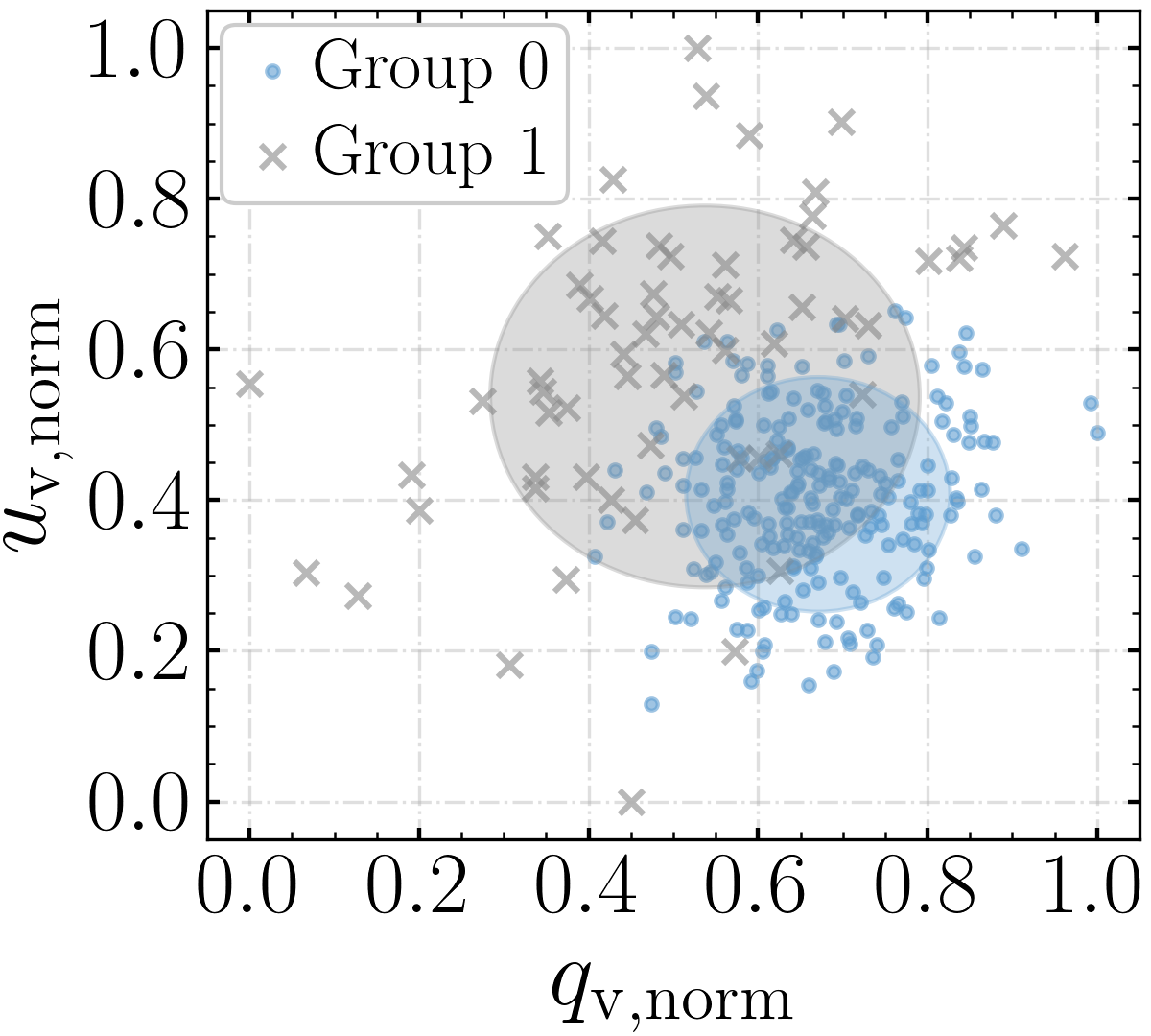}{.245\linewidth}{}
                  \fig{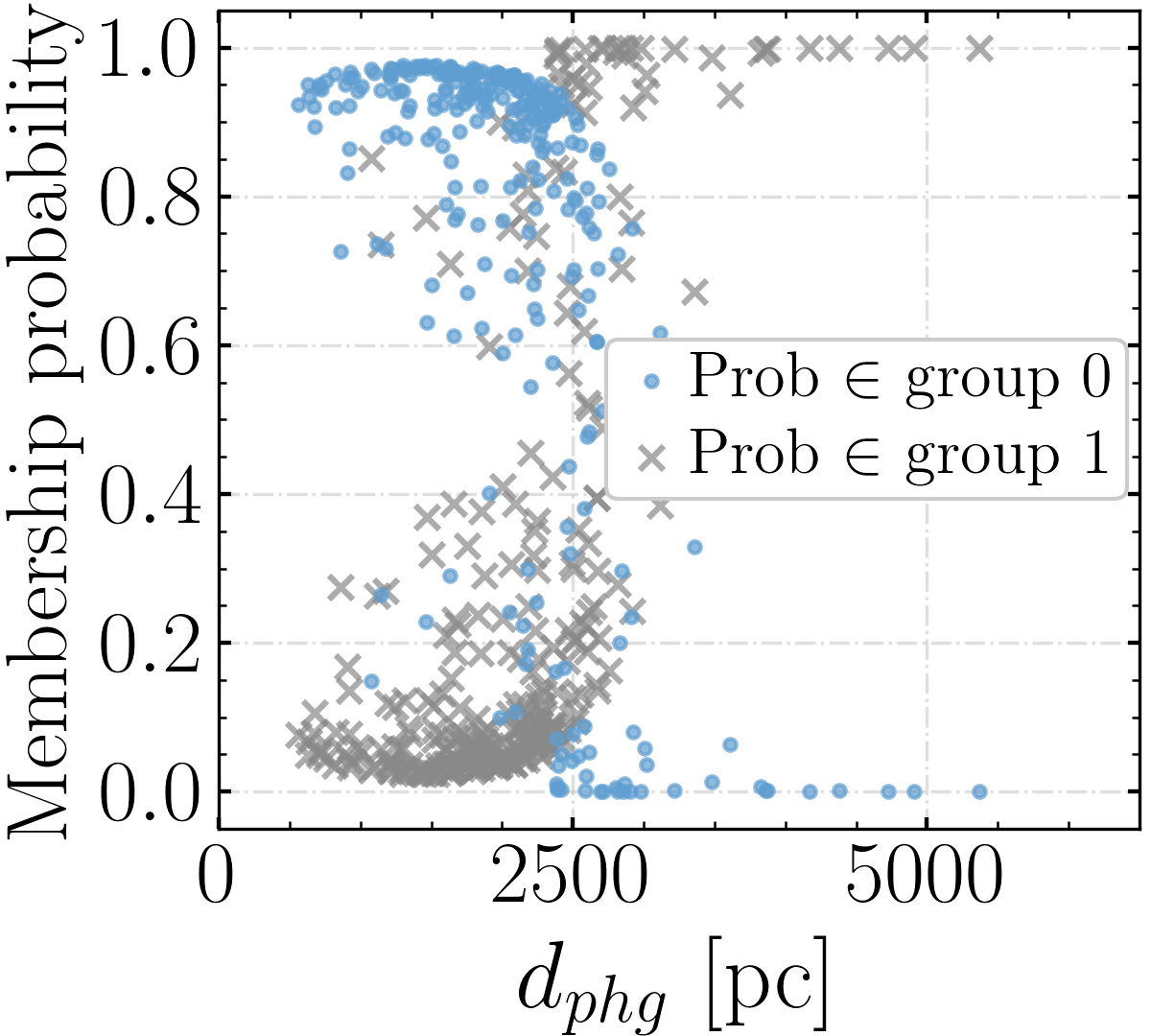}{.245\linewidth}{}
                  \fig{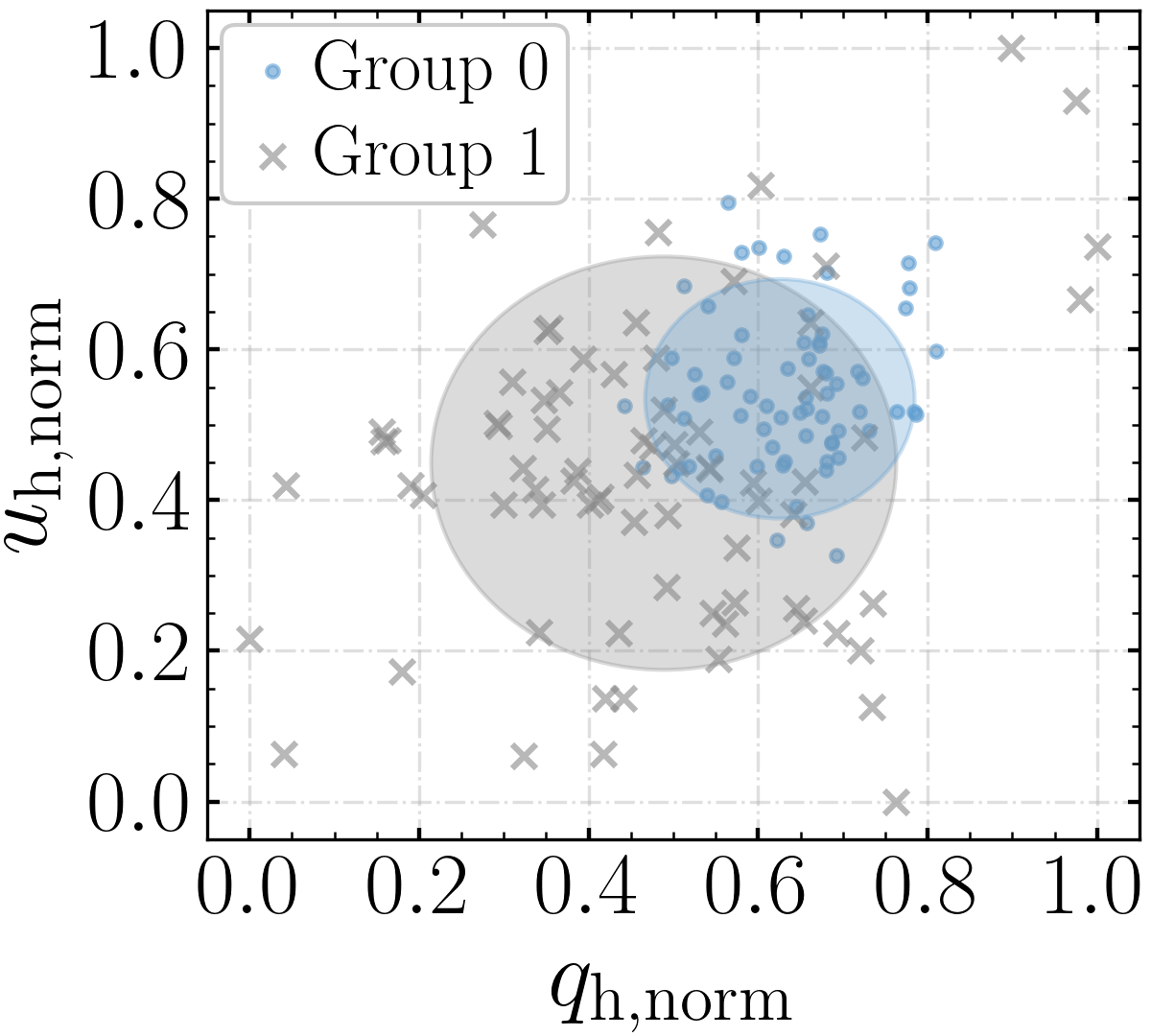}{.245\linewidth}{}
                  \fig{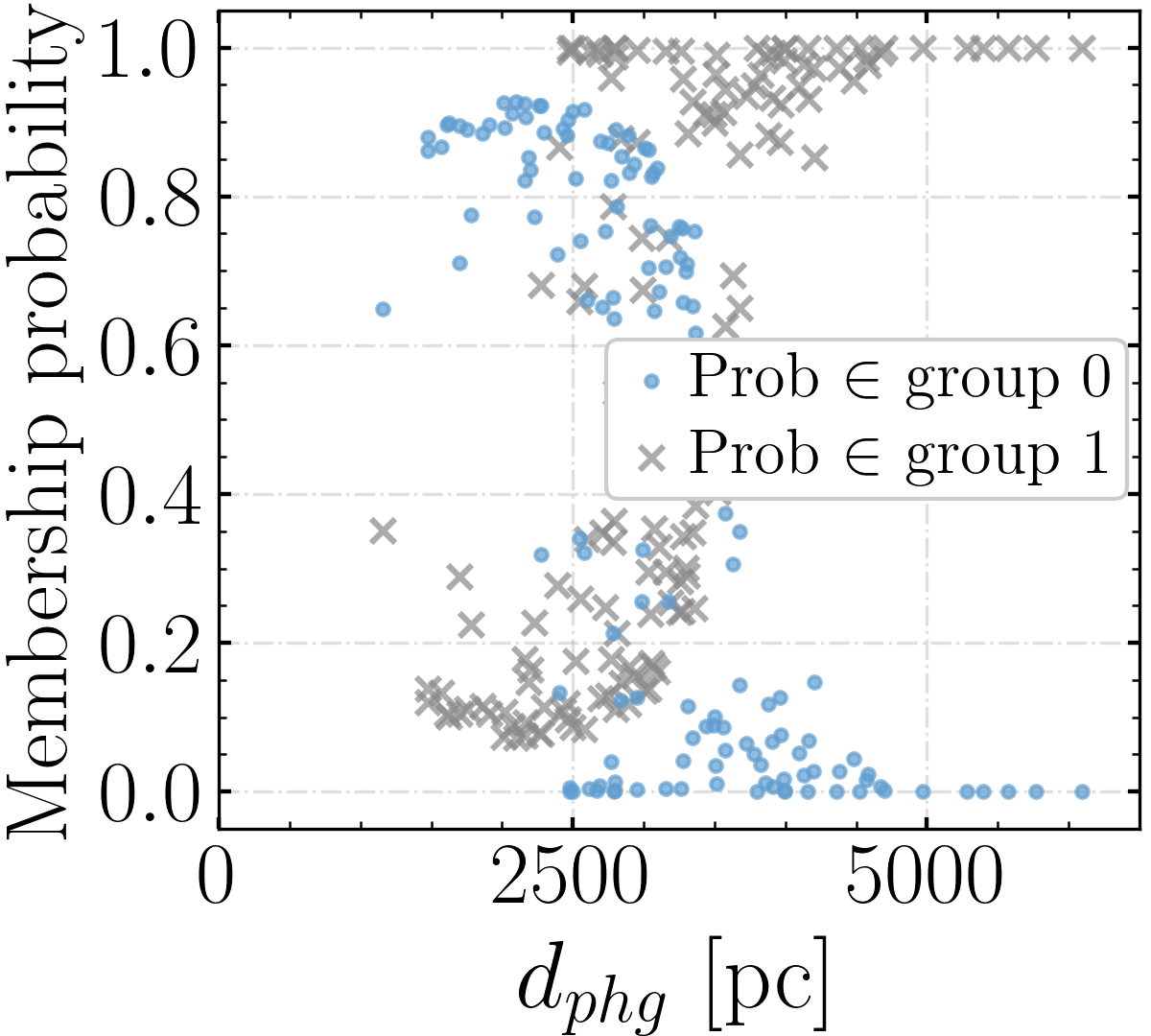}{.245\linewidth}{}
                  }\vspace{-0.8cm}
        \gridline{
                  \fig{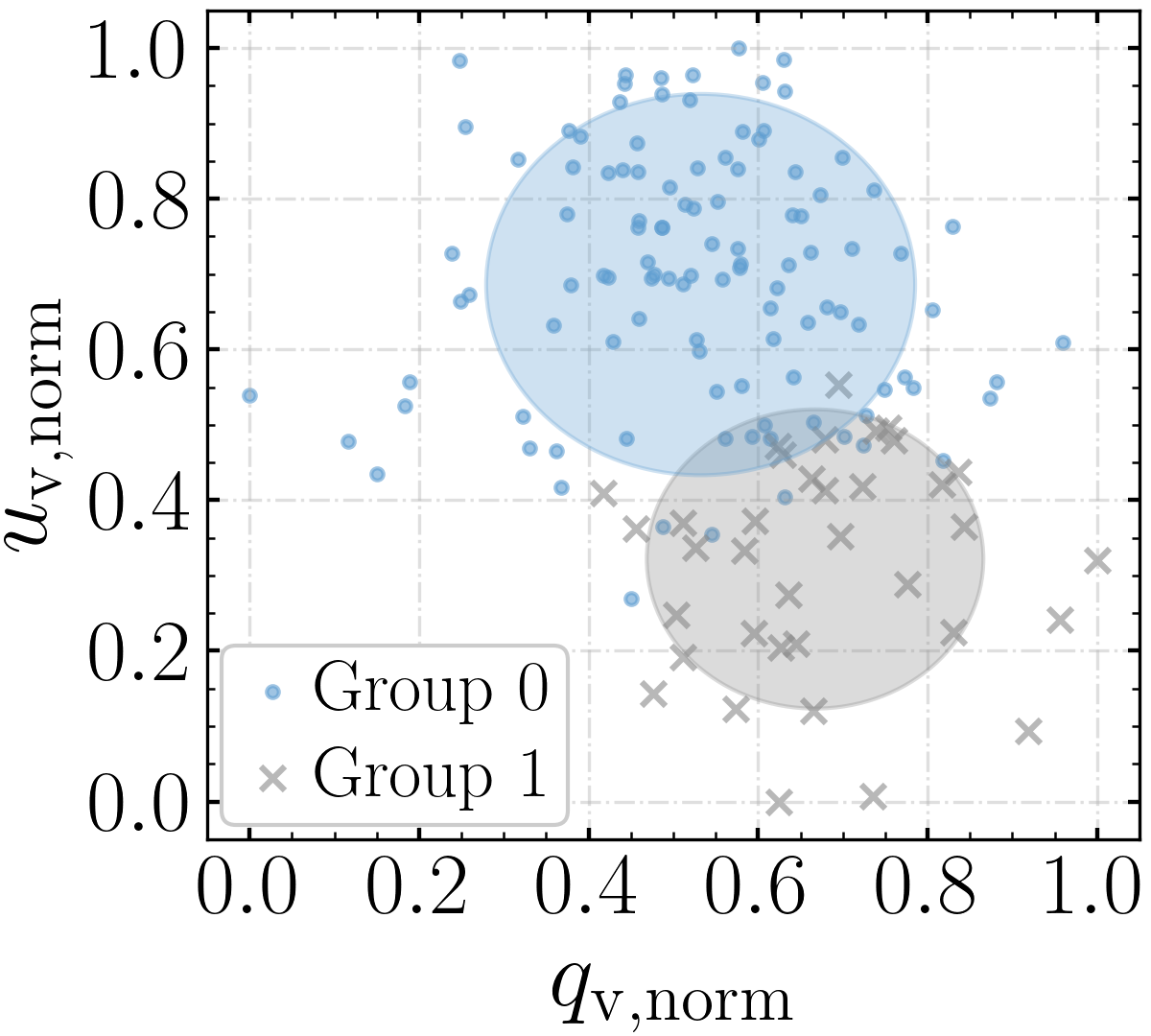}{.245\linewidth}{}
                  \fig{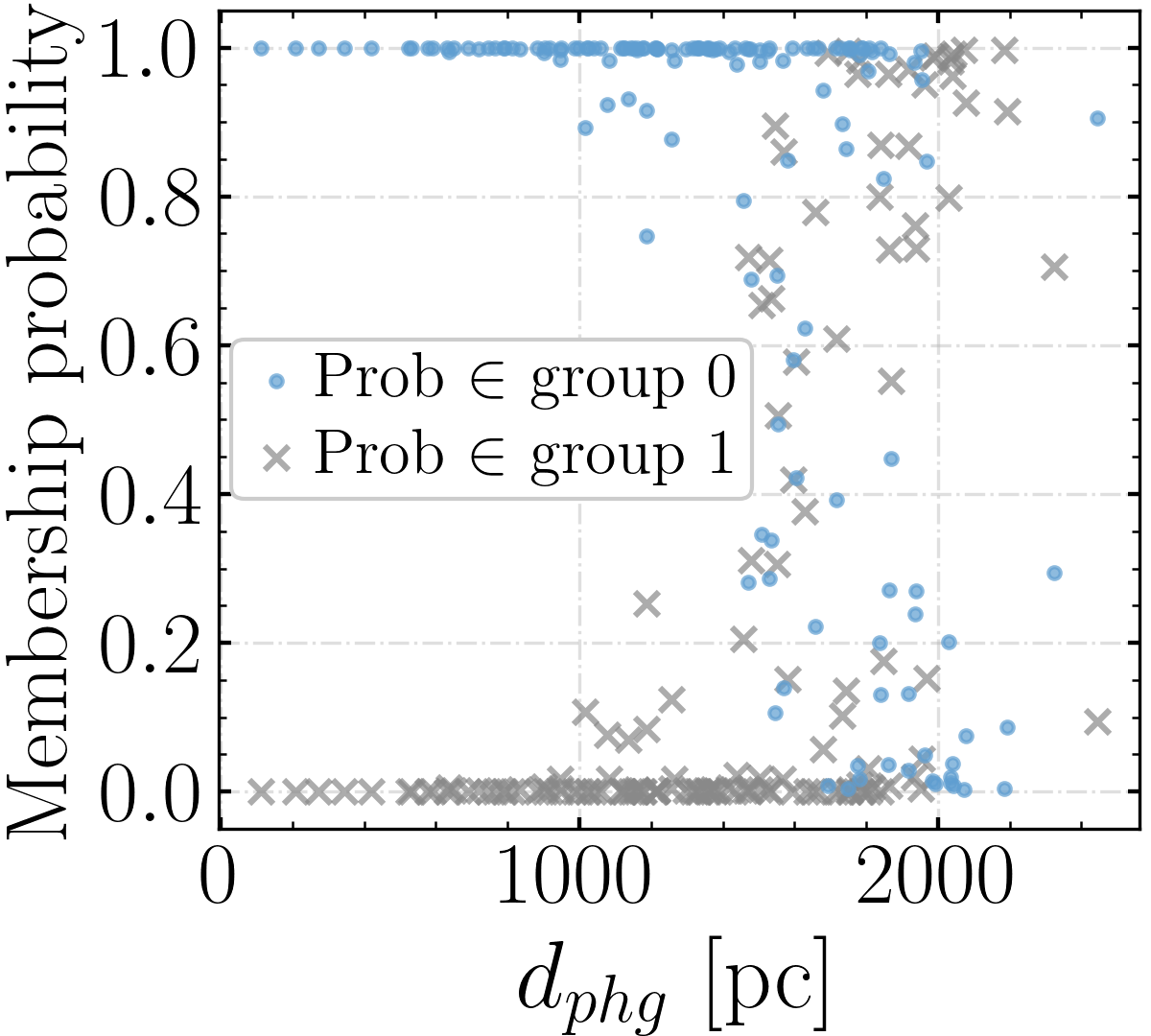}{.245\linewidth}{}
                  \fig{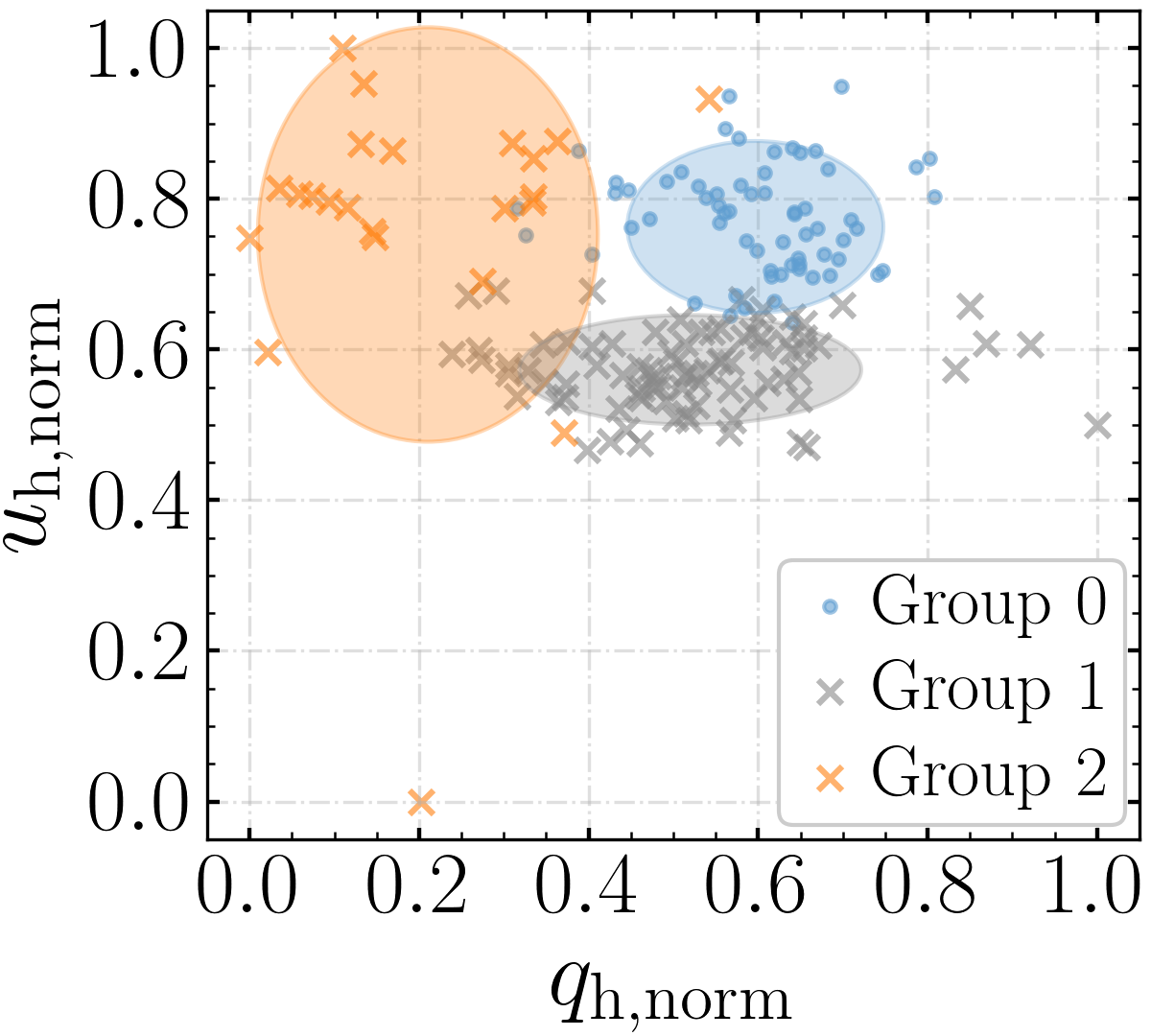}{.245\linewidth}{}
                  \fig{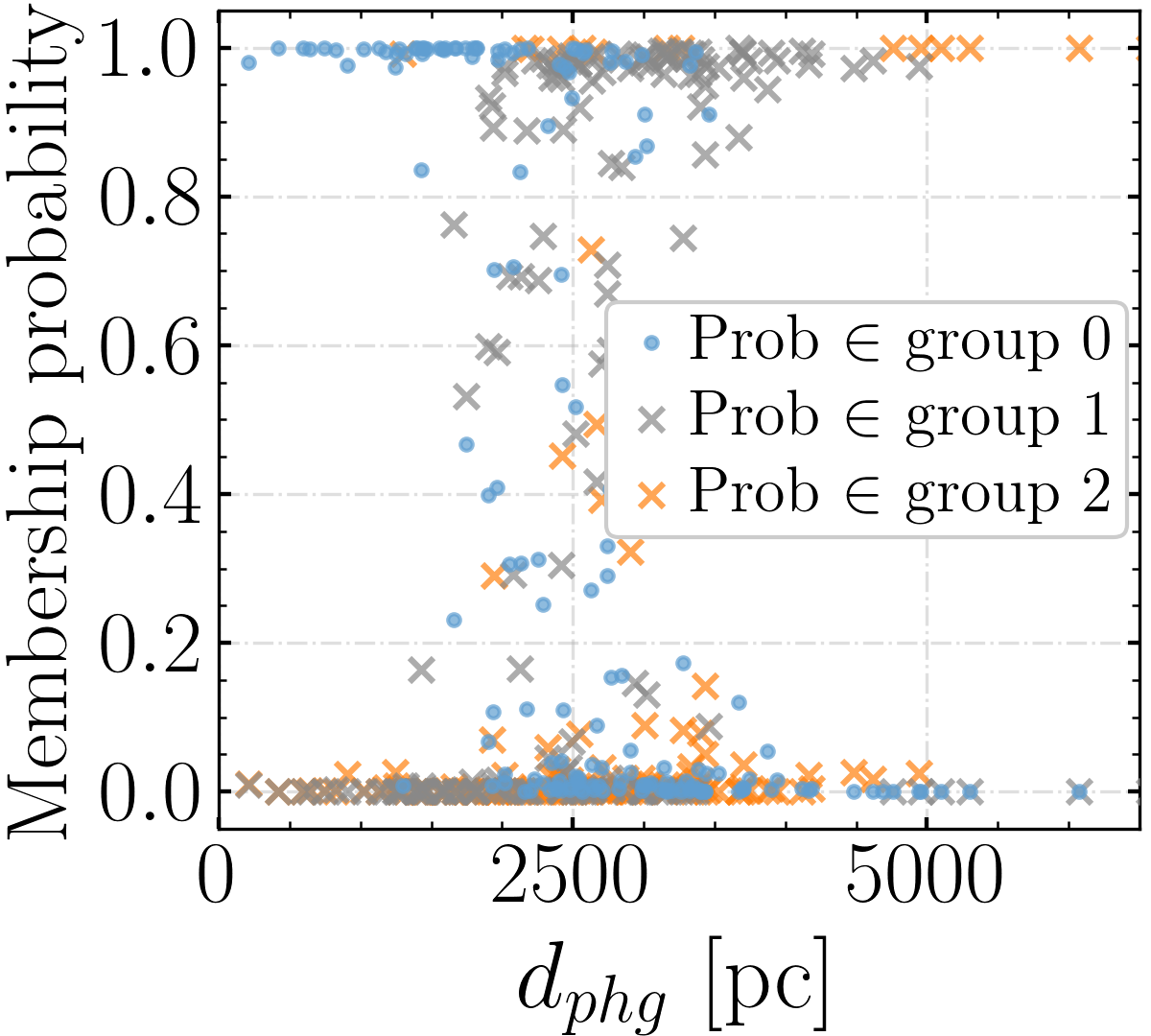}{.245\linewidth}{}
                  }\vspace{-0.8cm}
        \gridline{
                  \fig{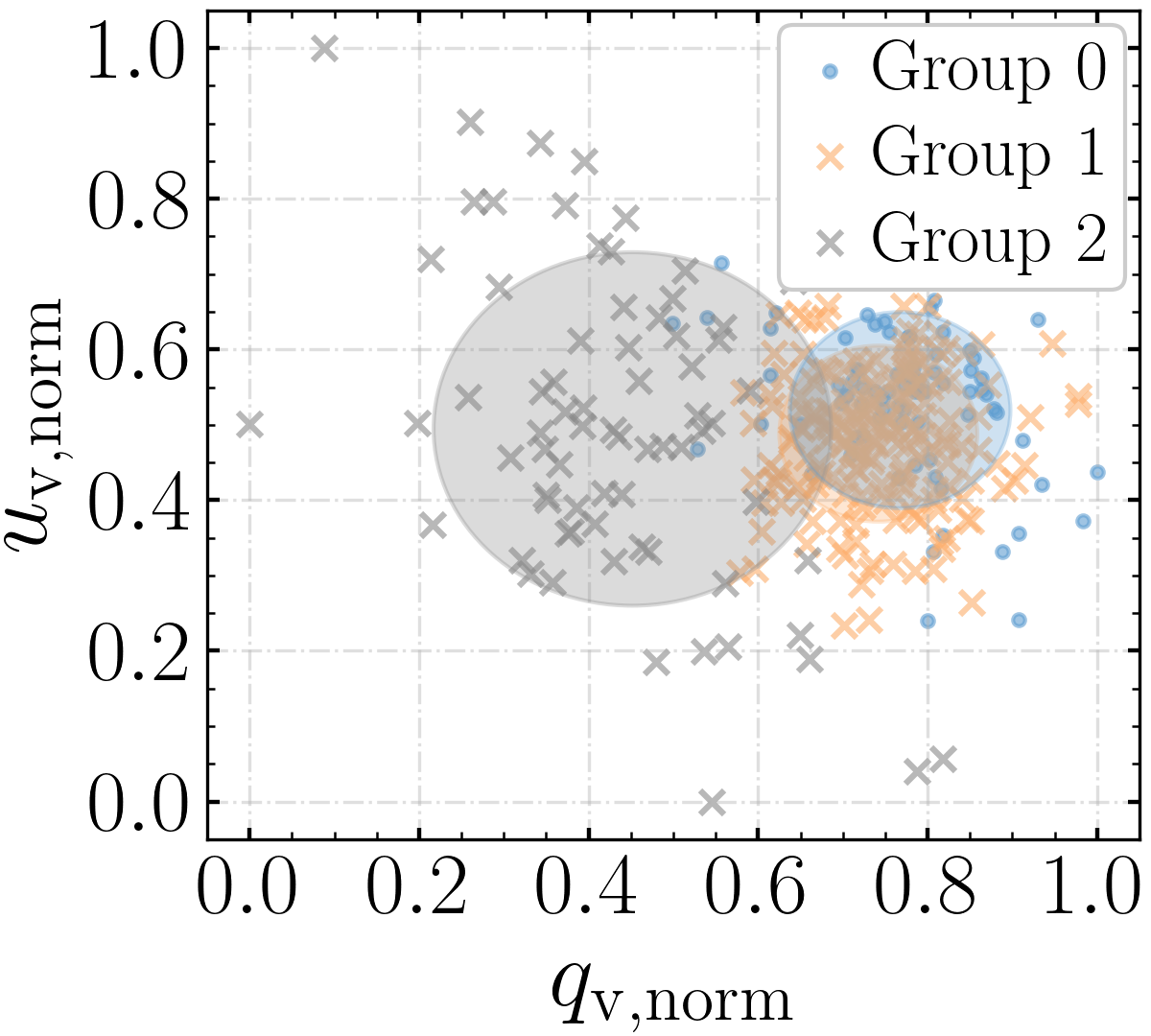}{.245\linewidth}{}
                  \fig{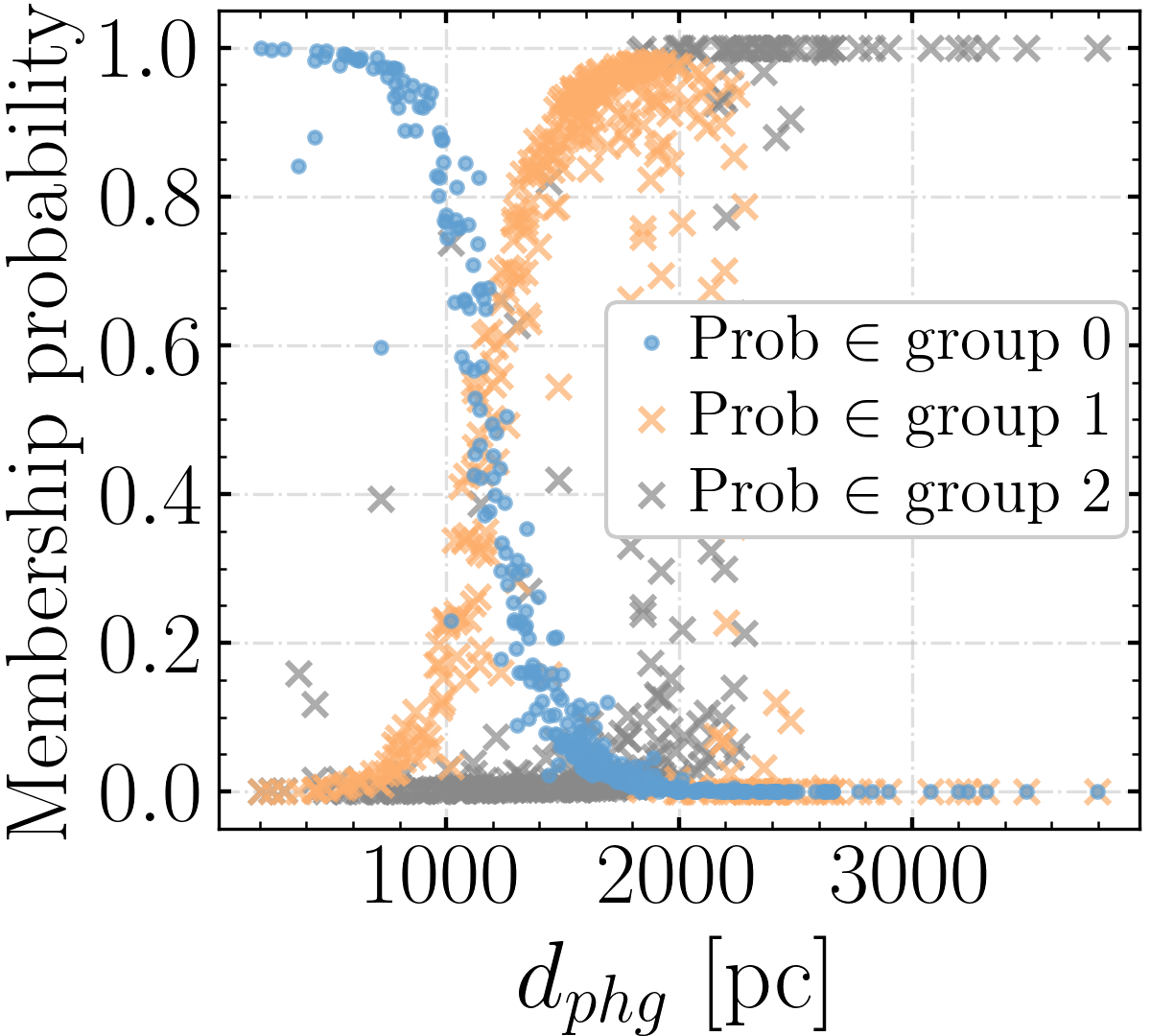}{.245\linewidth}{}
                  \fig{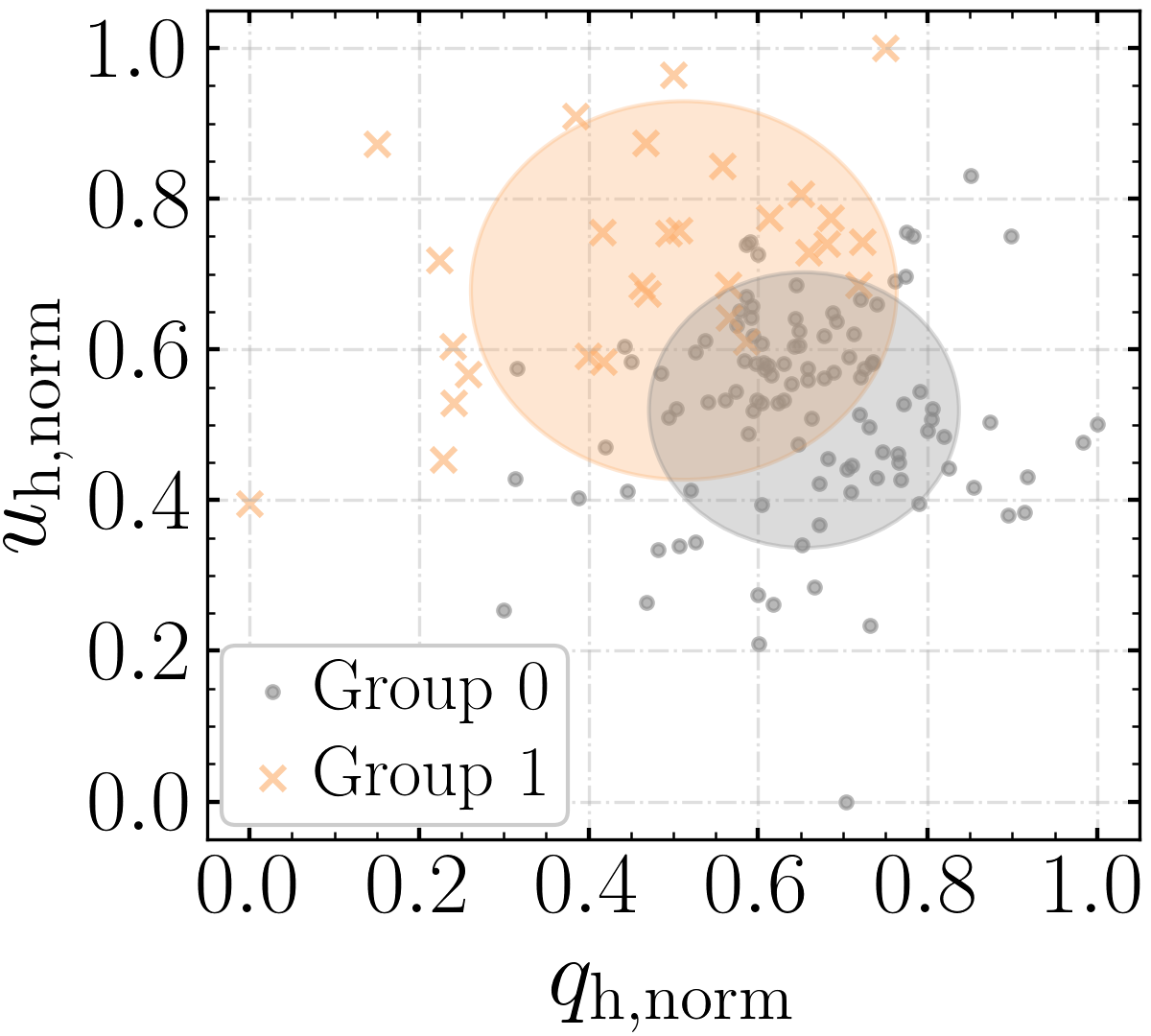}{.245\linewidth}{}
                  \fig{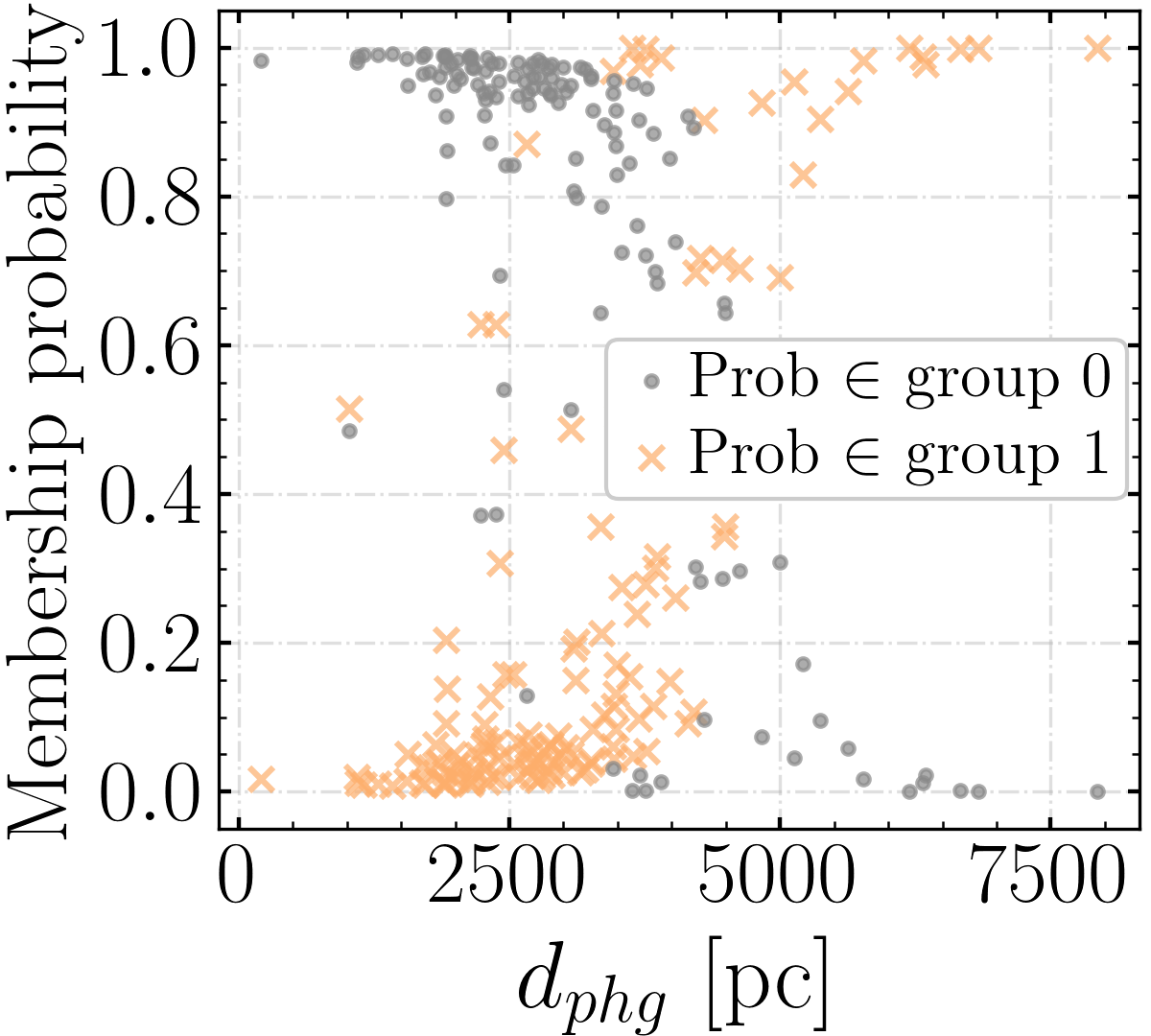}{.245\linewidth}{}
                  }\vspace{-0.8cm}
        \caption{GMM results with optical (the two left columns) and NIR (the two right columns) polarization in \textit{C5} (top), \textit{C45} (middle), and \textit{C50} (bottom). The first and third columns are the group separations in the normalized $q-u$ space. The ellipses represent the Gaussian components. The second and fourth columns are the group membership probability as a function of photo-geometric distance. The colors correspond to the components of Figures~\ref{fig:GMM_resul_C5} to~\ref{fig:GMM_resul_C50}.
        \label{fig:Append_GMM_result}}
    \end{figure*}

    Figure~\ref{fig:Append_GMM_result} shows the group separation in the $q$-$u$ space and the membership probability as a function of distance for each solution. The ellipses illustrate the 2D shapes and orientations of the individual Gaussian components obtained with the models from Table~\ref{tab:GMM_params}.  We used the diagnostics of Figure~\ref{fig:Append_GMM_result} to confirm that the solutions obtained from the set of initial parameters minimizing the BIC (Table~\ref{tab:GMM_params}) were well-defined in the $q$-$u$ space. Moreover, we ensured that the probabilities were above $80\%$ in the majority of the measurements within each component (Figure~\ref{fig:Append_GMM_result}).
    
    %
    \begin{deluxetable}{cccc|ccc}
    \setlength{\tabcolsep}{2pt}
        \tablecaption{Parameters that minimized the BIC in the GMM clustering. \label{tab:GMM_params}}
        \tablehead{
        \colhead{Region} & \multicolumn{3}{c}{Optical} & \multicolumn{3}{c}{NIR}\\
        \colhead{ } & \colhead{N} & \colhead{Cov. Model} &\colhead{BIC} & \colhead{N} & \colhead{Cov. Model} & \colhead{BIC}
        }
        \colnumbers
        \startdata
        \textit{C5}  &  2  & spherical & $-923.0$  & 2 & spherical & $-274.0$ \\
        \textit{C45}  & 2  & spherical & $-155.6$  & 3 & diagonal  & $-566.0$\tablenotemark{\dag} \\
        \textit{C50}  & 3  & spherical & $-1316.5$ & 2 & spherical & $-324.5$ \\
        \enddata
        \tablecomments{Columns: (1) Designation of the region; (2) and (5) the number of components; (3) and (6) the covariance model; (4) and (7) the minimum BIC number obtained with the parameters of the two columns on the right, respectively.}
        \tablenotetext{\dag}{Second smallest BIC value. The minimum is $-570.1$.}
    \end{deluxetable}
    %

\clearpage


\bibliography{IPSV_GMF_tomo_V_IR}{}

\newcommand{\noop}[1]{}
\begin{thebibliography}{}
\expandafter\ifx\csname natexlab\endcsname\relax\def\natexlab#1{#1}\fi
\providecommand{\url}[1]{\href{#1}{#1}}
\providecommand{\dodoi}[1]{doi:~\href{http://doi.org/#1}{\nolinkurl{#1}}}
\providecommand{\doeprint}[1]{\href{http://ascl.net/#1}{\nolinkurl{http://ascl.net/#1}}}
\providecommand{\doarXiv}[1]{\href{https://arxiv.org/abs/#1}{\nolinkurl{https://arxiv.org/abs/#1}}}

\bibitem[{H. Akaike(1998)Akaike}]{Akaike_1998}
Akaike, H. 1998, Information Theory and an Extension of the Maximum Likelihood Principle, ed. E.~Parzen, K.~Tanabe, \& G.~Kitagawa (New York, NY: Springer New York), 199--213, \dodoi{10.1007/978-1-4612-1694-0_15}

\bibitem[{M.~I.~R. {Alves} {et~al.}(2018){Alves}, {Boulanger}, {Ferri{\`e}re}, \& {Montier}}]{Alves_2018}
{Alves}, M.~I.~R., {Boulanger}, F., {Ferri{\`e}re}, K., \& {Montier}, L. 2018, \bibinfo{title}{{The Local Bubble: a magnetic veil to our Galaxy},} \aap, 611, L5, \dodoi{10.1051/0004-6361/201832637}

\bibitem[{F. {Anders} {et~al.}(2022){Anders}, {Khalatyan}, {Queiroz}, {Chiappini}, {Ard{\`e}vol}, {Casamiquela}, {Figueras}, {Jim{\'e}nez-Arranz}, {Jordi}, {Mongui{\'o}}, {Romero-G{\'o}mez}, {Altamirano}, {Antoja}, {Assaad}, {Cantat-Gaudin}, {Castro-Ginard}, {Enke}, {Girardi}, {Guiglion}, {Khan}, {Luri}, {Miglio}, {Minchev}, {Ramos}, {Santiago}, \& {Steinmetz}}]{Anders_2022}
{Anders}, F., {Khalatyan}, A., {Queiroz}, A.~B.~A., {et~al.} 2022, \bibinfo{title}{{Photo-astrometric distances, extinctions, and astrophysical parameters for Gaia EDR3 stars brighter than G = 18.5},} \aap, 658, A91, \dodoi{10.1051/0004-6361/202142369}

\bibitem[{B.~G. {Andersson} \& S.~B. {Potter}(2005){Andersson} \& {Potter}}]{Andersson_Potter_CoalSack_2005}
{Andersson}, B.~G., \& {Potter}, S.~B. 2005, \bibinfo{title}{{A high sampling-density polarization study of the Southern Coalsack},} \mnras, 356, 1088, \dodoi{10.1111/j.1365-2966.2004.08538.x}

\bibitem[{B.~G. {Andersson} \& S.~B. {Potter}(2007){Andersson} \& {Potter}}]{Andersson_Potter_2007}
{Andersson}, B.~G., \& {Potter}, S.~B. 2007, \bibinfo{title}{{Observational Constraints on Interstellar Grain Alignment},} \apj, 665, 369, \dodoi{10.1086/519755}

\bibitem[{Y. {Angarita} {et~al.}(2024){Angarita}, {Versteeg}, {Haverkorn}, {Marchal}, {Rodrigues}, {Magalh{\~a}es}, {Santos-Lima}, \& {Kawabata}}]{Angarita_2024}
{Angarita}, Y., {Versteeg}, M.~J.~F., {Haverkorn}, M., {et~al.} 2024, \bibinfo{title}{{Interstellar Polarization Survey. IV. Characterizing the Magnetic Field Strength and Turbulent Dispersion Using Optical Starlight Polarization in the Diffuse Interstellar Medium},} \aj, 168, 47, \dodoi{10.3847/1538-3881/ad4b14}

\bibitem[{Y. {Angarita} {et~al.}(2023){Angarita}, {Versteeg}, {Haverkorn}, {Rodrigues}, {Magalh{\~a}es}, {Santos-Lima}, \& {Kawabata}}]{Angarita_2023}
{Angarita}, Y., {Versteeg}, M.~J.~F., {Haverkorn}, M., {et~al.} 2023, \bibinfo{title}{{Interstellar Polarization Survey. III. Relation between Optical Polarization and Reddening in the General Interstellar Medium},} \aj, 166, 34, \dodoi{10.3847/1538-3881/acdc1e}

\bibitem[{ {\!\!Astropy Collaboration} {et~al.}(2013){\!\!Astropy Collaboration}, {Robitaille}, {Tollerud}, {Greenfield}, {Droettboom}, {Bray}, {Aldcroft}, {Davis}, {Ginsburg}, {Price-Whelan}, {Kerzendorf}, {Conley}, {Crighton}, {Barbary}, {Muna}, {Ferguson}, {Grollier}, {Parikh}, {Nair}, {Unther}, {Deil}, {Woillez}, {Conseil}, {Kramer}, {Turner}, {Singer}, {Fox}, {Weaver}, {Zabalza}, {Edwards}, {Azalee Bostroem}, {Burke}, {Casey}, {Crawford}, {Dencheva}, {Ely}, {Jenness}, {Labrie}, {Lim}, {Pierfederici}, {Pontzen}, {Ptak}, {Refsdal}, {Servillat}, \& {Streicher}}]{Astropy_Collaboration_2013}
{\!\!Astropy Collaboration}, {Robitaille}, T.~P., {Tollerud}, E.~J., {et~al.} 2013, \bibinfo{title}{{Astropy: A community Python package for astronomy},} \aap, 558, A33, \dodoi{10.1051/0004-6361/201322068}

\bibitem[{ {\!\!Astropy Collaboration} {et~al.}(2018){\!\!Astropy Collaboration}, {Price-Whelan}, {Sip{\H{o}}cz}, {G{\"u}nther}, {Lim}, {Crawford}, {Conseil}, {Shupe}, {Craig}, {Dencheva}, {Ginsburg}, {VanderPlas}, {Bradley}, {P{\'e}rez-Su{\'a}rez}, {de Val-Borro}, {Aldcroft}, {Cruz}, {Robitaille}, {Tollerud}, {Ardelean}, {Babej}, {Bach}, {Bachetti}, {Bakanov}, {Bamford}, {Barentsen}, {Barmby}, {Baumbach}, {Berry}, {Biscani}, {Boquien}, {Bostroem}, {Bouma}, {Brammer}, {Bray}, {Breytenbach}, {Buddelmeijer}, {Burke}, {Calderone}, {Cano Rodr{\'\i}guez}, {Cara}, {Cardoso}, {Cheedella}, {Copin}, {Corrales}, {Crichton}, {D'Avella}, {Deil}, {Depagne}, {Dietrich}, {Donath}, {Droettboom}, {Earl}, {Erben}, {Fabbro}, {Ferreira}, {Finethy}, {Fox}, {Garrison}, {Gibbons}, {Goldstein}, {Gommers}, {Greco}, {Greenfield}, {Groener}, {Grollier}, {Hagen}, {Hirst}, {Homeier}, {Horton}, {Hosseinzadeh}, {Hu}, {Hunkeler}, {Ivezi{\'c}}, {Jain}, {Jenness}, {Kanarek}, {Kendrew}, {Kern}, {Kerzendorf}, {Khvalko}, {King}, {Kirkby},
  {Kulkarni}, {Kumar}, {Lee}, {Lenz}, {Littlefair}, {Ma}, {Macleod}, {Mastropietro}, {McCully}, {Montagnac}, {Morris}, {Mueller}, {Mumford}, {Muna}, {Murphy}, {Nelson}, {Nguyen}, {Ninan}, {N{\"o}the}, {Ogaz}, {Oh}, {Parejko}, {Parley}, {Pascual}, {Patil}, {Patil}, {Plunkett}, {Prochaska}, {Rastogi}, {Reddy Janga}, {Sabater}, {Sakurikar}, {Seifert}, {Sherbert}, {Sherwood-Taylor}, {Shih}, {Sick}, {Silbiger}, {Singanamalla}, {Singer}, {Sladen}, {Sooley}, {Sornarajah}, {Streicher}, {Teuben}, {Thomas}, {Tremblay}, {Turner}, {Terr{\'o}n}, {van Kerkwijk}, {de la Vega}, {Watkins}, {Weaver}, {Whitmore}, {Woillez}, {Zabalza}, \& {Astropy Contributors}}]{Astropy_Collaboration_2018}
{\!\!Astropy Collaboration}, {Price-Whelan}, A.~M., {Sip{\H{o}}cz}, B.~M., {et~al.} 2018, \bibinfo{title}{{The Astropy Project: Building an Open-science Project and Status of the v2.0 Core Package},} \aj, 156, 123, \dodoi{10.3847/1538-3881/aabc4f}

\bibitem[{C.~A.~L. {Bailer-Jones} {et~al.}(2021){Bailer-Jones}, {Rybizki}, {Fouesneau}, {Demleitner}, \& {Andrae}}]{Bailer_Jones_2021}
{Bailer-Jones}, C.~A.~L., {Rybizki}, J., {Fouesneau}, M., {Demleitner}, M., \& {Andrae}, R. 2021, \bibinfo{title}{{Estimating Distances from Parallaxes. V. Geometric and Photogeometric Distances to 1.47 Billion Stars in Gaia Early Data Release 3},} \aj, 161, 147, \dodoi{10.3847/1538-3881/abd806}

\bibitem[{R. {Beck}(2003){Beck}}]{Beck_review_2003}
{Beck}, R. 2003, \bibinfo{title}{{Magnetic fields in the Milky Way and other spiral galaxies},} arXiv e-prints, astro, \dodoi{10.48550/arXiv.astro-ph/0310287}

\bibitem[{A. {Berdyugin} {et~al.}(2014){Berdyugin}, {Piirola}, \& {Teerikorpi}}]{Berdyugin_2014}
{Berdyugin}, A., {Piirola}, V., \& {Teerikorpi}, P. 2014, \bibinfo{title}{{Interstellar polarization at high galactic latitudes from distant stars. VIII. Patterns related to the local dust and gas shells from observations of \raisebox{-0.5ex}\textasciitilde3600 stars},} \aap, 561, A24, \dodoi{10.1051/0004-6361/201322604}

\bibitem[{A. {Berdyugin} {et~al.}(1995){Berdyugin}, {Snare}, \& {Teerikorpi}}]{Berdyugin_1995}
{Berdyugin}, A., {Snare}, M.~O., \& {Teerikorpi}, P. 1995, \bibinfo{title}{{Interstellar polarization at high galactic latitudes from distant stars. I. First results for Z<=600pc.},} \aap, 294, 568

\bibitem[{A. {Beresnyak} \& A. {Lazarian}(2019){Beresnyak} \& {Lazarian}}]{Beresnyak_2019}
{Beresnyak}, A., \& {Lazarian}, A. 2019, {Turbulence in Magnetohydrodynamics} (Berlin, Boston: De Gruyter), \dodoi{10.1515/9783110263282}

\bibitem[{B. Cabral \& L. Leedom(1993)Cabral \& Leedom}]{Cabral_LIC_1993}
Cabral, B., \& Leedom, L. 1993, \bibinfo{title}{Special Interest Group on GRAPHics and Interactive Techniques Proceedings,} Special Interest Group on GRAPHics and Interactive Techniques Proceedings, 263

\bibitem[{D.~P. {Clemens} {et~al.}(2012){Clemens}, {Pinnick}, {Pavel}, \& {Taylor}}]{Clemens_GPIPS_2012}
{Clemens}, D.~P., {Pinnick}, A.~F., {Pavel}, M.~D., \& {Taylor}, B.~W. 2012, \bibinfo{title}{{The Galactic Plane Infrared Polarization Survey (GPIPS)},} \apjs, 200, 19, \dodoi{10.1088/0067-0049/200/2/19}

\bibitem[{D.~P. {Clemens} {et~al.}(2020){Clemens}, {Cashman}, {Cerny}, {El-Batal}, {Jameson}, {Marchwinski}, {Montgomery}, {Pavel}, {Pinnick}, \& {Taylor}}]{Clemens_GPIPS-DR4_2020}
{Clemens}, D.~P., {Cashman}, L.~R., {Cerny}, C., {et~al.} 2020, \bibinfo{title}{{The Galactic Plane Infrared Polarization Survey (GPIPS): Data Release 4},} \apjs, 249, 23, \dodoi{10.3847/1538-4365/ab9f30}

\bibitem[{S. {Codina-Landaberry} \& A.~M. {Magalhaes}(1976){Codina-Landaberry} \& {Magalhaes}}]{Codina_Magalhaes_1976}
{Codina-Landaberry}, S., \& {Magalhaes}, A.~M. 1976, \bibinfo{title}{{On the polarizing interstellar dust.},} \aap, 49, 407

\bibitem[{Y. {Doi} {et~al.}(2024){Doi}, {Nakamura}, {Kawabata}, {Matsumura}, {Akitaya}, {Coud{\'e}}, {Rodrigues}, {Kwon}, {Tamura}, {Tahani}, {Magalh{\~a}es}, {Santos-Lima}, {Angarita}, {Versteeg}, {Haverkorn}, {Hasegawa}, {Sadavoy}, {Arzoumanian}, \& {Bastien}}]{Doi_2024}
{Doi}, Y., {Nakamura}, K., {Kawabata}, K.~S., {et~al.} 2024, \bibinfo{title}{{Tomographic Imaging of the Sagittarius Spiral Arm's Magnetic Field Structure},} \apj, 961, 13, \dodoi{10.3847/1538-4357/ad0fe2}

\bibitem[{D. {Durand} \& J.~A. {Greenwood}(1958){Durand} \& {Greenwood}}]{Durand_Greenwood_1958}
{Durand}, D., \& {Greenwood}, J.~A. 1958, \bibinfo{title}{{Modifications of the Rayleigh Test for Uniformity in Analysis of Two-Dimensional Orientation Data},} Journal of Geology, 66, 229, \dodoi{10.1086/626501}

\bibitem[{G. {Edenhofer} {et~al.}(2024){Edenhofer}, {Zucker}, {Frank}, {Saydjari}, {Speagle}, {Finkbeiner}, \& {En{\ss}lin}}]{Edenhofer_dustmap_2023}
{Edenhofer}, G., {Zucker}, C., {Frank}, P., {et~al.} 2024, \bibinfo{title}{{A parsec-scale Galactic 3D dust map out to 1.25 kpc from the Sun},} \aap, 685, A82, \dodoi{10.1051/0004-6361/202347628}

\bibitem[{R.~S. {Ellis} \& D.~J. {Axon}(1978){Ellis} \& {Axon}}]{Ellis_Axon_1978}
{Ellis}, R.~S., \& {Axon}, D.~J. 1978, \bibinfo{title}{{Stellar Polarization and the Structure of the Magnetic Field of Our Galaxy},} \apss, 54, 425, \dodoi{10.1007/BF00639446}

\bibitem[{K. {Ferri{\`e}re}(2015){Ferri{\`e}re}}]{Ferriere_2015}
{Ferri{\`e}re}, K. 2015, in Journal of Physics Conference Series, Vol. 577, Journal of Physics Conference Series, 012008, \dodoi{10.1088/1742-6596/577/1/012008}

\bibitem[{P. {Fosalba} {et~al.}(2002){Fosalba}, {Lazarian}, {Prunet}, \& {Tauber}}]{Fosalba_2002}
{Fosalba}, P., {Lazarian}, A., {Prunet}, S., \& {Tauber}, J.~A. 2002, \bibinfo{title}{{Statistical Properties of Galactic Starlight Polarization},} \apj, 564, 762, \dodoi{10.1086/324297}

\bibitem[{M. {Fouesneau} {et~al.}(2023){Fouesneau}, {Fr{\'e}mat}, {Andrae}, {Korn}, {Soubiran}, {Kordopatis}, {Vallenari}, {Heiter}, {Creevey}, {Sarro}, {de Laverny}, {Lanzafame}, {Lobel}, {Sordo}, {Rybizki}, {Slezak}, {{\'A}lvarez}, {Drimmel}, {Garabato}, {Delchambre}, {Bailer-Jones}, {Hatzidimitriou}, {Lorca}, {Le Fustec}, {Pailler}, {Mary}, {Robin}, {Utrilla}, {Abreu Aramburu}, {Bakker}, {Bellas-Velidis}, {Bijaoui}, {Blomme}, {Bouret}, {Brouillet}, {Brugaletta}, {Burlacu}, {Carballo}, {Casamiquela}, {Chaoul}, {Chiavassa}, {Contursi}, {Cooper}, {Dafonte}, {Demouchy}, {Dharmawardena}, {Garc{\'\i}a-Lario}, {Garc{\'\i}a-Torres}, {Gomez}, {Gonz{\'a}lez-Santamar{\'\i}a}, {Jean-Antoine Piccolo}, {Kontizas}, {Lebreton}, {Licata}, {Lindstr{\o}m}, {Livanou}, {Magdaleno Romeo}, {Manteiga}, {Marocco}, {Martayan}, {Marshall}, {Nicolas}, {Ordenovic}, {Palicio}, {Pallas-Quintela}, {Pichon}, {Poggio}, {Recio-Blanco}, {Riclet}, {Santove{\~n}a}, {Schultheis}, {Segol}, {Silvelo}, {Smart}, {S{\"u}veges}, {Th{\'e}venin},
  {Torralba Elipe}, {Ulla}, {van Dillen}, {Zhao}, \& {Zorec}}]{GaiaDR3_Fouesneau_StellarParams_2023}
{Fouesneau}, M., {Fr{\'e}mat}, Y., {Andrae}, R., {et~al.} 2023, \bibinfo{title}{{Gaia Data Release 3. Apsis. II. Stellar parameters},} \aap, 674, A28, \dodoi{10.1051/0004-6361/202243919}

\bibitem[{L.~A. {Fowler} \& M. {Harwit}(1974){Fowler} \& {Harwit}}]{Fowler_1974}
{Fowler}, L.~A., \& {Harwit}, M. 1974, \bibinfo{title}{{Incremental polarization of starlight at different locations in the galaxy},} \mnras, 167, 227, \dodoi{10.1093/mnras/167.1.227}

\bibitem[{ {\!\!Gaia Collaboration} {et~al.}(2018){\!\!Gaia Collaboration}, {Brown}, {Vallenari}, {Prusti}, {de Bruijne}, {Babusiaux}, {Bailer-Jones}, {Biermann}, {Evans}, {Eyer}, {Jansen}, {Jordi}, {Klioner}, {Lammers}, {Lindegren}, {Luri}, {Mignard}, {Panem}, {Pourbaix}, {Randich}, {Sartoretti}, {Siddiqui}, {Soubiran}, {van Leeuwen}, {Walton}, {Arenou}, {Bastian}, {Cropper}, {Drimmel}, {Katz}, {Lattanzi}, {Bakker}, {Cacciari}, {Casta{\~n}eda}, {Chaoul}, {Cheek}, {De Angeli}, {Fabricius}, {Guerra}, {Holl}, {Masana}, {Messineo}, {Mowlavi}, {Nienartowicz}, {Panuzzo}, {Portell}, {Riello}, {Seabroke}, {Tanga}, {Th{\'e}venin}, {Gracia-Abril}, {Comoretto}, {Garcia-Reinaldos}, {Teyssier}, {Altmann}, {Andrae}, {Audard}, {Bellas-Velidis}, {Benson}, {Berthier}, {Blomme}, {Burgess}, {Busso}, {Carry}, {Cellino}, {Clementini}, {Clotet}, {Creevey}, {Davidson}, {De Ridder}, {Delchambre}, {Dell'Oro}, {Ducourant}, {Fern{\'a}ndez-Hern{\'a}ndez}, {Fouesneau}, {Fr{\'e}mat}, {Galluccio}, {Garc{\'\i}a-Torres},
  {Gonz{\'a}lez-N{\'u}{\~n}ez}, {Gonz{\'a}lez-Vidal}, {Gosset}, {Guy}, {Halbwachs}, {Hambly}, {Harrison}, {Hern{\'a}ndez}, {Hestroffer}, {Hodgkin}, {Hutton}, {Jasniewicz}, {Jean-Antoine-Piccolo}, {Jordan}, {Korn}, {Krone-Martins}, {Lanzafame}, {Lebzelter}, {L{\"o}ffler}, {Manteiga}, {Marrese}, {Mart{\'\i}n-Fleitas}, {Moitinho}, {Mora}, {Muinonen}, {Osinde}, {Pancino}, {Pauwels}, {Petit}, {Recio-Blanco}, {Richards}, {Rimoldini}, {Robin}, {Sarro}, {Siopis}, {Smith}, {Sozzetti}, {S{\"u}veges}, {Torra}, {van Reeven}, {Abbas}, {Abreu Aramburu}, {Accart}, {Aerts}, {Altavilla}, {{\'A}lvarez}, {Alvarez}, {Alves}, {Anderson}, {Andrei}, {Anglada Varela}, {Antiche}, {Antoja}, {Arcay}, {Astraatmadja}, {Bach}, {Baker}, {Balaguer-N{\'u}{\~n}ez}, {Balm}, {Barache}, {Barata}, {Barbato}, {Barblan}, {Barklem}, {Barrado}, {Barros}, {Barstow}, {Bartholom{\'e} Mu{\~n}oz}, {Bassilana}, {Becciani}, {Bellazzini}, {Berihuete}, {Bertone}, {Bianchi}, {Bienaym{\'e}}, {Blanco-Cuaresma}, {Boch}, {Boeche}, {Bombrun}, {Borrachero},
  {Bossini}, {Bouquillon}, {Bourda}, {Bragaglia}, {Bramante}, {Breddels}, {Bressan}, {Brouillet}, {Br{\"u}semeister}, {Brugaletta}, {Bucciarelli}, {Burlacu}, {Busonero}, {Butkevich}, {Buzzi}, {Caffau}, {Cancelliere}, {Cannizzaro}, {Cantat-Gaudin}, {Carballo}, {Carlucci}, {Carrasco}, {Casamiquela}, {Castellani}, {Castro-Ginard}, {Charlot}, {Chemin}, {Chiavassa}, {Cocozza}, {Costigan}, {Cowell}, {Crifo}, {Crosta}, {Crowley}, {Cuypers}, {Dafonte}, {Damerdji}, {Dapergolas}, {David}, {David}, {de Laverny}, {De Luise}, {De March}, {de Martino}, {de Souza}, {de Torres}, {Debosscher}, {del Pozo}, {Delbo}, {Delgado}, {Delgado}, {Di Matteo}, {Diakite}, {Diener}, {Distefano}, {Dolding}, {Drazinos}, {Dur{\'a}n}, {Edvardsson}, {Enke}, {Eriksson}, {Esquej}, {Eynard Bontemps}, {Fabre}, {Fabrizio}, {Faigler}, {Falc{\~a}o}, {Farr{\`a}s Casas}, {Federici}, {Fedorets}, {Fernique}, {Figueras}, {Filippi}, {Findeisen}, {Fonti}, {Fraile}, {Fraser}, {Fr{\'e}zouls}, {Gai}, {Galleti}, {Garabato}, {Garc{\'\i}a-Sedano}, {Garofalo},
  {Garralda}, {Gavel}, {Gavras}, {Gerssen}, {Geyer}, {Giacobbe}, {Gilmore}, {Girona}, {Giuffrida}, {Glass}, {Gomes}, {Granvik}, {Gueguen}, {Guerrier}, {Guiraud}, {Guti{\'e}rrez-S{\'a}nchez}, {Haigron}, {Hatzidimitriou}, {Hauser}, {Haywood}, {Heiter}, {Helmi}, {Heu}, {Hilger}, {Hobbs}, {Hofmann}, {Holland}, {Huckle}, {Hypki}, {Icardi}, {Jan{\ss}en}, {Jevardat de Fombelle}, {Jonker}, {Juh{\'a}sz}, {Julbe}, {Karampelas}, {Kewley}, {Klar}, {Kochoska}, {Kohley}, {Kolenberg}, {Kontizas}, {Kontizas}, {Koposov}, {Kordopatis}, {Kostrzewa-Rutkowska}, {Koubsky}, {Lambert}, {Lanza}, {Lasne}, {Lavigne}, {Le Fustec}, {Le Poncin-Lafitte}, {Lebreton}, {Leccia}, {Leclerc}, {Lecoeur-Taibi}, {Lenhardt}, {Leroux}, {Liao}, {Licata}, {Lindstr{\o}m}, {Lister}, {Livanou}, {Lobel}, {L{\'o}pez}, {Managau}, {Mann}, {Mantelet}, {Marchal}, {Marchant}, {Marconi}, {Marinoni}, {Marschalk{\'o}}, {Marshall}, {Martino}, {Marton}, {Mary}, {Massari}, {Matijevi{\v{c}}}, {Mazeh}, {McMillan}, {Messina}, {Michalik}, {Millar}, {Molina}, {Molinaro},
  {Moln{\'a}r}, {Montegriffo}, {Mor}, {Morbidelli}, {Morel}, {Morris}, {Mulone}, {Muraveva}, {Musella}, {Nelemans}, {Nicastro}, {Noval}, {O'Mullane}, {Ord{\'e}novic}, {Ord{\'o}{\~n}ez-Blanco}, {Osborne}, {Pagani}, {Pagano}, {Pailler}, {Palacin}, {Palaversa}, {Panahi}, {Pawlak}, {Piersimoni}, {Pineau}, {Plachy}, {Plum}, {Poggio}, {Poujoulet}, {Pr{\v{s}}a}, {Pulone}, {Racero}, {Ragaini}, {Rambaux}, {Ramos-Lerate}, {Regibo}, {Reyl{\'e}}, {Riclet}, {Ripepi}, {Riva}, {Rivard}, {Rixon}, {Roegiers}, {Roelens}, {Romero-G{\'o}mez}, {Rowell}, {Royer}, {Ruiz-Dern}, {Sadowski}, {Sagrist{\`a} Sell{\'e}s}, {Sahlmann}, {Salgado}, {Salguero}, {Sanna}, {Santana-Ros}, {Sarasso}, {Savietto}, {Schultheis}, {Sciacca}, {Segol}, {Segovia}, {S{\'e}gransan}, {Shih}, {Siltala}, {Silva}, {Smart}, {Smith}, {Solano}, {Solitro}, {Sordo}, {Soria Nieto}, {Souchay}, {Spagna}, {Spoto}, {Stampa}, {Steele}, {Steidelm{\"u}ller}, {Stephenson}, {Stoev}, {Suess}, {Surdej}, {Szabados}, {Szegedi-Elek}, {Tapiador}, {Taris}, {Tauran}, {Taylor},
  {Teixeira}, {Terrett}, {Teyssandier}, {Thuillot}, {Titarenko}, {Torra Clotet}, {Turon}, {Ulla}, {Utrilla}, {Uzzi}, {Vaillant}, {Valentini}, {Valette}, {van Elteren}, {Van Hemelryck}, {van Leeuwen}, {Vaschetto}, {Vecchiato}, {Veljanoski}, {Viala}, {Vicente}, {Vogt}, {von Essen}, {Voss}, {Votruba}, {Voutsinas}, {Walmsley}, {Weiler}, {Wertz}, {Wevers}, {Wyrzykowski}, {Yoldas}, {{\v{Z}}erjal}, {Ziaeepour}, {Zorec}, {Zschocke}, {Zucker}, {Zurbach}, \& {Zwitter}}]{Gaia_Collaboration_2018}
{\!\!Gaia Collaboration}, {Brown}, A.~G.~A., {Vallenari}, A., {et~al.} 2018, \bibinfo{title}{{Gaia Data Release 2. Summary of the contents and survey properties},} \aap, 616, A1, \dodoi{10.1051/0004-6361/201833051}

\bibitem[{ {\!\!Gaia Collaboration} {et~al.}(2021){\!\!Gaia Collaboration}, {Brown}, {Vallenari}, {Prusti}, {de Bruijne}, {Babusiaux}, {Biermann}, {Creevey}, {Evans}, {Eyer}, {Hutton}, {Jansen}, {Jordi}, {Klioner}, {Lammers}, {Lindegren}, {Luri}, {Mignard}, {Panem}, {Pourbaix}, {Randich}, {Sartoretti}, {Soubiran}, {Walton}, {Arenou}, {Bailer-Jones}, {Bastian}, {Cropper}, {Drimmel}, {Katz}, {Lattanzi}, {van Leeuwen}, {Bakker}, {Cacciari}, {Casta{\~n}eda}, {De Angeli}, {Ducourant}, {Fabricius}, {Fouesneau}, {Fr{\'e}mat}, {Guerra}, {Guerrier}, {Guiraud}, {Jean-Antoine Piccolo}, {Masana}, {Messineo}, {Mowlavi}, {Nicolas}, {Nienartowicz}, {Pailler}, {Panuzzo}, {Riclet}, {Roux}, {Seabroke}, {Sordo}, {Tanga}, {Th{\'e}venin}, {Gracia-Abril}, {Portell}, {Teyssier}, {Altmann}, {Andrae}, {Bellas-Velidis}, {Benson}, {Berthier}, {Blomme}, {Brugaletta}, {Burgess}, {Busso}, {Carry}, {Cellino}, {Cheek}, {Clementini}, {Damerdji}, {Davidson}, {Delchambre}, {Dell'Oro}, {Fern{\'a}ndez-Hern{\'a}ndez}, {Galluccio},
  {Garc{\'\i}a-Lario}, {Garcia-Reinaldos}, {Gonz{\'a}lez-N{\'u}{\~n}ez}, {Gosset}, {Haigron}, {Halbwachs}, {Hambly}, {Harrison}, {Hatzidimitriou}, {Heiter}, {Hern{\'a}ndez}, {Hestroffer}, {Hodgkin}, {Holl}, {Jan{\ss}en}, {Jevardat de Fombelle}, {Jordan}, {Krone-Martins}, {Lanzafame}, {L{\"o}ffler}, {Lorca}, {Manteiga}, {Marchal}, {Marrese}, {Moitinho}, {Mora}, {Muinonen}, {Osborne}, {Pancino}, {Pauwels}, {Petit}, {Recio-Blanco}, {Richards}, {Riello}, {Rimoldini}, {Robin}, {Roegiers}, {Rybizki}, {Sarro}, {Siopis}, {Smith}, {Sozzetti}, {Ulla}, {Utrilla}, {van Leeuwen}, {van Reeven}, {Abbas}, {Abreu Aramburu}, {Accart}, {Aerts}, {Aguado}, {Ajaj}, {Altavilla}, {{\'A}lvarez}, {{\'A}lvarez Cid-Fuentes}, {Alves}, {Anderson}, {Anglada Varela}, {Antoja}, {Audard}, {Baines}, {Baker}, {Balaguer-N{\'u}{\~n}ez}, {Balbinot}, {Balog}, {Barache}, {Barbato}, {Barros}, {Barstow}, {Bartolom{\'e}}, {Bassilana}, {Bauchet}, {Baudesson-Stella}, {Becciani}, {Bellazzini}, {Bernet}, {Bertone}, {Bianchi}, {Blanco-Cuaresma}, {Boch},
  {Bombrun}, {Bossini}, {Bouquillon}, {Bragaglia}, {Bramante}, {Breedt}, {Bressan}, {Brouillet}, {Bucciarelli}, {Burlacu}, {Busonero}, {Butkevich}, {Buzzi}, {Caffau}, {Cancelliere}, {C{\'a}novas}, {Cantat-Gaudin}, {Carballo}, {Carlucci}, {Carnerero}, {Carrasco}, {Casamiquela}, {Castellani}, {Castro-Ginard}, {Castro Sampol}, {Chaoul}, {Charlot}, {Chemin}, {Chiavassa}, {Cioni}, {Comoretto}, {Cooper}, {Cornez}, {Cowell}, {Crifo}, {Crosta}, {Crowley}, {Dafonte}, {Dapergolas}, {David}, {David}, {de Laverny}, {De Luise}, {De March}, {De Ridder}, {de Souza}, {de Teodoro}, {de Torres}, {del Peloso}, {del Pozo}, {Delbo}, {Delgado}, {Delgado}, {Delisle}, {Di Matteo}, {Diakite}, {Diener}, {Distefano}, {Dolding}, {Eappachen}, {Edvardsson}, {Enke}, {Esquej}, {Fabre}, {Fabrizio}, {Faigler}, {Fedorets}, {Fernique}, {Fienga}, {Figueras}, {Fouron}, {Fragkoudi}, {Fraile}, {Franke}, {Gai}, {Garabato}, {Garcia-Gutierrez}, {Garc{\'\i}a-Torres}, {Garofalo}, {Gavras}, {Gerlach}, {Geyer}, {Giacobbe}, {Gilmore}, {Girona},
  {Giuffrida}, {Gomel}, {Gomez}, {Gonzalez-Santamaria}, {Gonz{\'a}lez-Vidal}, {Granvik}, {Guti{\'e}rrez-S{\'a}nchez}, {Guy}, {Hauser}, {Haywood}, {Helmi}, {Hidalgo}, {Hilger}, {H{\l}adczuk}, {Hobbs}, {Holland}, {Huckle}, {Jasniewicz}, {Jonker}, {Juaristi Campillo}, {Julbe}, {Karbevska}, {Kervella}, {Khanna}, {Kochoska}, {Kontizas}, {Kordopatis}, {Korn}, {Kostrzewa-Rutkowska}, {Kruszy{\'n}ska}, {Lambert}, {Lanza}, {Lasne}, {Le Campion}, {Le Fustec}, {Lebreton}, {Lebzelter}, {Leccia}, {Leclerc}, {Lecoeur-Taibi}, {Liao}, {Licata}, {Lindstr{\o}m}, {Lister}, {Livanou}, {Lobel}, {Madrero Pardo}, {Managau}, {Mann}, {Marchant}, {Marconi}, {Marcos Santos}, {Marinoni}, {Marocco}, {Marshall}, {Martin Polo}, {Mart{\'\i}n-Fleitas}, {Masip}, {Massari}, {Mastrobuono-Battisti}, {Mazeh}, {McMillan}, {Messina}, {Michalik}, {Millar}, {Mints}, {Molina}, {Molinaro}, {Moln{\'a}r}, {Montegriffo}, {Mor}, {Morbidelli}, {Morel}, {Morris}, {Mulone}, {Munoz}, {Muraveva}, {Murphy}, {Musella}, {Noval}, {Ord{\'e}novic}, {Orr{\`u}},
  {Osinde}, {Pagani}, {Pagano}, {Palaversa}, {Palicio}, {Panahi}, {Pawlak}, {Pe{\~n}alosa Esteller}, {Penttil{\"a}}, {Piersimoni}, {Pineau}, {Plachy}, {Plum}, {Poggio}, {Poretti}, {Poujoulet}, {Pr{\v{s}}a}, {Pulone}, {Racero}, {Ragaini}, {Rainer}, {Raiteri}, {Rambaux}, {Ramos}, {Ramos-Lerate}, {Re Fiorentin}, {Regibo}, {Reyl{\'e}}, {Ripepi}, {Riva}, {Rixon}, {Robichon}, {Robin}, {Roelens}, {Rohrbasser}, {Romero-G{\'o}mez}, {Rowell}, {Royer}, {Rybicki}, {Sadowski}, {Sagrist{\`a} Sell{\'e}s}, {Sahlmann}, {Salgado}, {Salguero}, {Samaras}, {Sanchez Gimenez}, {Sanna}, {Santove{\~n}a}, {Sarasso}, {Schultheis}, {Sciacca}, {Segol}, {Segovia}, {S{\'e}gransan}, {Semeux}, {Shahaf}, {Siddiqui}, {Siebert}, {Siltala}, {Slezak}, {Smart}, {Solano}, {Solitro}, {Souami}, {Souchay}, {Spagna}, {Spoto}, {Steele}, {Steidelm{\"u}ller}, {Stephenson}, {S{\"u}veges}, {Szabados}, {Szegedi-Elek}, {Taris}, {Tauran}, {Taylor}, {Teixeira}, {Thuillot}, {Tonello}, {Torra}, {Torra}, {Turon}, {Unger}, {Vaillant}, {van Dillen}, {Vanel},
  {Vecchiato}, {Viala}, {Vicente}, {Voutsinas}, {Weiler}, {Wevers}, {Wyrzykowski}, {Yoldas}, {Yvard}, {Zhao}, {Zorec}, {Zucker}, {Zurbach}, \& {Zwitter}}]{Gaia_Collaboration_2021b}
{\!\!Gaia Collaboration}, {Brown}, A.~G.~A., {Vallenari}, A., {et~al.} 2021, \bibinfo{title}{{Gaia Early Data Release 3. Summary of the contents and survey properties},} \aap, 649, A1, \dodoi{10.1051/0004-6361/202039657}

\bibitem[{ {\!\!Gaia Collaboration} {et~al.}(2023){\!\!Gaia Collaboration}, {Vallenari}, {Brown}, {Prusti}, {de Bruijne}, {Arenou}, {Babusiaux}, {Biermann}, {Creevey}, {Ducourant}, {Evans}, {Eyer}, {Guerra}, {Hutton}, {Jordi}, {Klioner}, {Lammers}, {Lindegren}, {Luri}, {Mignard}, {Panem}, {Pourbaix}, {Randich}, {Sartoretti}, {Soubiran}, {Tanga}, {Walton}, {Bailer-Jones}, {Bastian}, {Drimmel}, {Jansen}, {Katz}, {Lattanzi}, {van Leeuwen}, {Bakker}, {Cacciari}, {Casta{\~n}eda}, {De Angeli}, {Fabricius}, {Fouesneau}, {Fr{\'e}mat}, {Galluccio}, {Guerrier}, {Heiter}, {Masana}, {Messineo}, {Mowlavi}, {Nicolas}, {Nienartowicz}, {Pailler}, {Panuzzo}, {Riclet}, {Roux}, {Seabroke}, {Sordo}, {Th{\'e}venin}, {Gracia-Abril}, {Portell}, {Teyssier}, {Altmann}, {Andrae}, {Audard}, {Bellas-Velidis}, {Benson}, {Berthier}, {Blomme}, {Burgess}, {Busonero}, {Busso}, {C{\'a}novas}, {Carry}, {Cellino}, {Cheek}, {Clementini}, {Damerdji}, {Davidson}, {de Teodoro}, {Nu{\~n}ez Campos}, {Delchambre}, {Dell'Oro}, {Esquej},
  {Fern{\'a}ndez-Hern{\'a}ndez}, {Fraile}, {Garabato}, {Garc{\'\i}a-Lario}, {Gosset}, {Haigron}, {Halbwachs}, {Hambly}, {Harrison}, {Hern{\'a}ndez}, {Hestroffer}, {Hodgkin}, {Holl}, {Jan{\ss}en}, {Jevardat de Fombelle}, {Jordan}, {Krone-Martins}, {Lanzafame}, {L{\"o}ffler}, {Marchal}, {Marrese}, {Moitinho}, {Muinonen}, {Osborne}, {Pancino}, {Pauwels}, {Recio-Blanco}, {Reyl{\'e}}, {Riello}, {Rimoldini}, {Roegiers}, {Rybizki}, {Sarro}, {Siopis}, {Smith}, {Sozzetti}, {Utrilla}, {van Leeuwen}, {Abbas}, {{\'A}brah{\'a}m}, {Abreu Aramburu}, {Aerts}, {Aguado}, {Ajaj}, {Aldea-Montero}, {Altavilla}, {{\'A}lvarez}, {Alves}, {Anders}, {Anderson}, {Anglada Varela}, {Antoja}, {Baines}, {Baker}, {Balaguer-N{\'u}{\~n}ez}, {Balbinot}, {Balog}, {Barache}, {Barbato}, {Barros}, {Barstow}, {Bartolom{\'e}}, {Bassilana}, {Bauchet}, {Becciani}, {Bellazzini}, {Berihuete}, {Bernet}, {Bertone}, {Bianchi}, {Binnenfeld}, {Blanco-Cuaresma}, {Blazere}, {Boch}, {Bombrun}, {Bossini}, {Bouquillon}, {Bragaglia}, {Bramante}, {Breedt},
  {Bressan}, {Brouillet}, {Brugaletta}, {Bucciarelli}, {Burlacu}, {Butkevich}, {Buzzi}, {Caffau}, {Cancelliere}, {Cantat-Gaudin}, {Carballo}, {Carlucci}, {Carnerero}, {Carrasco}, {Casamiquela}, {Castellani}, {Castro-Ginard}, {Chaoul}, {Charlot}, {Chemin}, {Chiaramida}, {Chiavassa}, {Chornay}, {Comoretto}, {Contursi}, {Cooper}, {Cornez}, {Cowell}, {Crifo}, {Cropper}, {Crosta}, {Crowley}, {Dafonte}, {Dapergolas}, {David}, {David}, {de Laverny}, {De Luise}, {De March}, {De Ridder}, {de Souza}, {de Torres}, {del Peloso}, {del Pozo}, {Delbo}, {Delgado}, {Delisle}, {Demouchy}, {Dharmawardena}, {Di Matteo}, {Diakite}, {Diener}, {Distefano}, {Dolding}, {Edvardsson}, {Enke}, {Fabre}, {Fabrizio}, {Faigler}, {Fedorets}, {Fernique}, {Fienga}, {Figueras}, {Fournier}, {Fouron}, {Fragkoudi}, {Gai}, {Garcia-Gutierrez}, {Garcia-Reinaldos}, {Garc{\'\i}a-Torres}, {Garofalo}, {Gavel}, {Gavras}, {Gerlach}, {Geyer}, {Giacobbe}, {Gilmore}, {Girona}, {Giuffrida}, {Gomel}, {Gomez}, {Gonz{\'a}lez-N{\'u}{\~n}ez},
  {Gonz{\'a}lez-Santamar{\'\i}a}, {Gonz{\'a}lez-Vidal}, {Granvik}, {Guillout}, {Guiraud}, {Guti{\'e}rrez-S{\'a}nchez}, {Guy}, {Hatzidimitriou}, {Hauser}, {Haywood}, {Helmer}, {Helmi}, {Sarmiento}, {Hidalgo}, {Hilger}, {H{\l}adczuk}, {Hobbs}, {Holland}, {Huckle}, {Jardine}, {Jasniewicz}, {Jean-Antoine Piccolo}, {Jim{\'e}nez-Arranz}, {Jorissen}, {Juaristi Campillo}, {Julbe}, {Karbevska}, {Kervella}, {Khanna}, {Kontizas}, {Kordopatis}, {Korn}, {K{\'o}sp{\'a}l}, {Kostrzewa-Rutkowska}, {Kruszy{\'n}ska}, {Kun}, {Laizeau}, {Lambert}, {Lanza}, {Lasne}, {Le Campion}, {Lebreton}, {Lebzelter}, {Leccia}, {Leclerc}, {Lecoeur-Taibi}, {Liao}, {Licata}, {Lindstr{\o}m}, {Lister}, {Livanou}, {Lobel}, {Lorca}, {Loup}, {Madrero Pardo}, {Magdaleno Romeo}, {Managau}, {Mann}, {Manteiga}, {Marchant}, {Marconi}, {Marcos}, {Marcos Santos}, {Mar{\'\i}n Pina}, {Marinoni}, {Marocco}, {Marshall}, {Martin Polo}, {Mart{\'\i}n-Fleitas}, {Marton}, {Mary}, {Masip}, {Massari}, {Mastrobuono-Battisti}, {Mazeh}, {McMillan}, {Messina}, {Michalik},
  {Millar}, {Mints}, {Molina}, {Molinaro}, {Moln{\'a}r}, {Monari}, {Mongui{\'o}}, {Montegriffo}, {Montero}, {Mor}, {Mora}, {Morbidelli}, {Morel}, {Morris}, {Muraveva}, {Murphy}, {Musella}, {Nagy}, {Noval}, {Oca{\~n}a}, {Ogden}, {Ordenovic}, {Osinde}, {Pagani}, {Pagano}, {Palaversa}, {Palicio}, {Pallas-Quintela}, {Panahi}, {Payne-Wardenaar}, {Pe{\~n}alosa Esteller}, {Penttil{\"a}}, {Pichon}, {Piersimoni}, {Pineau}, {Plachy}, {Plum}, {Poggio}, {Pr{\v{s}}a}, {Pulone}, {Racero}, {Ragaini}, {Rainer}, {Raiteri}, {Rambaux}, {Ramos}, {Ramos-Lerate}, {Re Fiorentin}, {Regibo}, {Richards}, {Rios Diaz}, {Ripepi}, {Riva}, {Rix}, {Rixon}, {Robichon}, {Robin}, {Robin}, {Roelens}, {Rogues}, {Rohrbasser}, {Romero-G{\'o}mez}, {Rowell}, {Royer}, {Ruz Mieres}, {Rybicki}, {Sadowski}, {S{\'a}ez N{\'u}{\~n}ez}, {Sagrist{\`a} Sell{\'e}s}, {Sahlmann}, {Salguero}, {Samaras}, {Sanchez Gimenez}, {Sanna}, {Santove{\~n}a}, {Sarasso}, {Schultheis}, {Sciacca}, {Segol}, {Segovia}, {S{\'e}gransan}, {Semeux}, {Shahaf}, {Siddiqui}, {Siebert},
  {Siltala}, {Silvelo}, {Slezak}, {Slezak}, {Smart}, {Snaith}, {Solano}, {Solitro}, {Souami}, {Souchay}, {Spagna}, {Spina}, {Spoto}, {Steele}, {Steidelm{\"u}ller}, {Stephenson}, {S{\"u}veges}, {Surdej}, {Szabados}, {Szegedi-Elek}, {Taris}, {Taylor}, {Teixeira}, {Tolomei}, {Tonello}, {Torra}, {Torra}, {Torralba Elipe}, {Trabucchi}, {Tsounis}, {Turon}, {Ulla}, {Unger}, {Vaillant}, {van Dillen}, {van Reeven}, {Vanel}, {Vecchiato}, {Viala}, {Vicente}, {Voutsinas}, {Weiler}, {Wevers}, {Wyrzykowski}, {Yoldas}, {Yvard}, {Zhao}, {Zorec}, {Zucker}, \& {Zwitter}}]{GaiaDR3_Collaboration_2023}
{\!\!Gaia Collaboration}, {Vallenari}, A., {Brown}, A.~G.~A., {et~al.} 2023, \bibinfo{title}{{Gaia Data Release 3. Summary of the content and survey properties},} \aap, 674, A1, \dodoi{10.1051/0004-6361/202243940}

\bibitem[{G.~M. {Green} {et~al.}(2019){Green}, {Schlafly}, {Zucker}, {Speagle}, \& {Finkbeiner}}]{Green_2019}
{Green}, G.~M., {Schlafly}, E., {Zucker}, C., {Speagle}, J.~S., \& {Finkbeiner}, D. 2019, \bibinfo{title}{{A 3D Dust Map Based on Gaia, Pan-STARRS 1, and 2MASS},} \apj, 887, 93, \dodoi{10.3847/1538-4357/ab5362}

\bibitem[{J.~S. {Hall}(1949){Hall}}]{Hall_1949}
{Hall}, J.~S. 1949, \bibinfo{title}{{Observations of the Polarized Light from Stars},} Science, 109, 166, \dodoi{10.1126/science.109.2825.166}

\bibitem[{C.~R. {Harris} {et~al.}(2020){Harris}, Millman, van~der Walt, Gommers, Virtanen, Cournapeau, Wieser, Taylor, Berg, Smith, Kern, Picus, Hoyer, van Kerkwijk, Brett, Haldane, del R{\'{i}}o, Wiebe, Peterson, G{\'{e}}rard-Marchant, Sheppard, Reddy, Weckesser, Abbasi, Gohlke, \& Oliphant}]{Harris_numpy_2020}
{Harris}, C.~R., Millman, K.~J., van~der Walt, S.~J., {et~al.} 2020, \bibinfo{title}{Array programming with {NumPy},} Nature, 585, 357, \dodoi{10.1038/s41586-020-2649-2}

\bibitem[{H. {Hatano} {et~al.}(2013){Hatano}, {Nishiyama}, {Kurita}, {Kanai}, {Nakajima}, {Nagata}, {Tamura}, {Kandori}, {Kato}, {Sato}, {Yoshikawa}, {Suenaga}, \& {Sato}}]{Hatano_2013}
{Hatano}, H., {Nishiyama}, S., {Kurita}, M., {et~al.} 2013, \bibinfo{title}{{The Efficiency and Wavelength Dependence of Near-infrared Interstellar Polarization toward the Galactic Center},} \aj, 145, 105, \dodoi{10.1088/0004-6256/145/4/105}

\bibitem[{M. {Haverkorn}(2015){Haverkorn}}]{Haverkorn_2015}
{Haverkorn}, M. 2015, in Astrophysics and Space Science Library, Vol. 407, Magnetic Fields in Diffuse Media, ed. A.~{Lazarian}, E.~M. {de Gouveia Dal Pino}, \& C.~{Melioli}, 483, \dodoi{10.1007/978-3-662-44625-6_17}

\bibitem[{C. {Heiles}(1996){Heiles}}]{Heiles_1996}
{Heiles}, C. 1996, \bibinfo{title}{{The Local Direction and Curvature of the Galactic Magnetic Field Derived from Starlight Polarization},} \apj, 462, 316, \dodoi{10.1086/177153}

\bibitem[{W.~A. {Hiltner}(1949){Hiltner}}]{Hiltner_1949}
{Hiltner}, W.~A. 1949, \bibinfo{title}{{Polarization of Light from Distant Stars by Interstellar Medium},} Science, 109, 165, \dodoi{10.1126/science.109.2825.165}

\bibitem[{J.~H. {Hough} {et~al.}(1988){Hough}, {Sato}, {Tamura}, {Yamashita}, {McFadean}, {Rouse}, {Whittet}, {Kaifu}, {Suzuki}, {Nagata}, {Gatley}, \& {Bailey}}]{Hough_1988}
{Hough}, J.~H., {Sato}, S., {Tamura}, M., {et~al.} 1988, \bibinfo{title}{{Spectropolarimetry of the 3-mu.m ice band in Elias 16 (Taurus dark cloud).},} \mnras, 230, 107, \dodoi{10.1093/mnras/230.1.107}

\bibitem[{J.~D. {Hunter}(2007){Hunter}}]{Hunter_Matplotlib_2007}
{Hunter}, J.~D. 2007, \bibinfo{title}{Matplotlib: A 2D graphics environment,} Computing in Science \& Engineering, 9, 90, \dodoi{10.1109/MCSE.2007.55}

\bibitem[{D.~L. {Jow} {et~al.}(2018){Jow}, {Hill}, {Scott}, {Soler}, {Martin}, {Devlin}, {Fissel}, \& {Poidevin}}]{Jow_2018}
{Jow}, D.~L., {Hill}, R., {Scott}, D., {et~al.} 2018, \bibinfo{title}{{An application of an optimal statistic for characterizing relative orientations},} \mnras, 474, 1018, \dodoi{10.1093/mnras/stx2736}

\bibitem[{R. {Lallement} {et~al.}(2022){Lallement}, {Vergely}, {Babusiaux}, \& {Cox}}]{Lallement_2022}
{Lallement}, R., {Vergely}, J.~L., {Babusiaux}, C., \& {Cox}, N.~L.~J. 2022, \bibinfo{title}{{Updated Gaia-2MASS 3D maps of Galactic interstellar dust},} \aap, 661, A147, \dodoi{10.1051/0004-6361/202142846}

\bibitem[{R. {Lallement} {et~al.}(2003){Lallement}, {Welsh}, {Vergely}, {Crifo}, \& {Sfeir}}]{Lallement_2003}
{Lallement}, R., {Welsh}, B.~Y., {Vergely}, J.~L., {Crifo}, F., \& {Sfeir}, D. 2003, \bibinfo{title}{{3D mapping of the dense interstellar gas around the Local Bubble},} \aap, 411, 447, \dodoi{10.1051/0004-6361:20031214}

\bibitem[{A. {Lazarian} \& T. {Hoang}(2007){Lazarian} \& {Hoang}}]{Lazarian_Hoang_2007}
{Lazarian}, A., \& {Hoang}, T. 2007, \bibinfo{title}{{Radiative torques: analytical model and basic properties},} \mnras, 378, 910, \dodoi{10.1111/j.1365-2966.2007.11817.x}

\bibitem[{R.~H. {Leike} {et~al.}(2020){Leike}, {Glatzle}, \& {En{\ss}lin}}]{Leike_2020}
{Leike}, R.~H., {Glatzle}, M., \& {En{\ss}lin}, T.~A. 2020, \bibinfo{title}{{Resolving nearby dust clouds},} \aap, 639, A138, \dodoi{10.1051/0004-6361/202038169}

\bibitem[{D. {Lenz} {et~al.}(2017){Lenz}, {Hensley}, \& {Dor{\'e}}}]{Lenz_2017}
{Lenz}, D., {Hensley}, B.~S., \& {Dor{\'e}}, O. 2017, \bibinfo{title}{{A New, Large-scale Map of Interstellar Reddening Derived from H I Emission},} \apj, 846, 38, \dodoi{10.3847/1538-4357/aa84af}

\bibitem[{J.~L. {Leroy}(1999){Leroy}}]{Leroy_1999}
{Leroy}, J.~L. 1999, \bibinfo{title}{{Interstellar dust and magnetic field at the boundaries of the Local Bubble. Analysis of polarimetric data in the light of HIPPARCOS parallaxes},} \aap, 346, 955

\bibitem[{L. {Lindegren} {et~al.}(2021{\natexlab{a}}){Lindegren}, {Klioner}, {Hern{\'a}ndez}, {Bombrun}, {Ramos-Lerate}, {Steidelm{\"u}ller}, {Bastian}, {Biermann}, {de Torres}, {Gerlach}, {Geyer}, {Hilger}, {Hobbs}, {Lammers}, {McMillan}, {Stephenson}, {Casta{\~n}eda}, {Davidson}, {Fabricius}, {Gracia-Abril}, {Portell}, {Rowell}, {Teyssier}, {Torra}, {Bartolom{\'e}}, {Clotet}, {Garralda}, {Gonz{\'a}lez-Vidal}, {Torra}, {Abbas}, {Altmann}, {Anglada Varela}, {Balaguer-N{\'u}{\~n}ez}, {Balog}, {Barache}, {Becciani}, {Bernet}, {Bertone}, {Bianchi}, {Bouquillon}, {Brown}, {Bucciarelli}, {Busonero}, {Butkevich}, {Buzzi}, {Cancelliere}, {Carlucci}, {Charlot}, {Cioni}, {Crosta}, {Crowley}, {del Peloso}, {del Pozo}, {Drimmel}, {Esquej}, {Fienga}, {Fraile}, {Gai}, {Garcia-Reinaldos}, {Guerra}, {Hambly}, {Hauser}, {Jan{\ss}en}, {Jordan}, {Kostrzewa-Rutkowska}, {Lattanzi}, {Liao}, {Licata}, {Lister}, {L{\"o}ffler}, {Marchant}, {Masip}, {Mignard}, {Mints}, {Molina}, {Mora}, {Morbidelli}, {Murphy}, {Pagani}, {Panuzzo},
  {Pe{\~n}alosa Esteller}, {Poggio}, {Re Fiorentin}, {Riva}, {Sagrist{\`a} Sell{\'e}s}, {Sanchez Gimenez}, {Sarasso}, {Sciacca}, {Siddiqui}, {Smart}, {Souami}, {Spagna}, {Steele}, {Taris}, {Utrilla}, {van Reeven}, \& {Vecchiato}}]{Lindegren_Gaiaedr3_astrometric_2021}
{Lindegren}, L., {Klioner}, S.~A., {Hern{\'a}ndez}, J., {et~al.} 2021{\natexlab{a}}, \bibinfo{title}{{Gaia Early Data Release 3. The astrometric solution},} \aap, 649, A2, \dodoi{10.1051/0004-6361/202039709}

\bibitem[{L. {Lindegren} {et~al.}(2021{\natexlab{b}}){Lindegren}, {Bastian}, {Biermann}, {Bombrun}, {de Torres}, {Gerlach}, {Geyer}, {Hern{\'a}ndez}, {Hilger}, {Hobbs}, {Klioner}, {Lammers}, {McMillan}, {Ramos-Lerate}, {Steidelm{\"u}ller}, {Stephenson}, \& {van Leeuwen}}]{Lindegren_Gaiaedr3_Plx_bias_2021}
{Lindegren}, L., {Bastian}, U., {Biermann}, M., {et~al.} 2021{\natexlab{b}}, \bibinfo{title}{{Gaia Early Data Release 3. Parallax bias versus magnitude, colour, and position},} \aap, 649, A4, \dodoi{10.1051/0004-6361/202039653}

\bibitem[{W. {Liu} {et~al.}(2017){Liu}, {Chiao}, {Collier}, {Cravens}, {Galeazzi}, {Koutroumpa}, {Kuntz}, {Lallement}, {Lepri}, {McCammon}, {Morgan}, {Porter}, {Snowden}, {Thomas}, {Uprety}, {Ursino}, \& {Walsh}}]{Liu_LBW_2017}
{Liu}, W., {Chiao}, M., {Collier}, M.~R., {et~al.} 2017, \bibinfo{title}{{The Structure of the Local Hot Bubble},} \apj, 834, 33, \dodoi{10.3847/1538-4357/834/1/33}

\bibitem[{S. {Lloyd} \& M.~O. {Harwit}(1973){Lloyd} \& {Harwit}}]{Lloyd_Harwit_1973}
{Lloyd}, S., \& {Harwit}, M.~O. 1973, in Interstellar Dust and Related Topics, ed. J.~M. {Greenberg} \& H.~C. {van de Hulst}, Vol.~52, 203

\bibitem[{X. {Luri} {et~al.}(2018){Luri}, {Brown}, {Sarro}, {Arenou}, {Bailer-Jones}, {Castro-Ginard}, {de Bruijne}, {Prusti}, {Babusiaux}, \& {Delgado}}]{Luri_GaiaDR2_plx_2018}
{Luri}, X., {Brown}, A.~G.~A., {Sarro}, L.~M., {et~al.} 2018, \bibinfo{title}{{Gaia Data Release 2. Using Gaia parallaxes},} \aap, 616, A9, \dodoi{10.1051/0004-6361/201832964}

\bibitem[{A.~M. {Magalh{\~a}es} {et~al.}(2005){Magalh{\~a}es}, {Pereyra}, {Melgarejo}, {de Matos}, {Carciofi}, {Benedito}, {Valentim}, {Vidotto}, {da Silva}, {de Souza}, {Faria}, \& {Gabriel}}]{Magalhaes_2005}
{Magalh{\~a}es}, A.~M., {Pereyra}, A., {Melgarejo}, R., {et~al.} 2005, in Astronomical Society of the Pacific Conference Series, Vol. 343, Astronomical Polarimetry: Current Status and Future Directions, ed. A.~{Adamson}, C.~{Aspin}, C.~{Davis}, \& T.~{Fujiyoshi}, 305

\bibitem[{G.~J. {McLachlan} \& K.~E. {Basford}(1988){McLachlan} \& {Basford}}]{McLachlan_1988}
{McLachlan}, G.~J., \& {Basford}, K.~E. 1988, {Mixture models. Inference and applications to clustering}

\bibitem[{I. {Medan} \& B.~G. {Andersson}(2019){Medan} \& {Andersson}}]{Medan_Andersson_2019}
{Medan}, I., \& {Andersson}, B.~G. 2019, \bibinfo{title}{{Magnetic Field Strengths and Variations in Grain Alignment in the Local Bubble Wall},} \apj, 873, 87, \dodoi{10.3847/1538-4357/ab063c}

\bibitem[{S. {Nishiyama} {et~al.}(2009){Nishiyama}, {Tamura}, {Hatano}, {Kanai}, {Kurita}, {Sato}, {Matsunaga}, {Nagata}, {Nagayama}, {Kandori}, {Nakajima}, {Kusakabe}, {Sato}, {Hough}, {Sugitani}, \& {Okuda}}]{Nishiyama_2009}
{Nishiyama}, S., {Tamura}, M., {Hatano}, H., {et~al.} 2009, \bibinfo{title}{{Magnetic Field Configuration at the Galactic Center Investigated by Wide Field Near-Infrared Polarimetry},} \apj, 690, 1648, \dodoi{10.1088/0004-637X/690/2/1648}

\bibitem[{G.~V. {Panopoulou} {et~al.}(2021){Panopoulou}, {Dickinson}, {Readhead}, {Pearson}, \& {Peel}}]{Panopoulou_2021}
{Panopoulou}, G.~V., {Dickinson}, C., {Readhead}, A.~C.~S., {Pearson}, T.~J., \& {Peel}, M.~W. 2021, \bibinfo{title}{{Revisiting the Distance to Radio Loops I and IV Using Gaia and Radio/Optical Polarization Data},} \apj, 922, 210, \dodoi{10.3847/1538-4357/ac273f}

\bibitem[{G.~V. {Panopoulou} {et~al.}(2019){Panopoulou}, {Tassis}, {Skalidis}, {Blinov}, {Liodakis}, {Pavlidou}, {Potter}, {Ramaprakash}, {Readhead}, \& {Wehus}}]{Panopoulou_tomography_2019}
{Panopoulou}, G.~V., {Tassis}, K., {Skalidis}, R., {et~al.} 2019, \bibinfo{title}{{Demonstration of Magnetic Field Tomography with Starlight Polarization toward a Diffuse Sightline of the ISM},} \apj, 872, 56, \dodoi{10.3847/1538-4357/aafdb2}

\bibitem[{F. {Patat} {et~al.}(2010){Patat}, {Maund}, {Benetti}, {Botticella}, {Cappellaro}, {Harutyunyan}, \& {Turatto}}]{Patat_2010}
{Patat}, F., {Maund}, J.~R., {Benetti}, S., {et~al.} 2010, \bibinfo{title}{{VLT spectropolarimetry of the optical transient in NGC 300. Evidence of asymmetry in the circumstellar dust},} \aap, 510, A108, \dodoi{10.1051/0004-6361/200913083}

\bibitem[{K. {Pattle} {et~al.}(2023){Pattle}, {Fissel}, {Tahani}, {Liu}, \& {Ntormousi}}]{Pattle_2023}
{Pattle}, K., {Fissel}, L., {Tahani}, M., {Liu}, T., \& {Ntormousi}, E. 2023, in Astronomical Society of the Pacific Conference Series, Vol. 534, Protostars and Planets VII, ed. S.~{Inutsuka}, Y.~{Aikawa}, T.~{Muto}, K.~{Tomida}, \& M.~{Tamura}, 193, \dodoi{10.48550/arXiv.2203.11179}

\bibitem[{M.~D. {Pavel}(2014){Pavel}}]{Pavel_2014}
{Pavel}, M.~D. 2014, \bibinfo{title}{{Using Red Clump Stars to Decompose the Galactic Magnetic Field with Distance},} \aj, 148, 49, \dodoi{10.1088/0004-6256/148/3/49}

\bibitem[{M.~D. {Pavel} {et~al.}(2012){Pavel}, {Clemens}, \& {Pinnick}}]{Pavel_2012}
{Pavel}, M.~D., {Clemens}, D.~P., \& {Pinnick}, A.~F. 2012, \bibinfo{title}{{Testing Galactic Magnetic Field Models Using Near-infrared Polarimetry},} \apj, 749, 71, \dodoi{10.1088/0004-637X/749/1/71}

\bibitem[{F. {Pedregosa} {et~al.}(2011){Pedregosa}, {Varoquaux}, {Gramfort}, {Michel}, {Thirion}, {Grisel}, {Blondel}, {M{\"u}ller}, {Nothman}, {Louppe}, {Prettenhofer}, {Weiss}, {Dubourg}, {Vanderplas}, {Passos}, {Cournapeau}, {Brucher}, {Perrot}, \& {Duchesnay}}]{Pedregosa_Scikitlearn_2011}
{Pedregosa}, F., {Varoquaux}, G., {Gramfort}, A., {et~al.} 2011, \bibinfo{title}{{Scikit-learn: Machine Learning in Python},} Journal of Machine Learning Research, 12, 2825, \dodoi{10.48550/arXiv.1201.0490}

\bibitem[{V. {Pelgrims} {et~al.}(2020){Pelgrims}, {Ferri{\`e}re}, {Boulanger}, {Lallement}, \& {Montier}}]{Pelgrims_2020}
{Pelgrims}, V., {Ferri{\`e}re}, K., {Boulanger}, F., {Lallement}, R., \& {Montier}, L. 2020, \bibinfo{title}{{Modeling the magnetized Local Bubble from dust data},} \aap, 636, A17, \dodoi{10.1051/0004-6361/201937157}

\bibitem[{V. {Pelgrims} {et~al.}(2023){Pelgrims}, {Panopoulou}, {Tassis}, {Pavlidou}, {Basyrov}, {Blinov}, {Gjerl{\ensuremath{\varnothing}}w}, {Kiehlmann}, {Mandarakas}, {Papadaki}, {Skalidis}, {Tsouros}, {Anche}, {Eriksen}, {Ghosh}, {Kypriotakis}, {Maharana}, {Ntormousi}, {Pearson}, {Potter}, {Ramaprakash}, {Readhead}, \& {Wehus}}]{Pelgrims_2023}
{Pelgrims}, V., {Panopoulou}, G.~V., {Tassis}, K., {et~al.} 2023, \bibinfo{title}{{Starlight-polarization-based tomography of the magnetized ISM: PASIPHAE's line-of-sight inversion method},} \aap, 670, A164, \dodoi{10.1051/0004-6361/202244625}

\bibitem[{V. {Pelgrims} {et~al.}(2024){Pelgrims}, {Mandarakas}, {Skalidis}, {Tassis}, {Panopoulou}, {Pavlidou}, {Blinov}, {Kiehlmann}, {Clark}, {Hensley}, {Romanopoulos}, {Basyrov}, {Eriksen}, {Falalaki}, {Ghosh}, {Gjerl{\o}w}, {Kypriotakis}, {Maharana}, {Papadaki}, {Pearson}, {Potter}, {Ramaprakash}, {Readhead}, \& {Wehus}}]{Pelgrims_2024}
{Pelgrims}, V., {Mandarakas}, N., {Skalidis}, R., {et~al.} 2024, \bibinfo{title}{{The first degree-scale starlight-polarization-based tomography map of the magnetized interstellar medium},} arXiv e-prints, arXiv:2404.10821, \dodoi{10.48550/arXiv.2404.10821}

\bibitem[{ {\!\!Planck Collaboration III} {et~al.}(2020){\!\!Planck Collaboration III}, {Aghanim}, {Akrami}, {Ashdown}, {Aumont}, {Baccigalupi}, {Ballardini}, {Banday}, {Barreiro}, {Bartolo}, {Basak}, {Benabed}, {Bernard}, {Bersanelli}, {Bielewicz}, {Bond}, {Borrill}, {Bouchet}, {Boulanger}, {Bucher}, {Burigana}, {Calabrese}, {Cardoso}, {Carron}, {Challinor}, {Chiang}, {Colombo}, {Combet}, {Couchot}, {Crill}, {Cuttaia}, {de Bernardis}, {de Rosa}, {de Zotti}, {Delabrouille}, {Delouis}, {Di Valentino}, {Diego}, {Dor{\'e}}, {Douspis}, {Ducout}, {Dupac}, {Efstathiou}, {Elsner}, {En{\ss}lin}, {Eriksen}, {Falgarone}, {Fantaye}, {Finelli}, {Frailis}, {Fraisse}, {Franceschi}, {Frolov}, {Galeotta}, {Galli}, {Ganga}, {G{\'e}nova-Santos}, {Gerbino}, {Ghosh}, {Gonz{\'a}lez-Nuevo}, {G{\'o}rski}, {Gratton}, {Gruppuso}, {Gudmundsson}, {Handley}, {Hansen}, {Henrot-Versill{\'e}}, {Herranz}, {Hivon}, {Huang}, {Jaffe}, {Jones}, {Karakci}, {Keih{\"a}nen}, {Keskitalo}, {Kiiveri}, {Kim}, {Kisner}, {Krachmalnicoff}, {Kunz},
  {Kurki-Suonio}, {Lagache}, {Lamarre}, {Lasenby}, {Lattanzi}, {Lawrence}, {Levrier}, {Liguori}, {Lilje}, {Lindholm}, {L{\'o}pez-Caniego}, {Ma}, {Mac{\'\i}as-P{\'e}rez}, {Maggio}, {Maino}, {Mandolesi}, {Mangilli}, {Martin}, {Mart{\'\i}nez-Gonz{\'a}lez}, {Matarrese}, {Mauri}, {McEwen}, {Melchiorri}, {Mennella}, {Migliaccio}, {Miville-Desch{\^e}nes}, {Molinari}, {Moneti}, {Montier}, {Morgante}, {Moss}, {Mottet}, {Natoli}, {Pagano}, {Paoletti}, {Partridge}, {Patanchon}, {Patrizii}, {Perdereau}, {Perrotta}, {Pettorino}, {Piacentini}, {Puget}, {Rachen}, {Reinecke}, {Remazeilles}, {Renzi}, {Rocha}, {Roudier}, {Salvati}, {Sandri}, {Savelainen}, {Scott}, {Sirignano}, {Sirri}, {Spencer}, {Sunyaev}, {Suur-Uski}, {Tauber}, {Tavagnacco}, {Tenti}, {Toffolatti}, {Tomasi}, {Tristram}, {Trombetti}, {Valiviita}, {Vansyngel}, {Van Tent}, {Vibert}, {Vielva}, {Villa}, {Vittorio}, {Wandelt}, {Wehus}, \& {Zonca}}]{Planck-Collaboration_2020}
{\!\!Planck Collaboration III}, {Aghanim}, N., {Akrami}, Y., {et~al.} 2020, \bibinfo{title}{{Planck 2018 results. III. High Frequency Instrument data processing and frequency maps},} \aap, 641, A3, \dodoi{10.1051/0004-6361/201832909}

\bibitem[{ {\!\!Planck Collaboration Int. XIX} {et~al.}(2015){\!\!Planck Collaboration Int. XIX}, {Ade}, {Aghanim}, {Alina}, {Alves}, {Armitage-Caplan}, {Arnaud}, {Arzoumanian}, {Ashdown}, {Atrio-Barandela}, {Aumont}, {Baccigalupi}, {Banday}, {Barreiro}, {Battaner}, {Benabed}, {Benoit-L{\'e}vy}, {Bernard}, {Bersanelli}, {Bielewicz}, {Bock}, {Bond}, {Borrill}, {Bouchet}, {Boulanger}, {Bracco}, {Burigana}, {Butler}, {Cardoso}, {Catalano}, {Chamballu}, {Chary}, {Chiang}, {Christensen}, {Colombi}, {Colombo}, {Combet}, {Couchot}, {Coulais}, {Crill}, {Curto}, {Cuttaia}, {Danese}, {Davies}, {Davis}, {de Bernardis}, {de Gouveia Dal Pino}, {de Rosa}, {de Zotti}, {Delabrouille}, {D{\'e}sert}, {Dickinson}, {Diego}, {Donzelli}, {Dor{\'e}}, {Douspis}, {Dunkley}, {Dupac}, {Efstathiou}, {En{\ss}lin}, {Eriksen}, {Falgarone}, {Ferri{\`e}re}, {Finelli}, {Forni}, {Frailis}, {Fraisse}, {Franceschi}, {Galeotta}, {Ganga}, {Ghosh}, {Giard}, {Giraud-H{\'e}raud}, {Gonz{\'a}lez-Nuevo}, {G{\'o}rski}, {Gregorio}, {Gruppuso}, {Guillet},
  {Hansen}, {Harrison}, {Helou}, {Hern{\'a}ndez-Monteagudo}, {Hildebrandt}, {Hivon}, {Hobson}, {Holmes}, {Hornstrup}, {Huffenberger}, {Jaffe}, {Jaffe}, {Jones}, {Juvela}, {Keih{\"a}nen}, {Keskitalo}, {Kisner}, {Kneissl}, {Knoche}, {Kunz}, {Kurki-Suonio}, {Lagache}, {L{\"a}hteenm{\"a}ki}, {Lamarre}, {Lasenby}, {Lawrence}, {Leahy}, {Leonardi}, {Levrier}, {Liguori}, {Lilje}, {Linden-V{\o}rnle}, {L{\'o}pez-Caniego}, {Lubin}, {Mac{\'\i}as-P{\'e}rez}, {Maffei}, {Magalh{\~a}es}, {Maino}, {Mandolesi}, {Maris}, {Marshall}, {Martin}, {Mart{\'\i}nez-Gonz{\'a}lez}, {Masi}, {Matarrese}, {Mazzotta}, {Melchiorri}, {Mendes}, {Mennella}, {Migliaccio}, {Miville-Desch{\^e}nes}, {Moneti}, {Montier}, {Morgante}, {Mortlock}, {Munshi}, {Murphy}, {Naselsky}, {Nati}, {Natoli}, {Netterfield}, {Noviello}, {Novikov}, {Novikov}, {Oxborrow}, {Pagano}, {Pajot}, {Paladini}, {Paoletti}, {Pasian}, {Pearson}, {Perdereau}, {Perotto}, {Perrotta}, {Piacentini}, {Piat}, {Pietrobon}, {Plaszczynski}, {Poidevin}, {Pointecouteau}, {Polenta}, {Popa},
  {Pratt}, {Prunet}, {Puget}, {Rachen}, {Reach}, {Rebolo}, {Reinecke}, {Remazeilles}, {Renault}, {Ricciardi}, {Riller}, {Ristorcelli}, {Rocha}, {Rosset}, {Roudier}, {Rubi{\~n}o-Mart{\'\i}n}, {Rusholme}, {Sandri}, {Savini}, {Scott}, {Spencer}, {Stolyarov}, {Stompor}, {Sudiwala}, {Sutton}, {Suur-Uski}, {Sygnet}, {Tauber}, {Terenzi}, {Toffolatti}, {Tomasi}, {Tristram}, {Tucci}, {Umana}, {Valenziano}, {Valiviita}, {Van Tent}, {Vielva}, {Villa}, {Wade}, {Wandelt}, {Zacchei}, \& {Zonca}}]{Planck-Collaboration_XIX_2015}
{\!\!Planck Collaboration Int. XIX}, {Ade}, P.~A.~R., {Aghanim}, N., {et~al.} 2015, \bibinfo{title}{{Planck intermediate results. XIX. An overview of the polarized thermal emission from Galactic dust},} \aap, 576, A104, \dodoi{10.1051/0004-6361/201424082}

\bibitem[{ {\!\!Planck Collaboration Int. XLVIII} {et~al.}(2016){\!\!Planck Collaboration Int. XLVIII}, {Aghanim}, {Ashdown}, {Aumont}, {Baccigalupi}, {Ballardini}, {Banday}, {Barreiro}, {Bartolo}, {Basak}, {Benabed}, {Bernard}, {Bersanelli}, {Bielewicz}, {Bonavera}, {Bond}, {Borrill}, {Bouchet}, {Boulanger}, {Burigana}, {Calabrese}, {Cardoso}, {Carron}, {Chiang}, {Colombo}, {Comis}, {Couchot}, {Coulais}, {Crill}, {Curto}, {Cuttaia}, {de Bernardis}, {de Zotti}, {Delabrouille}, {Di Valentino}, {Dickinson}, {Diego}, {Dor{\'e}}, {Douspis}, {Ducout}, {Dupac}, {Dusini}, {Elsner}, {En{\ss}lin}, {Eriksen}, {Falgarone}, {Fantaye}, {Finelli}, {Forastieri}, {Frailis}, {Fraisse}, {Franceschi}, {Frolov}, {Galeotta}, {Galli}, {Ganga}, {G{\'e}nova-Santos}, {Gerbino}, {Ghosh}, {Giraud-H{\'e}raud}, {Gonz{\'a}lez-Nuevo}, {G{\'o}rski}, {Gruppuso}, {Gudmundsson}, {Hansen}, {Helou}, {Henrot-Versill{\'e}}, {Herranz}, {Hivon}, {Huang}, {Jaffe}, {Jones}, {Keih{\"a}nen}, {Keskitalo}, {Kiiveri}, {Kisner}, {Krachmalnicoff}, {Kunz},
  {Kurki-Suonio}, {Lamarre}, {Langer}, {Lasenby}, {Lattanzi}, {Lawrence}, {Le Jeune}, {Levrier}, {Lilje}, {Lilley}, {Lindholm}, {L{\'o}pez-Caniego}, {Ma}, {Mac{\'\i}as-P{\'e}rez}, {Maggio}, {Maino}, {Mandolesi}, {Mangilli}, {Maris}, {Martin}, {Mart{\'\i}nez-Gonz{\'a}lez}, {Matarrese}, {Mauri}, {McEwen}, {Melchiorri}, {Mennella}, {Migliaccio}, {Miville-Desch{\^e}nes}, {Molinari}, {Moneti}, {Montier}, {Morgante}, {Moss}, {Natoli}, {Oxborrow}, {Pagano}, {Paoletti}, {Patanchon}, {Perdereau}, {Perotto}, {Pettorino}, {Piacentini}, {Plaszczynski}, {Polastri}, {Polenta}, {Puget}, {Rachen}, {Racine}, {Reinecke}, {Remazeilles}, {Renzi}, {Rocha}, {Rosset}, {Rossetti}, {Roudier}, {Rubi{\~n}o-Mart{\'\i}n}, {Ruiz-Granados}, {Salvati}, {Sandri}, {Savelainen}, {Scott}, {Sirignano}, {Sirri}, {Soler}, {Spencer}, {Suur-Uski}, {Tauber}, {Tavagnacco}, {Tenti}, {Toffolatti}, {Tomasi}, {Tristram}, {Trombetti}, {Valiviita}, {Van Tent}, {Vielva}, {Villa}, {Vittorio}, {Wandelt}, {Wehus}, {Zacchei}, \&
  {Zonca}}]{Planck-Collaboration_2016}
{\!\!Planck Collaboration Int. XLVIII}, {Aghanim}, N., {Ashdown}, M., {et~al.} 2016, \bibinfo{title}{{Planck intermediate results. XLVIII. Disentangling Galactic dust emission and cosmic infrared background anisotropies},} \aap, 596, A109, \dodoi{10.1051/0004-6361/201629022}

\bibitem[{ {\!\!Planck Collaboration XII} {et~al.}(2020){\!\!Planck Collaboration XII}, {Aghanim}, {Akrami}, {Alves}, {Ashdown}, {Aumont}, {Baccigalupi}, {Ballardini}, {Banday}, {Barreiro}, {Bartolo}, {Basak}, {Benabed}, {Bernard}, {Bersanelli}, {Bielewicz}, {Bock}, {Bond}, {Borrill}, {Bouchet}, {Boulanger}, {Bracco}, {Bucher}, {Burigana}, {Calabrese}, {Cardoso}, {Carron}, {Chary}, {Chiang}, {Colombo}, {Combet}, {Crill}, {Cuttaia}, {de Bernardis}, {de Zotti}, {Delabrouille}, {Delouis}, {Di Valentino}, {Dickinson}, {Diego}, {Dor{\'e}}, {Douspis}, {Ducout}, {Dupac}, {Efstathiou}, {Elsner}, {En{\ss}lin}, {Eriksen}, {Falgarone}, {Fantaye}, {Fernandez-Cobos}, {Ferri{\`e}re}, {Finelli}, {Forastieri}, {Frailis}, {Fraisse}, {Franceschi}, {Frolov}, {Galeotta}, {Galli}, {Ganga}, {G{\'e}nova-Santos}, {Gerbino}, {Ghosh}, {Gonz{\'a}lez-Nuevo}, {G{\'o}rski}, {Gratton}, {Green}, {Gruppuso}, {Gudmundsson}, {Guillet}, {Handley}, {Hansen}, {Helou}, {Herranz}, {Hivon}, {Huang}, {Jaffe}, {Jones}, {Keih{\"a}nen}, {Keskitalo},
  {Kiiveri}, {Kim}, {Krachmalnicoff}, {Kunz}, {Kurki-Suonio}, {Lagache}, {Lamarre}, {Lasenby}, {Lattanzi}, {Lawrence}, {Le Jeune}, {Levrier}, {Liguori}, {Lilje}, {Lindholm}, {L{\'o}pez-Caniego}, {Lubin}, {Ma}, {Mac{\'\i}as-P{\'e}rez}, {Maggio}, {Maino}, {Mandolesi}, {Mangilli}, {Marcos-Caballero}, {Maris}, {Martin}, {Mart{\'\i}nez-Gonz{\'a}lez}, {Matarrese}, {Mauri}, {McEwen}, {Melchiorri}, {Mennella}, {Migliaccio}, {Miville-Desch{\^e}nes}, {Molinari}, {Moneti}, {Montier}, {Morgante}, {Moss}, {Natoli}, {Pagano}, {Paoletti}, {Patanchon}, {Perrotta}, {Pettorino}, {Piacentini}, {Polastri}, {Polenta}, {Puget}, {Rachen}, {Reinecke}, {Remazeilles}, {Renzi}, {Ristorcelli}, {Rocha}, {Rosset}, {Roudier}, {Rubi{\~n}o-Mart{\'\i}n}, {Ruiz-Granados}, {Salvati}, {Sandri}, {Savelainen}, {Scott}, {Sirignano}, {Sunyaev}, {Suur-Uski}, {Tauber}, {Tavagnacco}, {Tenti}, {Toffolatti}, {Tomasi}, {Trombetti}, {Valiviita}, {Vansyngel}, {Van Tent}, {Vielva}, {Villa}, {Vittorio}, {Wandelt}, {Wehus}, {Zacchei}, \&
  {Zonca}}]{Planck-Collaboration_2018_20}
{\!\!Planck Collaboration XII}, {Aghanim}, N., {Akrami}, Y., {et~al.} 2020, \bibinfo{title}{{Planck 2018 results. XII. Galactic astrophysics using polarized dust emission},} \aap, 641, A12, \dodoi{10.1051/0004-6361/201833885}

\bibitem[{S. {Plaszczynski} {et~al.}(2014){Plaszczynski}, {Montier}, {Levrier}, \& {Tristram}}]{Plaszczynski_2014}
{Plaszczynski}, S., {Montier}, L., {Levrier}, F., \& {Tristram}, M. 2014, \bibinfo{title}{{A novel estimator of the polarization amplitude from normally distributed Stokes parameters},} \mnras, 439, 4048, \dodoi{10.1093/mnras/stu270}

\bibitem[{E.~A. {Ram{\'\i}rez} {et~al.}(2017){Ram{\'\i}rez}, {Magalh{\~a}es}, {Davidson}, {Pereyra}, \& {Rubinho}}]{Ramirez_2017}
{Ram{\'\i}rez}, E.~A., {Magalh{\~a}es}, A.~M., {Davidson}, James~W., J., {Pereyra}, A., \& {Rubinho}, M. 2017, \bibinfo{title}{{Solvepol: A Reduction Pipeline for Imaging Polarimetry Data},} \pasp, 129, 055001, \dodoi{10.1088/1538-3873/aa54a7}

\bibitem[{G. {Schwarz}(1978){Schwarz}}]{Schwarz_1978}
{Schwarz}, G. 1978, \bibinfo{title}{{Estimating the Dimension of a Model},} Annals of Statistics, 6, 461

\bibitem[{K. {Serkowski}(1962){Serkowski}}]{Serkowski_1962}
{Serkowski}, K. 1962, \bibinfo{title}{{Polarization of Starlight},} Advances in Astronomy and Astrophysics, 1, 289, \dodoi{10.1016/B978-1-4831-9919-1.50009-1}

\bibitem[{K. {Serkowski} {et~al.}(1975){Serkowski}, {Mathewson}, \& {Ford}}]{Serkowski_1975}
{Serkowski}, K., {Mathewson}, D.~S., \& {Ford}, V.~L. 1975, \bibinfo{title}{{Wavelength dependence of interstellar polarization and ratio of total to selective extinction.},} \apj, 196, 261, \dodoi{10.1086/153410}

\bibitem[{R. {Skalidis} \& V. {Pelgrims}(2019){Skalidis} \& {Pelgrims}}]{Skalidis_Pelgrims_2019}
{Skalidis}, R., \& {Pelgrims}, V. 2019, \bibinfo{title}{{Local Bubble contribution to the 353-GHz dust polarized emission},} \aap, 631, L11, \dodoi{10.1051/0004-6361/201936547}

\bibitem[{J.~D. {Soler} {et~al.}(2016){Soler}, {Alves}, {Boulanger}, {Bracco}, {Falgarone}, {Franco}, {Guillet}, {Hennebelle}, {Levrier}, {Martin}, \& {Miville-Desch{\^e}nes}}]{Soler_2016}
{Soler}, J.~D., {Alves}, F., {Boulanger}, F., {et~al.} 2016, \bibinfo{title}{{Magnetic field morphology in nearby molecular clouds as revealed by starlight and submillimetre polarization},} \aap, 596, A93, \dodoi{10.1051/0004-6361/201628996}

\bibitem[{N. {Uppal} {et~al.}(2024){Uppal}, {Ganesh}, {Pelgrims}, {Joshi}, \& {Sarkar}}]{Uppal_2024}
{Uppal}, N., {Ganesh}, S., {Pelgrims}, V., {Joshi}, S., \& {Sarkar}, M. 2024, \bibinfo{title}{{Linear polarization study of open clusters in the anticenter direction: Signature of the spiral arms},} \aap, 690, A49, \dodoi{10.1051/0004-6361/202449537}

\bibitem[{B. {Uyaniker} {et~al.}(2003){Uyaniker}, {Landecker}, {Gray}, \& {Kothes}}]{Uyaniker_2003}
{Uyaniker}, B., {Landecker}, T.~L., {Gray}, A.~D., \& {Kothes}, R. 2003, \bibinfo{title}{{Radio Polarization from the Galactic Plane in Cygnus},} \apj, 585, 785, \dodoi{10.1086/346234}

\bibitem[{J.~L. {Vergely} {et~al.}(2022){Vergely}, {Lallement}, \& {Cox}}]{Vergely_2022}
{Vergely}, J.~L., {Lallement}, R., \& {Cox}, N.~L.~J. 2022, \bibinfo{title}{{Three-dimensional extinction maps: Inverting inter-calibrated extinction catalogues},} \aap, 664, A174, \dodoi{10.1051/0004-6361/202243319}

\bibitem[{M.~J.~F. {Versteeg} {et~al.}(2024){Versteeg}, {Angarita}, {Magalh{\~a}es}, {Haverkorn}, {Rodrigues}, {Santos-Lima}, \& {Kawabata}}]{Versteeg_2024}
{Versteeg}, M.~J.~F., {Angarita}, Y., {Magalh{\~a}es}, A.~M., {et~al.} 2024, \bibinfo{title}{{Magnetic Fields in the Southern Coalsack and Beyond},} \aj, 167, 177, \dodoi{10.3847/1538-3881/ad2e08}

\bibitem[{M.~J.~F. {Versteeg} {et~al.}(2023){Versteeg}, {Magalh{\~a}es}, {Haverkorn}, {Angarita}, {Rodrigues}, {Santos-Lima}, \& {Kawabata}}]{Versteeg_2023}
{Versteeg}, M.~J.~F., {Magalh{\~a}es}, A.~M., {Haverkorn}, M., {et~al.} 2023, \bibinfo{title}{{Interstellar Polarization Survey. II. General Interstellar Medium},} \aj, 165, 87, \dodoi{10.3847/1538-3881/aca8fd}

\bibitem[{P. {Virtanen} {et~al.}(2020){Virtanen}, Gommers, Oliphant, Haberland, Reddy, Cournapeau, Burovski, Peterson, Weckesser, Bright, {van der Walt}, Brett, Wilson, Millman, Mayorov, Nelson, Jones, Kern, Larson, Carey, Polat, Feng, Moore, {VanderPlas}, Laxalde, Perktold, Cimrman, Henriksen, Quintero, Harris, Archibald, Ribeiro, Pedregosa, {van Mulbregt}, \& {SciPy 1.0 Contributors}}]{Virtanen_SciPy_2020}
{Virtanen}, P., Gommers, R., Oliphant, T.~E., {et~al.} 2020, \bibinfo{title}{{{SciPy} 1.0: Fundamental Algorithms for Scientific Computing in Python},} Nature Methods, 17, 261, \dodoi{10.1038/s41592-019-0686-2}

\bibitem[{D.~C.~B. {Whittet} {et~al.}(2001){Whittet}, {Gerakines}, {Hough}, \& {Shenoy}}]{Whittet_2001}
{Whittet}, D.~C.~B., {Gerakines}, P.~A., {Hough}, J.~H., \& {Shenoy}, S.~S. 2001, \bibinfo{title}{{Interstellar Extinction and Polarization in the Taurus Dark Clouds: The Optical Properties of Dust near the Diffuse/Dense Cloud Interface},} \apj, 547, 872, \dodoi{10.1086/318421}

\bibitem[{D.~C.~B. {Whittet} {et~al.}(1992){Whittet}, {Martin}, {Hough}, {Rouse}, {Bailey}, \& {Axon}}]{Whittet_1992}
{Whittet}, D.~C.~B., {Martin}, P.~G., {Hough}, J.~H., {et~al.} 1992, \bibinfo{title}{{Systematic Variations in the Wavelength Dependence of Interstellar Linear Polarization},} \apj, 386, 562, \dodoi{10.1086/171039}

\end{thebibliography}
\bibliographystyle{aasjournalv7}



\end{document}